\RequirePackage{fix-cm}

\documentclass[natbib,smallcondensed]{svjour3}       % onecolumn (second format)
\journalname{Springer book collection}
\smartqed  

\usepackage{graphicx}
\usepackage{epsfig}
\usepackage{amsmath}
\usepackage{amssymb}
\usepackage{natbib}
\usepackage{threeparttable} 
\usepackage{xspace}
\usepackage{color}
\usepackage{epstopdf}
\usepackage{gensymb}

\newcommand{\astrosat}{\textit{ASTROSAT}\xspace}
\newcommand{\athena}{\textit{Athena}\xspace}

\newcommand{\chan}{\textit{Chandra}\xspace}

\newcommand{\extp}{\textit{eXTP}\xspace}

\newcommand{\hitomi}{\textit{Hitomi}\xspace}
\newcommand{\hxmt}{\textit{HXMT}\xspace}

\newcommand{\nicer}{\textit{NICER}\xspace}

\newcommand{\rxte}{\textit{RXTE}\xspace}

\newcommand{\xmm}{\textit{XMM-Newton}\xspace}

\newcommand{\gaia}{\textit{Gaia}\xspace}
\newcommand{\xipe}{\textit{XIPE}\xspace}
\newcommand{\ixpe}{\textit{IXPE}\xspace}
\newcommand{\strobex}{\textit{STROBE-X}\xspace}
\newcommand{\eelt}{\textit{E-ELT}\xspace}
\newcommand{\tmt}{\textit{TMT}\xspace}
\newcommand{\lsst}{\textit{LSST}\xspace}
\newcommand{\xarm}{\textit{XARM}\xspace}
\newcommand{\ska}{\textit{SKA}\xspace}

\newcommand{\Msun}{\mathrm{M}_{\odot}}
\newcommand{\mdot}{\mathrm{M_{\odot}~yr}^{-1}}
\newcommand{\mdotgs}{\mathrm{g~s}^{-1}}

\newcommand{\flux}{\mathrm{erg~cm}^{-2}~\mathrm{s}^{-1}}

\newcommand{\mdens}{\mathrm{g~cm}^{-3}}

\renewcommand{\deg}{\ensuremath{^{\circ}}\xspace}

\newcommand{\mnras}{{MNRAS}}
\newcommand{\apj}{{ApJ}}
\newcommand{\apjs}{{ApJS}}
\newcommand{\apss}{{Ap\&SS}}
\newcommand{\apjl}{{ApJL}}
\newcommand{\aap}{{A\&A}}
\newcommand{\nat}{{Nature}}
\newcommand{\araa}{{ARAA}}

\newcommand{\aapr}{{A\&A Rev.}}
\newcommand{\prl}{{PRL}}
\newcommand{\prc}{{PhRvC}}
\newcommand{\prd}{{PhRvD}}

\newcommand{\pasj}{{PASJ}}
\newcommand{\ssr}{{Space Sci. Rev.}}
\newcommand{\procspie}{{Proc. SPIE}}
\newcommand{\physrep}{{Phys. Rev.}}

\newcommand{\be}{\begin{equation}}
\newcommand{\ee}{\end{equation}}
\newcommand{\bea}{\begin{eqnarray}}
\newcommand{\eea}{\end{eqnarray}}
\newcommand{\fc}{f_{\rm c}}

\newcommand{\kappae}{\kappa_{\rm e}}

\begin{document}

\title{Testing the equation of state of neutron stars with electromagnetic observations}

\author{N. Degenaar        
	\and
        V.F. Suleimanov 
}

\institute{N. Degenaar \at
Anton Pannekoek Institute, University of Amsterdam, Science Park 904, 1098 XH, Amsterdam, the Netherlands\\
\email{degenaar@uva.nl}        
		\and
           V.F. Suleimanov \at
Institut f\"{u}r Astronomie und Astrophysik, Kepler Center for Astro and Particle Physics, Universit\"{a}t T\"{u}bingen, Sand 1, D-72076 T\"{u}bingen, Germany\\
%\and
Kazan (Volga region) Federal University, Kremlevskaya str. 18, Kazan 420008, Russia\\
\email{suleimanov@astro.uni-tuebingen.de}  
}

\date{To appear as a chapter in the White Book of ``NewCompStar'' European COST Action MP1304}

\maketitle

\begin{abstract}
Neutron stars are the densest, directly observable stellar objects in the universe and serve as unique astrophysical laboratories to study the behavior of matter under extreme physical conditions. This book chapter is devoted to describing how electromagnetic observations, particularly at X-ray, optical and radio wavelengths, can be used to measure the mass and radius of neutron stars and how this leads to constraints on the equation of state of ultra-dense matter. Having accurate theoretical models to describe the astrophysical data is essential in this effort. We will review different methods to constrain neutron star masses and radii, discuss the main observational results and theoretical developments achieved over the past decade, and provide an outlook of how further progress can be made with new and upcoming ground-based and space-based observatories.

\end{abstract}

%%%%%%%%%%%%%%%%%%%%%%%%%
% ArXiv note 
%%%%%%%%%%%%%%%%%%%%%%%%%%
%\section*{Note accompanying this arXiv publication}\label{sec:arxiv}
%The review presented here is an adapted version of a book chapter that will appear in Springer book collection on Astronomy and Astrophysics Library. 

%\newpage
\tableofcontents

%%%%%%%%%%%%%%%%%%%%%%%%%
% INTRODUCTION 
%%%%%%%%%%%%%%%%%%%%%%%%%%
\newpage
\section{Introduction to this book chapter}\label{sec:thisbook}
It is probably not an exaggeration to state that there are more reviews and book chapters written on the equation of state of neutron stars than there are actual constraining measurements \citep[e.g.][to name a few in just the past 5 years]{lattimer2012,lattimer2016,ozel2013,ozel2016_review,watts2015_review,watts2016_review,sb16,miller2016_MRreview}. This is not surprising, however, as it merely underlines the immense interest in understanding the physics of ultra-dense matter that is not encountered on Earth. Such constraints can be uniquely obtained through electromagnetic observations of neutron stars, and this is therefore one of the prime pursuits in modern astrophysics. 

Everything around us is constructed of atoms, which themselves consist of electrons and nucleons (i.e. protons and neutrons). This familiar structure of matter is, however, disrupted when matter is compressed to densities that reach beyond the {\it nuclear saturation density} of $\rho_0 \approx 2.8\times 10^{14}~\mdens$. One of the main open questions in modern physics is how matter behaves at supra-nuclear densities, where particle interactions are governed by the strong force. For instance, does matter remain nucleonic or does it rather transition into more exotic forms of particles? What are the superfluid properties of matter compressed beyond the nuclear density? The detailed microphysics and particle interactions of dense matter result in a specific relation between the pressure and density of the bulk matter, which is called the {\it equation of state} (EOS). Elucidating the behavior of matter at high densities thus implies understanding the dense-matter EOS \citep[e.g.][]{lattimer2012}. 

Mathematically, it is inherently difficult to describe multiple-particle interactions at high densities and therefore there are no unique theoretical predictions for the dense-matter EOS. Fortunately, the behavior of matter at supra-nuclear densities can be probed through different kinds of experiments. In Earth-based laboratories, particle acceleration experiments are conducted to probe matter near the nuclear saturation density at very high temperatures. However, probing the properties of matter beyond the nuclear density at (relatively) low temperatures solely relies on astrophysical observations of neutron stars.

Neutron stars are the remnants of once massive stars that ended their life in a supernova explosion. A defining property of neutron stars is that these objects are very compact; while being roughly a factor of 1.5 more massive than our Sun, their radius is almost a factor of $\sim 10^5$ smaller \citep[$\approx 10$~km;][]{baade1934,wheeler1966,hewish1968}. This extreme compactness implies that the density of neutron stars is incredibly high and must, in fact, reach beyond the nuclear density. Being the densest, directly observable stellar objects in our universe, neutron stars thus serve as natural laboratories to study the behavior of matter at supra-nuclear densities. Neutron stars come in different classes, which provides various angles to test and constrain the dense-matter EOS (Section~\ref{subsec:classes}). 

Here, we aim to provide a student-level introduction into the many different ways in which we can obtain constraints on the neutron star EOS using observations of their electromagnetic radiation. This book chapter is organized as follows: after introducing some general concepts and background context in Section~\ref{sec:intro}, we review the various ways in which the dense-matter EOS of neutron stars can be constrained from electromagnetic observations in Section~\ref{sec:past}, including a discussion on the uncertainties and systematic biases that the different techniques are subject to. In Section~\ref{sec:future} we lay out what progress can be made in the future using new and upcoming facilities. A summary is provided in Section~\ref{sec:conclusions}.

\section{Basic concepts: Neutron stars and the dense matter equation of state}\label{sec:intro}

\subsection{The plethora of observable neutron stars}\label{subsec:classes}
As is clear from the diverse content of this book, there are several different observational manifestations of neutron stars \citep[e.g.][for a review]{kaspi2010}. At the very basic level, we can distinguish neutron stars that are isolated and those that are part of a binary star system. Furthermore, neutron stars are often characterized by their magnetic field strength ($B$) and rotation period ($P_s$), which is often connected to their age and their environment. 

Many neutron stars are located in a binary where they are accompanied by a main sequence or evolved star with a mass $M_{\rm c}\lesssim 1~\Msun$. In such a configuration, neutron stars can manifest themselves as a \textit{low-mass X-ray binary} (LMXB), when accreting gas from their companion star, or as a non-accreting radio pulsar. In the former class of objects, the energy emitted by the neutron star (at X-ray wavelengths) is powered by the accretion process, whereas in the latter the rotational energy of the pulsar is tapped (producing mostly radio emission). Both types of neutron stars are typically spinning rapidly, at millisecond periods, and are assumed to be spun up to such high speeds by gaining angular momentum via accretion \citep[e.g.][]{CST94,strohmayer1996,wijnands1998,burderi1999,archibald2009,papitto2013}. These two classes of binary neutron stars are thus thought to be evolutionary linked, with millisecond radio pulsars descending from LMXBs \citep[][]{alpar1982,bhattacharya1991}. The magnetic field of these neutron stars is typically low ($B \lesssim 10^9$~G), and is thought to have degraded by accretion \citep[e.g.][]{radhakrishnan1982,romani1990,bhattacharya1995,cumming2001}. Nevertheless, some neutron stars in LMXBs display coherent X-ray pulsations, which indicates that their magnetic field is strong enough to channel the accreted gas towards the magnetic poles \citep[][]{wijnands1998}. This sub-class of pulsating LMXBs, of which currently 20 are known, are referred to as \textit{accreting millisecond X-ray pulsars} (AMXPs; see \cite{patruno2017_spin} for a list of the first 19 and \cite{sanna2017_amxp} for the most recent discovery).

Whereas accretion in LMXBs typically proceeds via an accretion disk that is fed by the Roche-lobe overflowing donor star, a modest number of neutron stars is known to accrete from the stellar wind of their low-mass companion star \citep[Symbiotic X-ray binaries, SyXRBs; e.g.][]{chakrabarty1997,masetti2006,lu2012}. Neutron stars are also found to accrete from much more massive companions ($M_{\rm c}\gtrsim 10~\Msun$), and are then referred to as high-mass X-ray binaries (HMXBs). Depending on the type of the companion and the particular process of mass transfer, neutron star HMXBs can be further divided into different sub-classes \citep[e.g.][for a review]{reig2011}. A small number of intermediate-mass X-ray binaries (IMXBs, companion mass $M_{\rm c}\sim 1-10~\Msun$) are also known, but these are thought to be short-lived and quickly evolve into LMXBs \citep[e.g.][]{vandenheuvel1975,podsiadlowski2002,pfahl2003,tauris2006}. The neutron stars in SyXRBs, HMXBs and IMXBs are typically spinning much more slowly (seconds) than those in LMXBs and also have stronger magnetic fields ($B \sim 10^{11}-10^{13}$~G). A few slowly spinning radio pulsars with massive companion stars are known; these apparently have not been spun up yet by accretion (possibly due to their highly eccentric orbits) and might be the progenitors of HXMBs or IMXBs \citep[e.g.][]{johnston1992,kaspi1994,stairs2001}. 

Isolated neutron stars also come in various flavors \citep[e.g.][for a recent review]{kaspi2016}. For instance, one can distinguish young slowly spinning radio pulsars (including Rotating RAdio Transients, RRATs) and old millisecond radio pulsars (which likely originate from binaries but ablated their companion star). Furthermore, there are young and X-ray emitting (sometimes pulsating) neutron stars in the centers of supernova remnants (central compact objects, CCOs), nearby Dim Isolated Neutron Stars (DINS), and magnetars (originally further classified as Anomalous X-ray Pulsars, AXPs, and Soft Gamma Repeaters, SGRs). Several of these different classes of isolated neutron stars may be linked, possibly representing different stages in an overall thermal and magnetic field evolution \citep[e.g.][]{faucher2006,vigano2013,MPM15}.

\begin{figure}
\centering
\includegraphics[width=0.99\textwidth]{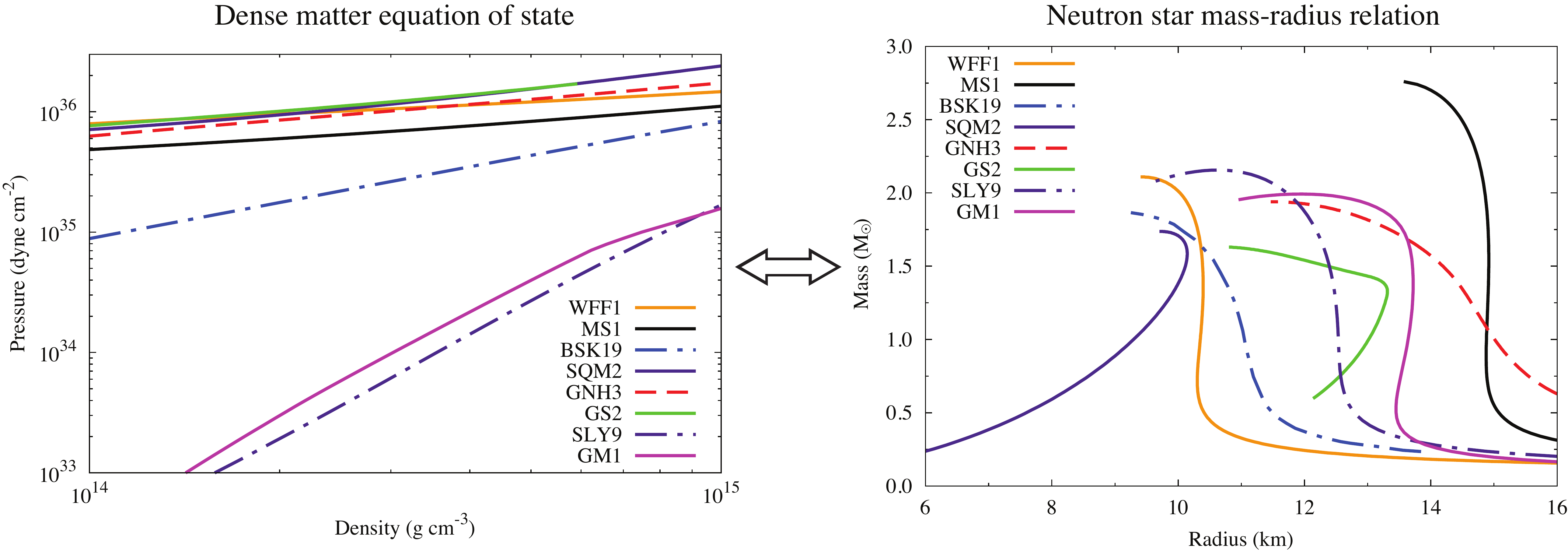}
\caption{Direct mapping between the EOS (left) and mass-radius relation for neutron stars (right). Shown are illustrative examples of different EOSs. Data for the following EOSs were taken from the compilation of \citet{ozel2016}: WFF1=\citet{wiringa1988}, MS1=\citet{muller1996}, BSK19=\citet{potekhin2013}, SQM2=\citet{prakash1995}, GNH3=\citet{glendenning1985}, and GS2=\citet{glendenning1999}. Added to this are two unified EOSs, SLY9 and GM1, from \citet{fortin2016}.%\footnote{http://xtreme.as.arizona.edu/NeutronStars}
%MPA1=\citet{muther1987},
%H4=\citet{lackey2006},
% GS2 = kaon condensates
% SQM2 = quark matter eos
} 
\label{fig:eosmr}
\end{figure}

%%%%%%%%%%
%% introduction to EOS and connection with neutron stars
%%%%%%%%%%%%%%%%%%%%%%%
\subsection{The dense-matter EOS and the connection with neutron stars}\label{subsec:eosmr}
The macroscopic properties of neutron stars, such as their maximum mass and corresponding radius, are determined by the details of the dense-matter EOS. Due to the complexity of the calculations and the inability to probe high densities via laboratory experiments, numerous EOSs with a wide range of different parameters have been constructed \citep[e.g.][see also Figure~\ref{fig:eosmr} left]{lattimer2012}. 

For each theoretical model describing the microphysics and strong-force interactions, i.e. for a given EOS of the bulk matter, the general relativistic structure equations (the Tolman-Oppenheimer-Volkoff equations) can be solved for an assumed central density, leading to a predicted mass ($M$) and radius ($R$) for the neutron star. Using a range of different central densities, the $M-R$ relation can be constructed for any given EOS, which is illustrated in Figure~\ref{fig:eosmr} (right). This unique mapping between the EOS and the basic properties of neutron stars implies that measuring a wide range of mass-radius pairs, with an accuracy of a few percent, can constrain the properties of ultra-dense matter through revise-engineering \citep[][]{lindblom1992}. However, it is very challenging to determine both the mass and radius (for a sizable sample of) neutron stars with high accuracy. Some of the most constraining results have therefore instead come from measuring extrema. 

For each EOS there is a maximum central density beyond which no stable configuration is possible, hence every EOS is characterized by a maximum neutron star mass (the Tolman-Oppenheimer-Volkoff limit). This is illustrated by Figure~\ref{fig:eosmr}, which displays the $M-R$ relation (right) for a selection of different EOSs (left). The EOS is effectively a measure for the compressibility of the matter, which is also referred to as its \textit{softness}: The shallower the rise in pressure with increasing density, the more compressible the matter is, i.e. the softer the EOS. Since for soft EOSs the matter is more compressible, these generally predict neutron stars with smaller maximum masses than harder EOSs. 

Looking at Figure~\ref{fig:eosmr} (right), we can see that finding a neutron star with a high mass of $M\gtrsim 2~\Msun$ can put tight constraints on the EOS by eliminating entire families of soft EOS models. On the other hand, for many EOSs there is a regime in which the radius remains relatively constant over a fairly large range of mass (see Figure~\ref{fig:eosmr} right), which implies that an accurate radius measurement can also provide important constraints. If in addition to the mass or the radius also the \textit{compactness} (i.e. the ratio of mass and radius) can be determined, degeneracies can be broken so that valuable constraints on the EOS can be obtained. The quest of probing the dense-matter EOS through neutron stars can be approached in several different ways, exploiting electromagnetic observations of both isolated neutron stars and neutrons stars that are located in binaries (Section~\ref{sec:past}).

\begin{figure}
\centering
\includegraphics[width=0.95\textwidth]{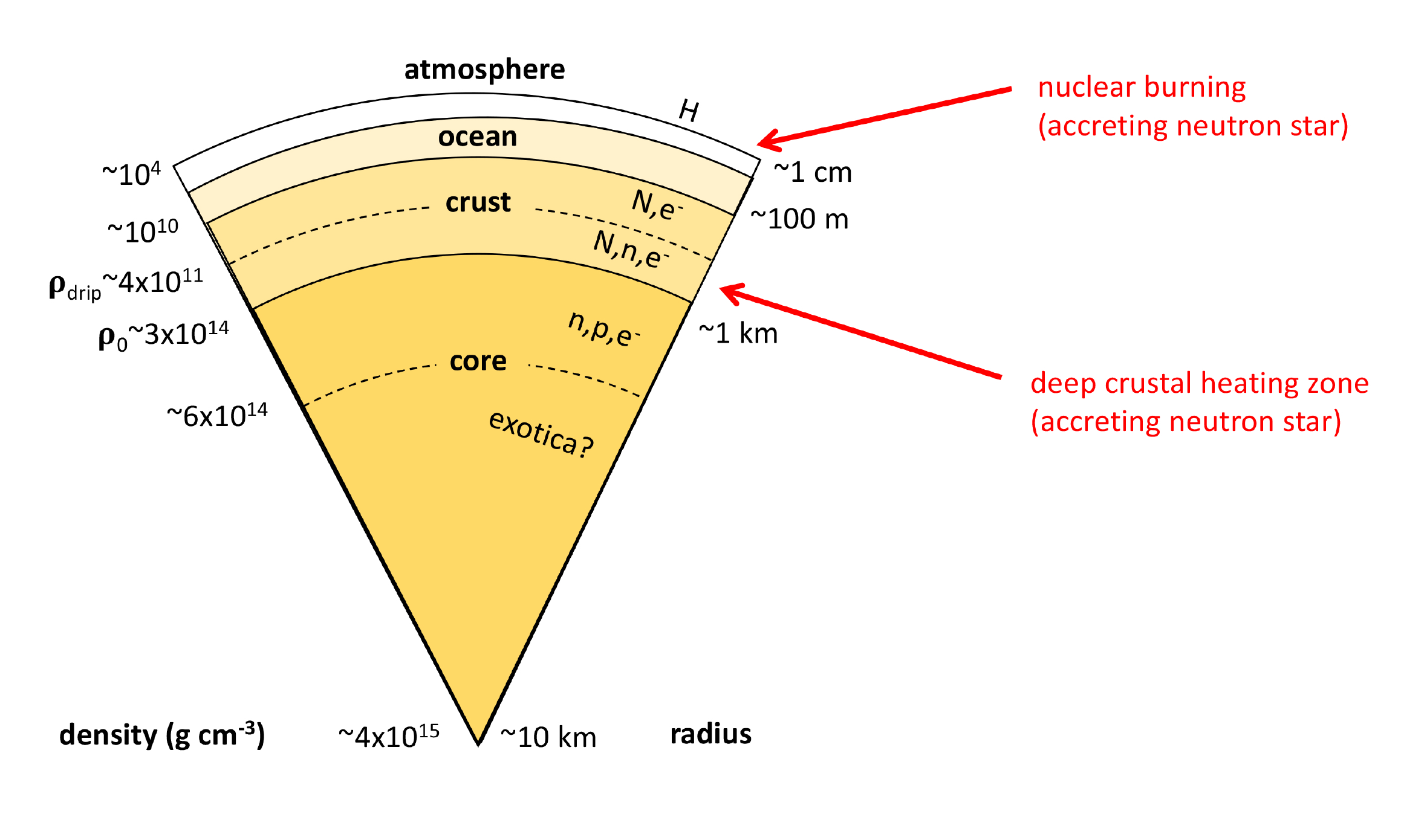}
\caption{
Schematic drawing of the structure of a neutron star (not to scale). Some indicative numbers for the size and density are given, and the main particle constituents are indicated. $H$ stands for hydrogen, $N$ for nuclei, $n$ for neutrons, $e^{-}$ for electrons and $p$ for protons. The dashed lines indicate dividing lines between the inner/outer crust and core. When a neutron star is accreting, nuclear burning (of the accreted gas) occurs on the surface of the neutron star and the bottom of the crust is severely heated due to density-driven fusion reactions. These zones are indicated with red text and arrows. 
} 
\label{fig:structure}
\end{figure}

\subsection{General structure of a neutron star}\label{subsec:structure}
Figure~\ref{fig:structure} shows a schematic drawing of the general structure of a neutron star. The density and pressure rise with increasing depth. The dense, liquid core lies beneath the solid crust, which is covered by a thin ocean/envelop and a very thin atmosphere. When a neutron star accretes, its structure and composition change (see Section~\ref{subsec:accretion}). Below we describe the structure in a bit more detail.

The \textit{core} makes up the largest part of the neutron star, containing approximately 99\% of the total mass, and may be subdivided into an outer and an inner part. The outer core occupies the density range $\rho \sim (1-2)\rho_0$, where matter consists mainly of degenerate neutrons and merely a few percent of protons and electrons. Both protons and neutrons are expected to be superfluid in the outer core. In the inner core of the neutron star, the density may become as high as $\rho \sim (10 - 15) \rho_0$, depending on the EOS model. Due to the growing Fermi energies it may become energetically favorable for more exotic particles, rather than the standard composition of $p$, $e^-$ and $n$, to occur at these high densities. For instance, neutrons may be replaced by hyperons, 
%\citep[e.g.][]{bethe1974}
electrons may be replaced by pions or kaons and form a (superfluid) Boson-Einstein condensate, 
% \citep[e.g.][]{sawyer1972,pandharipande1995}
and perhaps the density becomes even so high that the attractive force between quarks can be neglected so that the quarks become unconfined. The occurrence of exotica in the core leads to a softer EOS, since it relaxes the Fermi surface and degeneracy pressure.

The \textit{crust} typically covers about one tenth of the neutron star radius and can be subdivided into an inner and an outer part. The outer crust extends from the bottom of the atmosphere to the \textit{neutron drip density}, $\rho_{\mathrm{drip}} \approx 4.3 \times 10^{11}~\mdens$, and matter consists of electrons and ions. Due to the rise in electron Fermi energy, the nuclei suffer inverse $\beta$-decay and become more neutron-rich with increasing density. The inner crust covers the region from the neutron drip density, where neutrons start to drip out of the nuclei, to the nuclear density and is composed of electrons, free superfluid neutrons and neutron-rich nuclei. With increasing density the nuclei grow heavier and the number of neutrons residing in the free neutron fluid (rather than in nuclei) increases. Nuclei begin to dissolve and merge together around the crust-core interface. 

The \textit{envelop or ocean} refers to a $\sim$100-m thick layer that lies on top of the crust and extends to a density of $\rho \sim 10^{10}~\mdens$. If one wants to study the interior temperature of neutron stars, which gives additional information on the EOS (see Section~\ref{subsec:other}), it is essential to understand the properties of this layer: The envelop/ocean couples the sought-after temperature of the (isothermal) interior, $T_{\mathrm{B}}$, to the observable effective surface temperature, $T_{\mathrm{eff}}$ \citep[e.g.][]{gudmundsson1982,gudmundsson1983,PCY97,heyl1998,geppert2004}, which is then further modified by the atmosphere. 
%gudmundsson1982,gudmundsson1983,PCY97 for envelop composition effects
%heyl1998,geppert2004 for B-field effects

The \textit{atmosphere} covers the neutron star and is very thin ($\sim$ cm). It is expected to be composed of pure H, since heavier elements should sink on very short timescales \citep[e.g.][]{romani1987,bildsten1992,ozel2012}. However, for some isolated or quiescent neutron stars, observations suggest atmospheres composed of He or C (see Sections~\ref{subsubsec:qatmosphere} and~\ref{subsubsec:atmoscco}). Moreover, when a neutron star is actively accreting, the atmosphere composition should be more complex (see Section~\ref{subsubsec:atmosburst}). The atmosphere accounts for a negligible fraction of the total mass, but shapes the thermal photon spectrum emerging from the neutron star. It is therefore a crucial ingredient in modeling their surface emission, which provides means to constraint their mass and radius (Section~\ref{sec:surface}). 

Due to the very steep temperature gradient in the atmosphere of a neutron star, its spectrum looks very different from a black body. Since the atmosphere consists of ions and free electrons, the opacity will be dominated by free-free absorption (absorption of a photon by a free electron in the Coulomb field of an ion) and is therefore proportional to $\nu^{-3}$, where $\nu$ is the photon frequency. As a result of this strong dependence and the steep temperature gradient, high-energetic photons escape from deeper atmospheric layers (where the temperature is much higher), than low-energetic photons \citep[e.g.][]{rajagopalromani1996,Zavlin.etal:96}. For a particular temperature, the spectrum  from a neutron star is thus expected to be harder than a pure black body spectrum. One important implication is that fitting black body models to such modified spectra results in an overestimation of the effective temperature and hence the size of the emission region is underestimated \citep[e.g.][]{rutledge1999}. Other parameters that shape the emerging spectrum are the chemical composition of the atmosphere, the surface magnetic field and the surface gravity \citep[see e.g.][and references therein]{Zavlin.etal:96,ho2008,Ho:Heinke:09,servillat2012,potekhin2016}.

\subsection{The thermal evolution of neutron stars: Cooling and re-heating}\label{subsec:thermalevo}
Neutron stars are born in a supernova explosions, exhibiting very high temperatures of $T_{\rm B}\sim 10^{12}$~K. In absence of an internal furnace, however, the neutron star will rapidly cool via neutrino emissions from its dense core and photon emissions from its surface \citep[e.g.][for reviews]{YP04,page2006}. Initially the core cools faster and is thermally decoupled from the hotter crust, but after $\sim$1--100~yr (depending on the neutron star structure), the crust has thermally relaxed. The crust and core of the neutron star will then be in thermal equilibrium, having a uniform temperature \citep[e.g.][]{lattimer1994}. In the ocean/envelop and atmosphere, there always continues to be a steep temperature gradient. 

There are two factors that can alter the thermal evolution of a neutron star. Firstly, when a neutron star accretes gas from its surroundings, its composition changes and provides a site for nuclear reactions that release energy and (temporarily) disrupt the thermal balance between the core and the crust (see Section~\ref{subsec:accretion}). Secondly, in extreme cases the neutron star magnetic field can be a powerful source of energy that can alter the thermal balance. For instance, the decay of their magnetic field and fracturing of the crust by magnetic stresses can release significant amounts of energy that heat the outer layers of the neutron star \citep[e.g.][]{arras2004,pons2007,Pons.etal:09,aguilera2008,cooper2010}. These processes are effective only for very strong magnetic fields ($B\gtrsim 10^{14}$~G), i.e. only relevant for magnetars. Indeed, the temperature of several magnetars have been observed to change in response to periods of magnetic activity (see Section~\ref{subsec:other}).

\subsection{Accretion: Effect on the composition and thermal structure of neutron stars}\label{subsec:accretion}
When located in a binary, a neutron star may be able to accrete matter from a companion star. The transferred matter typically consists of H (or He), but as this gas accumulates on the surface of the neutron star, thermonuclear nuclear burning will transform it into heavier elements \citep[e.g.][]{schatz1999}. Under the weight of freshly accreted material, the burning ashes get pushed into the neutron star and eventually the original crust is fully replaced by an accreted one \citep[e.g.][]{sato1979,haensel1990_eos}. The temperatures involved are typically much lower ($T_{\rm B}\sim 10^7-10^8$~K) than in a newborn neutron star ($T_{\rm B}\sim 10^{12}$~K), and do not allow to overcome the nucleon Coulomb-barrier so  thermonuclear fusion reactions cannot take place. As a result, an accreted crust is constructed of smaller nuclei and a larger number of free neutrons than a non-accreted crust \citep[e.g.][]{haensel1990_eos}. 
Due to the larger number of free neutrons, at a given density (at $\rho> \rho_{\rm drip}= 4.3 \times 10^{11}~\mdens$) the pressure in an accreted crust will be higher than for a crust made of cold catalyzed matter. This implies that the EOS of the inner crust will be harder when a neutron star has experienced accretion, i.e. the crust will be slightly thicker \citep[e.g.][]{zdunik2017}. 

A non-equilibrium crust provides a site for nuclear reactions and hence a potential source of energy. An accreted matter element passes through a chain of non-equilibrium nuclear reactions as it is pushed deeper within the neutron star crust. This results in the release of a considerable amount of heat energy ($\approx 2$~MeV per accreted baryon). Most of this energy is released in the pycnonuclear reaction chains that occur deep in the inner crust at densities of $\rho \sim 10^{12}-10^{13}~\mdens$ (see Figure~\ref{fig:structure}). The total energy deposited in the crustal reactions is similar for different initial compositions of the burning ashes \citep[][]{haensel2003}.

The nuclear reactions induced in the crust due to accretion can reheat a neutron star and make it much hotter than an isolated neutron star of the same age. Indeed, \citet{brown1998} showed that deep crustal heating can efficiently maintain the core of an accreting neutron star at a temperature $T_{\rm B}\sim 5\times10^7-10^8$~K. The  temperature of neutron stars in LMXBs can be probed during periods of quiescence during which little or no accretion takes place \citep[see Section~\ref{subsec:thermal}; e.g.][]{brown1998,campana1998,wijnands2013,wijnands2017}.

%%%%%%%%%%%%%%%%%%%%%%%%%%
%%%%% SURFACE EMISSION
\subsection{Surface emission from neutron stars}\label{sec:surface}
Several techniques that are employed to constrain the neutron star EOS rely on detecting radiation directly from the stellar surface. The thermal radiation emitted by a stellar body depends on its temperature and radius. Therefore, if thermal radiation can be detected from (part of) the surface of a neutron star, this provides means to measure its radius. However, as gravitational effects come into play, the mass of the neutron star also enters the equations. Whereas in principle this complicates a simple measurement of the stellar radius, there may be ways to break degeneracies so that, in fact, both the mass and the radius can be determined.

The surface of a neutron star may not necessarily be emitting isotropically. Under certain circumstances, \textit{hotspots} may be observable: regions that are hotter than the bulk of the surface and are offset from the rotational pole of the neutron star. As the hotter region rotates around the star, the observer will see its emission modulated  (i.e. pulsed) at or near the stellar spin frequency. As we will see in this book chapter, both isotropic surface emission and hotspot emission can be employed to put constraints on the neutron star radius (and mass).

Typically, neutron stars have a temperature of $T_{\rm B}\sim 10^5$~K (for radio pulsars) to $T_{\rm B}\sim 10^7$~K (for LMXBs), which implies that the thermal surface radiation is emitted in the X-ray, (extreme) UV, or optical band. It is not obvious, however, that surface emission from a neutron star is detectable. In isolated neutron stars, for instance, the surface emission can be overpowered by non-thermal magnetospheric processes and in an X-ray binaries the gravitational energy released in the accretion process usually overwhelms the surface radiation from the neutron star. Fortunately there are situations in which thermal emission from the neutron star surface is directly observable. Firstly, during thermonuclear X-ray bursts, the neutron star surface briefly becomes brighter than the accretion flow due to the enormous amount of thermal energy liberated (Section~\ref{subsec:bursts}). Secondly, many neutron star LMXBs are transient and the neutron star surface may be visible during their quiescent episodes (Section~\ref{subsec:thermal}). Thirdly, in AMXPs the accretion flow is concentrated onto the magnetic poles, which creates observable hotspots (Section~\ref{subsec:timing}). Finally, thermal surface emission has also been detected for a number of (nearby) radio pulsars (Section~\ref{subsec:timing}), isolated neutron stars (Section~\ref{subsec:cco}), and magnetars (Section~\ref{subsubsec:magnetars}).

If the distance, $D$, is well known, measuring the observed thermal flux density of the neutron star, $F_{\nu}$, can be coupled to its intrinsic flux density ${\cal{F}}_{\nu (1+z)}$ via:
\be \label{norm} 
    F_\nu = {\cal{F}}_{\nu,\infty} \, K = {\cal{F}}_{\nu,\infty} \frac{R_\infty^2}{D^2} = \frac{{\cal{F}}_{\nu(1+z)}}{1+z} \frac{R^2}{D^2},
\ee 

where $K=(R_\infty/D)^2$ is a normalization factor. Due to their extreme compactness, neutron stars gravitationally bend their own surface emission \citep[e.g.][]{PFC83,psaltis2000}. Moreover, the emission is gravitationally lensed. The combined effect causes the observed angular size $R_\infty$ to be related to the true physical radius of the neutron star as $R_\infty=R(1+z)$, where $z$ is the gravitational redshift defined as:
\begin{equation} \label{eq:redshift_def}
    1+z= \left (1-\frac{2GM}{c^2R}  \right )^{-1/2}, 
\end{equation}

with $G$ the gravitational constant and $c$ the speed of light. The flux density seen by an observer in infinity is thus related to the intrinsic flux density as:
\be \label{spinf}
{\cal{F}}_{\nu,\infty} = \frac{{\cal{F}}_{\nu(1+z)}}{(1+z)^{3}}.
\ee

Since $z$ is significant for neutron stars, the stellar radius cannot be determined independently of its mass such that the outcome of the measurement is actually a strip along the mass-radius plane. Formally, further corrections need to be applied to account for the fast spin rates of neutron stars (see Section~\ref{subsec:timing}). For instance, due to Doppler boosting the approaching side of the neutron star will be brighter than the receding side. This asymmetry is further enhanced by frame dragging, as the space time differs from the simple Schwarzschild metric. Rapid rotation also causes oblateness (i.e. the radius at the equator is larger than at the poles). The combined effect implies that the surface emission is non-uniform \citep[e.g.][]{miller1998_2,braje2000,muno2003,PG03,cadeau2007,MLCB07}, and that any possible atomic lines from the atmosphere will be broadened \citep[][]{ozelpsaltis2003,bhattacharyya2006,chang2006,BPOJ12}. Nevertheless, even for the highest spins measured to date (few hundred Hz), the effects of rapid rotation are smaller than the typical systematic uncertainties involved in the EOS determinations \citep[e.g.][]{steiner2013}. Spin effects are often not taken into account. 

If the neutron star magnetic field is sufficiently strong, the surface temperature may be inhomogeneous and this hampers inferring the true physical radius \citep[][]{elshamouty2016}. Moreover, other emission processes \citep[e.g. from magnetospheric processes, shocks from a pulsar wind, or an accretion flow;][]{campana1998} may contaminate the surface emission. This complicates inferring physical radii and introduces systematic effects. 

Finally, and perhaps most importantly, the thermal radiation from the neutron star surface is shaped by the atmosphere (Section~\ref{subsec:structure}). Using the observed surface emission as a tool to measure neutron star parameters and infer information about its interior properties thus requires a careful consideration of the atmospheric properties.  Various classes of neutron stars are subject to different physical circumstances and therefore require their own mapping of the surface emission to the observed spectrum. Atmosphere models have thus been developed to cover a wide range of temperatures, ionization states, magnetic fields strengths, and compositions. This includes models for hydrogen atmospheres of cool, non-magnetic neutron stars \citep[$B\lesssim 10^8$~G and $T_{\mathrm{eff}}\sim 10^5-10^7$~K, applicable to quiescent LMXBs; e.g.][Section~\ref{subsubsec:qatmosphere}]{Zavlin.etal:96,Heinke.etal:06,haakonsen2012}, cool neutron stars with moderate magnetic field strengths \citep[$B\sim 10^{11}-10^{13}$~G, applicable to dim isolated neutron stars; e.g.][Section~\ref{subsubsec:atmoscco}]{shibanov1992,ho2008}, strongly magnetic neutron stars \citep[i.e. magnetars; e.g.][Section~\ref{subsubsec:magnetars}]{ozel2003,adelsberg2006}, C and He atmospheres \citep[e.g.][Sections~\ref{subsubsec:qatmosphere} and~\ref{subsubsec:atmoscco}]{Ho:Heinke:09}, hot bursting neutron stars with different metallicities \citep[e.g.][Section~\ref{subsubsec:atmosburst}]{SPW11,SPW12,Netal15,medin2016}, and weakly accreting neutron stars \citep[][]{zampieri1995}. Over the years, a large number of various grids of neutron star atmospheres and corresponding model spectra have been computed by many authors. Several of these are implemented into the X-ray spectral fitting package {\sc XSpec}, for instance {\sc nsa} and {\sc nsgrav} \citep[][]{Zavlin.etal:96}, {\sc nsatmos} \citep[][]{Heinke.etal:06}, {\sc nsx} \citep[][]{Ho:Heinke:09}, {\sc carbatm} and
{\sc hatm}  \citep[][]{Suleimanov.etal:14, Suleimanov.etal:17}.

%TABLE OF SOURCES/METHODS
\begin{table*}
\caption{Sources that have so far yielded the most constraining mass and/or radius limits.}\label{tab:sources}
\begin{threeparttable}       
\begin{tabular*}{0.99\textwidth}{@{\extracolsep{\fill}}lll}
\hline\noalign{\smallskip}
Approach & Sources & Reference  \\ %comments
\noalign{\smallskip}\hline\noalign{\smallskip}
X-ray bursters  & SAX J1748.9--2021 & \citet{guver2013,ozel2016} \\ %touchdown method -- NGC 6440
 & EXO 1745--248 & \citet{ozel:09,ozel2016} \\ %touchdown method  -- Terzan 5
 & KS 1731--260 & \citet{ozel2012_ks1731,ozel2016} \\ %touchdown method 
 & 4U 1608--52 & \citet{guver2010_4u1608,ozel2016} \\ %touchdown method and cooling tail method
 & & \citet{Poutanen.etal:14}\\
& 4U 1820--30 & \citet{guver2010,ozel2016} \\ %touchdown method  -- glob clust?
 & & \citet{suleimanov2017} \\ %cooling tail method  %shaposhnikov2004 -- method?
 %%
 % & 4U 1728--34 & \cite{majczyna2005} \\ %boundary layer spectrum, unphysical results
% &  4U 1728--34 & \citet{ozel2016} \\ %touchdown method, wrong?
 & 4U 1724--307 & \citet{SPW11,nattila16} \\ %cooling tail method for a long burst -- (Terzan 2)
  &  & \cite{ozel2016} \\ %touchdown method for short bursts
 & 4U 1702--429 & \citet{nattila16,nattila2017} \\ %cooling tail method 
 & SAX J1810.8--260 & \citet{nattila16,Sul.etal:17} \\ %cooling tail method 
 & GS 1826--24 & \citet{zamfir2012} \\ %cooling tail method 
Quiescent LMXBs & NGC 6397 U24 & \citet{guillot2011,Heinke.etal:14} \\
&  & \citet{ozel2016} \\
 & 47 Tuc X5 & \citet{heinke2003,Heinke.etal:06}  \\
 & & \citet{bogdanov2016}  \\
 & 47 Tuc X7 & \citet{heinke2003,Heinke.etal:06}  \\
 & & \citet{bogdanov2016}  \\
 & M28 source 26 & \citet{becker2003,servillat2012} \\
  &  & \citet{ozel2016} \\
 & NGC 2808 & \citet{webb2007,servillat2008} \\
 & M13 & \citet{gendre2003,webb2007}  \\
 &  & \citet{catuneanu2013,ozel2016}  \\
 & $\omega$ Cen & \citet{rutledge2002,gendre2003_omcen} \\
 &  & \citet{Heinke.etal:14,ozel2016} \\
 & M30 & \citet{lugger2007,guillot2014}\\
 &  & \citet{ozel2016}\\
 & NGC 6304 & \cite{ozel2016} \\
Radio pulsars & PSR J1614--2230 & \citet{demorest2010,fonseca2016} \\% high mass measurement from Shapiro delay 
 & PSR J0348+0432 & \citet{antoniadis2013}\\ %high mass measurement, combined with opticall
%
%Optical dynamical mass & PSR J0348+0432 & \citet{antoniadis2013}\\ %high mass measurement
% & Aql X-1 & \citet{matasanchez2017} \\
% & SAX J1808.4--3658 & \citet{wang2013} \\
% & PSR J1311--3430 & \citet{romani2012} \\
% & PSR B1957+20 & \citet{vankerkwijk2011} \\
\noalign{\smallskip}\hline
\end{tabular*}
%\begin{tablenotes}
%\item[]Note -- 
%\end{tablenotes}
\end{threeparttable}
\end{table*}

%%%%%%%%%%%%%%%%%%%%%%%%%%%%
% REVIEW PAST WORK 
%%%%%%%%%%%%%%%%%%%%%%%%%%%%
\section{EOS constraints from electromagnetic observations of neutron stars}\label{sec:past}
As explained in Section~\ref{subsec:eosmr}, the observable properties that directly constrain the EOS are the neutron star mass and radius. Significant constraints on the EOS at supra-nuclear densities could be realized when both the mass and radius of a single neutron star are deduced from observations, but this is very challenging in practice. In the next Sections we review the many different approaches that can be taken to constrain the neutron star EOS using radio, optical and X-ray observations (Sections 3.1--3.8), and how various methods can be combined for increased accuracy and checks for systematic biases (Section~\ref{subsec:combined}).

There are several radio pulsars in binaries for which very accurate mass estimates are available from measuring gravitational effects through radio pulsar timing (Section~\ref{subsec:pulsars}), combined with dynamical information on the donor star from optical observations (Section~\ref{subsec:dynamical}). However, so far there exists no accurate radius measurement for a radio pulsar that has its mass measured with high accuracy. 
Measuring neutron star radii relies on measuring radiation that comes directly from (part of) the neutron star surface (see Section~\ref{sec:surface}). This can be achieved with observations of accreting neutron stars through modeling thermonuclear explosions that occur on their surface (Section~\ref{subsec:bursts}) and studying their thermal glow during quiescent episodes (Section~\ref{subsec:thermal}), or by analyzing the pulse profiles of hotspots on the stellar surface (Section~\ref{subsec:timing}). Such hotspots may also occur on non-accreting binary radio pulsars, which allows to apply similar techniques to measure their radii. Furthermore, dim isolated neutron stars also emit thermal radiation that can be used to constrain their emission radii (Section~\ref{subsec:cco}). 

Apart from putting direct constraints on the EOS through measuring $M$ and $R$, a very high spin frequency can also provide stringent constraints on the dense matter EOS (Section~\ref{subsec:spin}). We also review a number of other approaches to obtain constraints on the neutron star core properties and EOS that are less developed at present (Section~\ref{subsec:other}), but can perhaps in the future lead to better constraints, especially when combined with other methods (Section~\ref{subsec:combined}).

%%%%%%%%%%%%%%%%%%%%%%%%%%
%%%%% Radio pulsars 
%
\subsection{Mass measurements of radio pulsars in binaries}\label{subsec:pulsars} 
Radio observations can be used to detect every rotation of a neutron star, a technique that is known as \textit{pulsar timing}. Such measurements provide very accurate constraints not only on the rotation period of the neutron star itself, but also on the orbital parameters of the binary. For dedicated reviews on neutron star masses obtained through pulsar timing techniques, and what advances will be brought by future instrumentation (see also Section~\ref{subsubsec:ska}), we refer to \citet{kramer2008,watts2015_review,ozel2016_review}. Here, we briefly summarize the basic concepts and results that are most relevant for constraining the neutron star EOS.

By accurately timing every rotation of a radio pulsar, the binary mass function can be determined from the Keplerian orbital parameters:
\begin{equation}
f_{\rm ns} = \left ( \frac{ 2 \pi}{P_{\rm orb}} \right)^2  \frac{ (a_{\rm ns} \sin i)^3}{G} = \frac{(M_{\rm c} \sin i)^3}{M_{\rm T}^2},
\end{equation}

where $P_{\rm orb}$ is the binary orbital period and  $a_{\rm ns} \sin i$ the the projection of the pulsar semi-major axis on the line of sight. This mass function has three unknown parameters: the angle $i$ between the line of sight and the direction orthogonal to the orbital plane, the mass of the companion star $M_{\rm c}$, and the total mass of the binary $M_{\rm T} = M_{\rm c} + M$. The neutron star mass $M$ can therefore not be obtained from timing its pulsations alone. However, in some occasions the projected semi-major axis of the companion's orbit ($a_{\rm c} \sin i$) can be measured: via radio pulsar timing in case of double pulsar binaries, or via phase-resolved optical spectroscopy if the companion is a white dwarf or main-sequence star. This provides a measurement of the mass ratio $q$ of the two binary components: 
\begin{equation}
q= \frac{ M}{M_{\rm c}} =  \frac{ (a_{\rm c} \sin i)^3}{(a_{\rm ns} \sin i)^3}.
\end{equation}

Nevertheless, further information is required still to pinpoint the mass of the neutron star. For instance, some binaries that are viewed nearly edge-on allow to constrain the inclination from eclipses (where one of the binary components is periodically obscured by the other). Furthermore, optical studies can occasionally yield an independent measurement of the mass of the companion star (see Section~\ref{subsec:dynamical}). Finally, for some radio pulsars relativistic effects are measurable that allow to close the set of equations and obtain the masses of the binary components \citep[e.g.][for details]{shao2015,watts2015_review,ozel2016_review}. For instance, for very eccentric binaries one can observe changes in the argument of periapsis, while for very compact binaries the orbital decay due to gravitational wave radiation can be measurable. In exceptional cases, when the binary is viewed nearly edge-on, it has been possible to measure a Shapiro delay arising from the pulses traveling through the gravitational potential of the companion star. If these various pieces of information are available, the masses of radio pulsars can be measured to very high precision \citep[e.g.][]{thorsett1993,thorsett1999,stairs2004,kiziltan2013}. 

Accurate mass measurements have so far been obtained for $\approx 40$ radio pulsars and span a range of $M\approx 1.2 -2.0~\Msun$ \citep[see][for a recent overview]{ozel2016_review}. Assembling a firm statistical sample of mass measurements is of high interest for a variety of astrophysics questions, e.g. the dynamics of mass transfer and its role in binary evolution, the detailed physics of supernova explosions, and the birth-mass distribution of neutron stars \citep[e.g.][]{ozel2012,kiziltan2013}. However, without radius determinations, these accurate mass measurements generally do not put any stringent constraints on the neutron star EOS. It is only the most extreme mass measurements that can provide interesting constraints, by ruling out EOSs that predict lower maximum masses (see Section~\ref{subsec:eosmr}). 

There are currently two radio pulsars with reliable mass measurements of $M\approx 2~\Msun$ that have allowed to put some  constraints on the neutron star EOS. The first is the millisecond radio pulsar PSR J1614--2230, for which the Shapiro delay can be measured and has led to a mass determination of $M=1.928 \pm 0.017~\Msun$ \citep[][]{demorest2010,fonseca2016}. The second is the millisecond radio pulsar PSR J0348+0432 for which the combination of radio pulse timing and optical studies of its white dwarf companion yielded a reliable mass estimate for both binary components, with $M=2.01\pm 0.03~\Msun$ measured for the neutron star \citep[][]{antoniadis2013}. A single extreme measurement can already provide very strong constraints on the neutron star EOS, primarily by constraining the nucleon interactions \citep[][]{Hebeler.etal:13}. However, the current extremes of $M\approx 2~\Msun$ do not necessarily rule out the occurrence of any exotic particles the core \citep[e.g.][]{oertel2015,fortin2016,fortin2017}.

The above approach of measuring neutron star masses require the radio pulsars to be located in binaries and hence cannot be applied to isolated radio pulsars. However, some young (isolated) pulsars display sudden and temporary increases in the neutron star's spin. These \textit{glitches} can be used to put some constraints on the superfluid properties and the masses of these neutron stars (see Section~\ref{subsubsec:glitches}).

%%%%%%%%%%%%%%%%%%%%%%%%%%
%%%%% neutron stars in binaries: dynamical mass measurements 
%%%%% 
\subsection{Optical dynamical mass measurements of neutron stars in binaries}\label{subsec:dynamical}
If a neutron star is in a binary and periodic changes in the line-of-sight velocity of the companion star ($K_2$) can be measured from its optical emission, then the mass function $f(M,M_{\rm c})$ can be constructed from Newton's laws of Keplerian motion, which provides a lower limit on the mass of the neutron star:
\begin{equation}\label{eq:massfunction}
M \geq f(M,M_{\rm c}) \equiv \frac{M^3 \sin^3 i}{(M_{\rm T})^2} = \frac{K_2^3 P_{orb}}{2 \pi G}
\end{equation}

where $M_{\rm T} = M_{\rm c} + M$ is again the total binary mass. Furthermore, $i$ is the inclination angle at which we view the binary orbit (with $i=0\deg$ implying face-on and $i=90\deg$ edge-on) and $P_{\rm orb}$ is the orbital period. When the orbital period can be measured along with the companion star's velocity variations, the mass function provides a lower limit on the neutron star mass; $M_c=0$ and $i=90\deg$ would yield $M = f$, but since $M_c > 0$, we have $M > f$ for any inclination angle. 

Additional constraints from other types of measurements can turn this lower limit into an actual mass measurement. So far, this has been most successful in combination with accurate timing of radio pulsars (Section~\ref{subsec:pulsars}). For instance, for PSR J1614--2230 the Shapiro delay yielded $i$ and $M_c$ and hence allowed for measuring the pulsar mass of  $M=1.928\pm 0.017~\Msun$ via equation~(\ref{eq:massfunction}) since $K_2$ (from optical observations of the companion) and $P_{\mathrm{orb}}$ (from pulsar timing) can also be determined. Another way to remove degeneracies is through the detection of atmospheric lines of the companion star, which allowed for the accurate and constraining $M=2.01\pm 0.03~\Msun$ mass measurement of PSR J0348+0432 \citep[][]{antoniadis2013}. The atmospheric lines detected from its white dwarf companion provide a second mass function, through the periodic variations in the central wavelength of these lines, and a mass measurement of the white dwarf via the gravitational redshift of the lines, hence allowing a determination of the mass of the pulsar.

Some binary radio pulsars are eclipsed, so-called \textit{black widow pulsars}, which allows for a constraint on the inclination and can thus lead to a measurement of the neutron star. Some of these have similar or even higher mass estimates as PSR J1614--2230 and PSR J0348+0432, albeit with larger systematic uncertainties. For instance, for the black widow pulsar PSR B1757+20 a mass of $M\approx 2.40 \pm 0.12~\Msun$ was estimated, but the binary inclination is highly uncertain and in fact allows for a mass as low as $M\approx 1.66~\Msun$ \citep[][]{vankerkwijk2011}. Furthermore, light curve fitting yielded an estimated mass of $M\approx 2.68 \pm 0.14~\Msun$ for the black widow PSR J1311--3430 \citep[][]{romani2012}, but the poor quality of the fit suggests the presence of un-modelled obscuration or emission; attempts to account for that result in a lower pulsar mass, down to $M\approx 1.8~\Msun$ \citep[][]{romani2015}. Whereas systematic uncertainties currently preclude from drawing firm conclusions, these methods can potentially be refined \citep[e.g.][]{romani2015}. Difficulties currently lie in the fact that the companions are oblated and even when irradiation is taken into account in the light curve modelling there is residual short time scale variability. Moreover, the surface is unevenly heated, and there is asymmetry in the light curves. All this significantly hinders a reliable determination of the binary inclination, which thus directly translates into uncertainties in the neutron star mass. Nevertheless the high mass estimates for these black widow pulsars are tantalising and foster the idea that neutron stars with $M> 2~\Msun$ may exist and will be found one day. If such high masses were to be confirmed that would firmly rule out several classes of EOSs. Note that in these studies timing of the radio pulses provides the binary parameters, but the mass measurements are obtained from modeling the companion's optical light curves (to constrain the orbital inclination) and optical spectroscopy (to measure the mass ratio). 

Similar studies can be performed for X-ray binaries. So far, neutron star masses have been obtained from optical observations (combined with other techniques) for $\approx$10 HMXBs and $\approx$7 LMXBs. Together these span a range of $M \approx 1.1-1.9~\Msun$, but these measurements are much less accurate than those of millisecond radio pulsars \citep[see][for an overview]{ozel2016_review}. The additional constraints required to obtain mass measurements for these objects can be obtained in different ways. For instance, eclipses are also observed for some high-inclination X-ray binaries, causing the X-ray emission near the neutron star to be periodically blocked by the donor star, in which case $i$ can be accurately determined from the duration of X-ray eclipses \citep[][]{horne1985}. Furthermore, X-ray pulse timing can directly constrain the orbit of the neutron star and hence its radial velocity semi-amplitude $K_1$. If the radial velocity semi-amplitude of the donor star ($K_2$) can also be measured from optical (or near-infrared) observations in quiescence, the mass ratio $q=M_{c} / M$ is obtained. Alternatively, this ratio can be determined from the broadening of absorption lines ($v \sin i$) from the companion star  \citep[e.g.][]{horne1986}. A promising source to obtain this combined information is the only eclipsing millisecond X-ray pulsar, Swift J1749.4--2807, but crowding has so far prevented to identify the quiescent near-infrared counterpart \citep[][]{jonker2013}.  In the future, X-ray polarization may be employed to constrain the inclination of some neutron star X-ray binaries via studies of their hotspot emission (see Section~\ref{subsec:polarization}).

%%%%%%%%%%%%%%%%%%%%%%%
%%%%% BURSTS 
%%%%%
\subsection{Radius constraints from thermonuclear X-ray bursts}\label{subsec:bursts}
\begin{figure}
\begin{center}
\resizebox{0.5\textwidth}{!}{%
  \includegraphics{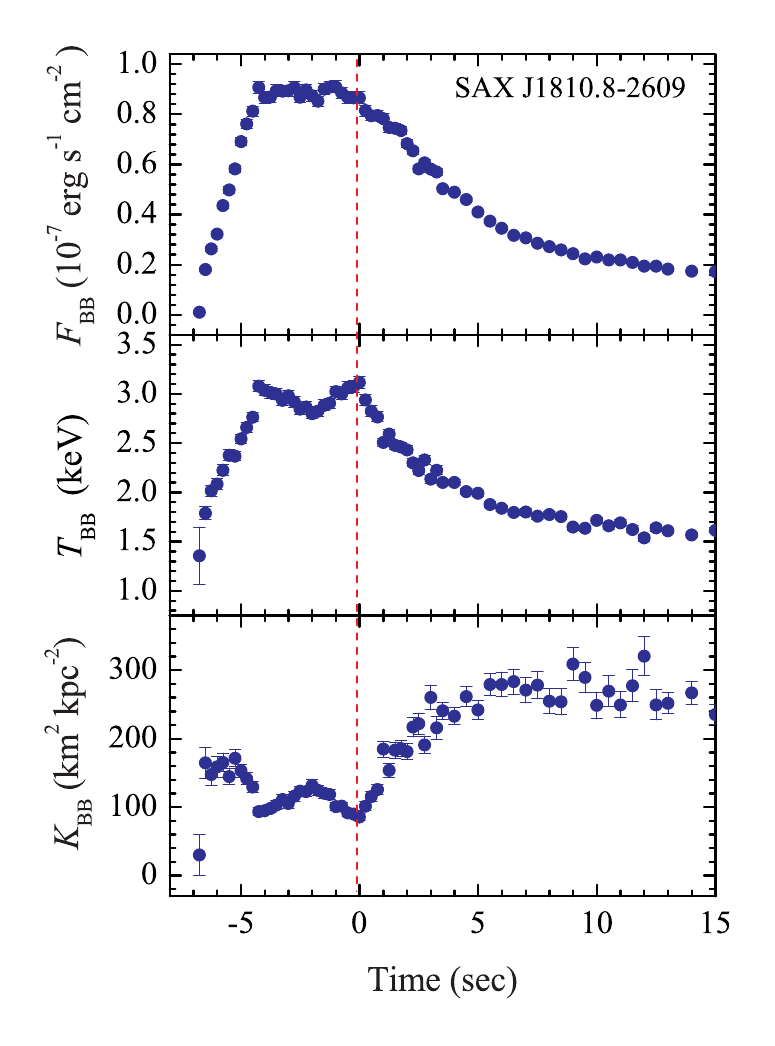}
}
\caption{Results of fitting of a PRE burst of the LMXB SAX\,J1810.08--2609 with a black-body model, showing the evolution of the black-body flux (top), temperature (middle) and normalization (bottom) along the X-ray burst. The touchdown point where the expanding atmosphere rejoins the neutron star surface, is thought to occur at $t=0$ (marked with the vertical dashed line). The part of the X-ray burst where the temperature is gradually and steadily decreasing (i.e. after the touchdown point), is referred to as the cooling tail.
}
\label{sv_fig1}  
\end{center}
\end{figure}

The X-ray luminosities of neutron star LMXBs in outburst vary in a wide range of $L_{\rm X} \sim 10^{35} -10^{38}$\,erg\,s$^{-1}$. This suggests that $\dot{M} = L_{\rm acc}R/GM\sim 10^{15} - 10^{18}$\,g\,s$^{-1}$ of mass is accreted \citep[where $L_{\rm acc} \sim 2\times L_{\rm X}$ is the bolometric accretion flux; e.g.][]{zand2007}. This gas accumulates on the neutron star surface where it undergoes thermonuclear burning. Under certain conditions, mainly determined by the local mass-accretion rate and thermal properties of the ocean/envelop, the nuclear burning is unstable and results in runaway energy production that is observable as a thermonuclear X-ray burst (also called Type-I X-ray bursts; shortly X-ray bursts from here on). Currently, $\sim$110 X-ray bursting neutron star LMXBs are known. %\footnote{Thermonuclear X-ray bursts have not been observed from other types of accreting neutron stars such as HMXBs, IMXBs or SyXRBs.} 
The majority of X-ray bursts have a duration of ten to hundreds of seconds and their spectra are generally well fitted with a black body model \citep{gallow08}, of which the temperature $T_{\rm BB}$ and normalization $K_{\rm BB}$ evolve in a characteristic way (see Figure~\ref{sv_fig1}). We refer to \citet{LvPT93} and \citet{SB06} for extensive reviews on X-ray bursts. Here, we focus on exploiting these events to constrain the neutron star properties.  

The X-ray bursting neutron stars are attractive candidates to constrain the EOS because during the X-ray bursts the neutron star surface becomes visible (Section~\ref{sec:surface}). In the following, we assume that the stellar surface is uniformly emitting and that there is no other emission process that can alter the observed spectral flux distribution. It may seem that this last assumption cannot be fulfilled since the accretion flow is a prominent source of radiation, but we show below that for several X-ray bursting neutron stars the accretion rates are low enough to ignore its contribution to the burst spectrum. 

There are two additional advantages that make X-ray bursts very suitable for constraining the neutron star parameters. Firstly, some X-ray bursts are so powerful that the radiation pressure causes the outer layers of the neutron star to expand, which is referred to as \textit{photospheric radius expansion} (PRE). This implies that in these layers the radiation pressure force $g_{\rm rad}= c^{-1}\int_0^\infty \kappa_\nu {\cal{F}_{\nu}}\,d\nu$ is larger than the local gravity $g=GM(1+z)/R^2$ and that that the luminosity exceeds the Eddington limit during the PRE burst phases. For every specific neutron star photosphere the value of this critical luminosity is unique, owing to the specific chemical composition and the difference in the opacities $\kappa_\nu$. Usually, the Eddington luminosity is determined using the Thomson electron scattering opacity:
\begin{equation} \label{eq:kappae}
\kappae  \approx 0.2\ (1+X) \ \mbox{cm}^2\ \mbox{g}^{-1},  
\end{equation}

where $X$ is the hydrogen mass fraction. The expression for the observed Eddington luminosity is then:
\be \label{ledd}
    L_{\rm Edd, \infty} = \frac{L_{\rm Edd}}{(1+z)^2} = \frac{4\pi GMc}{\kappae} (1+z)^{-1}, 
\ee

which corresponds to an observed bolometric Eddington flux of:
\be   \label{eddfl}
    F_{\rm Edd} = \frac{L_{\rm Edd, \infty}}{4\pi D^2}.
\ee

The Eddington flux is linked to the observable critical effective temperature $T_{\rm Edd,\infty}$, via:
\be \label{tedd}
     F_{\rm Edd} = \sigma_{\rm SB} T_{\rm Edd,\infty}^4,
\ee

where $\sigma_{\rm SB}$ is the Stefan Boltzmann constant. We note that the parameters defined by Equations\,(\ref{ledd}-\ref{tedd}) are the prerequisite for measuring the neutron star radius and mass. Unfortunately, however, it is not possible to determine exactly at which burst phase the observed flux equals the Eddington flux. The flux at the touchdown point, where the black-body temperature $T_{\rm BB}$ reaches a maximum and the black-body normalization has a local minimum (see Figure.\,\ref{sv_fig1}), is generally believed to be close to the Eddington flux. In one approach, the neutron star mass and radius are constrained by attempting to identify the Eddington flux of PRE bursts: this is often referred to as the \textit{touchdown method}.

The spectral evolution of X-ray bursts allows us to obtain an additional constraints on the neutron star mass and radius. Since these parameters cannot change, an acceptable model must be able to describe the X-ray burst spectrum at every phase, from the Eddington flux near to peak down to the much lower fluxes in the X-ray burst tail, with constant values of $M$ and $R$. Constraining the neutron star parameters via the spectral evolution of the X-ray burst is usually referred to as the \textit{cooling-tail method}.

During X-ray bursts, in principle both the apparent surface area and the Eddington limit of the neutron stars can be measured, hence the degeneracy between $M$ and $R$ can be broken using the equations laid out above. Although the approach is very promising, various systematic uncertainties (e.g. varying accretion emission during a burst, the emission radius, the detailed spectral properties, and emission anisotropies) need to be resolved before orthogonal constraints can reliably be obtained with high accuracy \citep[e.g.][see Section~\ref{subsubsec:burstbias}]{steiner2010,galloway2012,zamfir2012,intzand2013,degenaar2018}. 
%Pinpointing the pressure beyond the nuclear saturation density requires mass/radius measurements with a $\sim$5\% accuracy level

\subsubsection{Model atmospheres for X-ray bursts}\label{subsubsec:atmosburst}
\begin{figure}
\begin{center}
\resizebox{0.7\textwidth}{!}{%
  \includegraphics{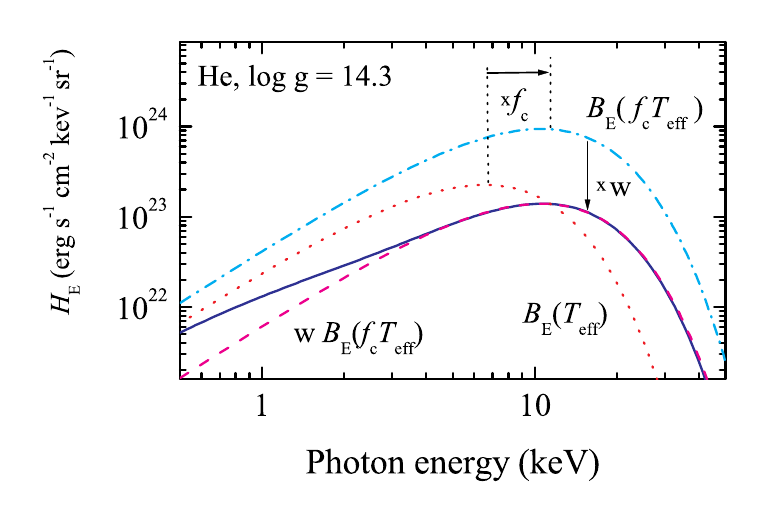}
}
\caption{Fits to model spectra (solid curve) with a diluted black body (dashed curve). Black-body spectra with temperatures $T_{\rm eff}$ (dotted curve) and $\fc\,T_{\rm eff}$ (dot-dashed curve) are also shown. The model spectrum of a pure 
helium atmosphere with $\log g = 14.3$ and $\ell=1.0$ is used. The fitting parameters are $\fc=1.61$ and $w=0.148$.
}
\label{sv_fig2}   
\end{center}
\end{figure}

The black-body fits give important initial information about an X-ray bursting neutron star, but this kind of fitting is not accurate enough to obtain reliable mass and radius measurements. Accurate model atmospheres of X-ray bursting neutron stars have to be computed for this aim to
obtain model flux spectral distributions ${\cal{F}}_{\nu}$ (hereafter spectra). The model atmosphere parameters are the effective temperature
$T_{\rm eff}$, the surface gravity $g$, and the chemical composition of the atmosphere. The very first models of hot neutron star 
atmospheres \citep{Londonetal:84, Londonetal:86, Lapidusetal:86, Ebisuzaki87, madej:91, Pavlov.etal:91} demonstrated that Compton scattering is the main source of opacity and that the energy exchange between high energy photons escaping from the deep and hot atmospheric layers with the relatively cold surface electrons establishes the equilibrium spectra. These are close to the diluted black-body spectra (Figure\,\ref{sv_fig2}). Therefore, there are two parameters describing every model spectrum, the dilution factor $w$ and the color correction factor
$\fc$:
\be \label{bbfit}
    {\cal{F}}_\nu \approx w\,\pi B_\nu(\fc\,T_{\rm eff}),
\ee

where $B_\nu$ is the Planck function. The color correction factor links the color temperature $T_{\mathrm{c}}$, obtained from fitting a spectrum distorted by atmospheric effects to a Planck spectrum, to the effective temperature $T_{\rm eff}$ (i.e. $f_c = T_{\mathrm{c}}/T_{\mathrm{eff}}$). We note that in first approximation, $w \approx \fc^{-1/4}$. However, approximation (\ref{bbfit}) actually doesn't conserve the bolometric flux
\be
     \int_0^{\infty}w\pi\,B_\nu(\fc\,T_{\rm eff})\,d\nu = w\fc^4\,\sigma_{\rm SB}T_{\rm eff}^4,
\ee
and therefore the bolometric flux obtained from the black-body fit has to be multiplied by the bolometric correction $w\fc^4$.

\begin{figure}
\begin{center}
\resizebox{1.\textwidth}{!}{%
  \includegraphics{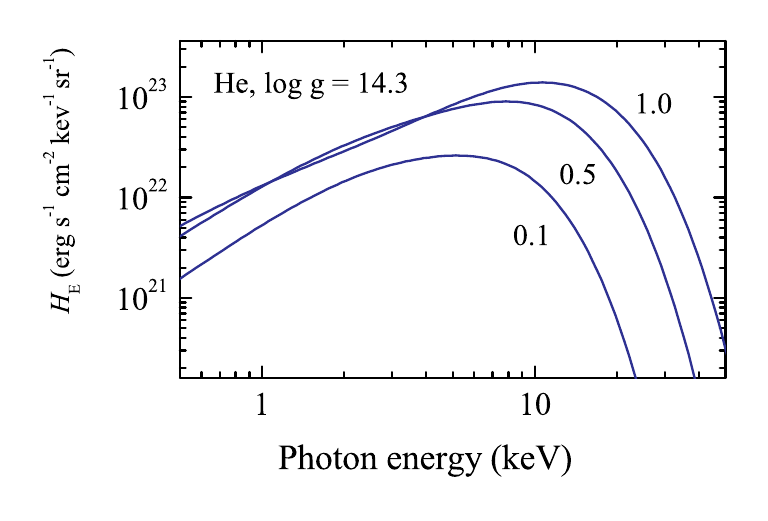}
  \includegraphics{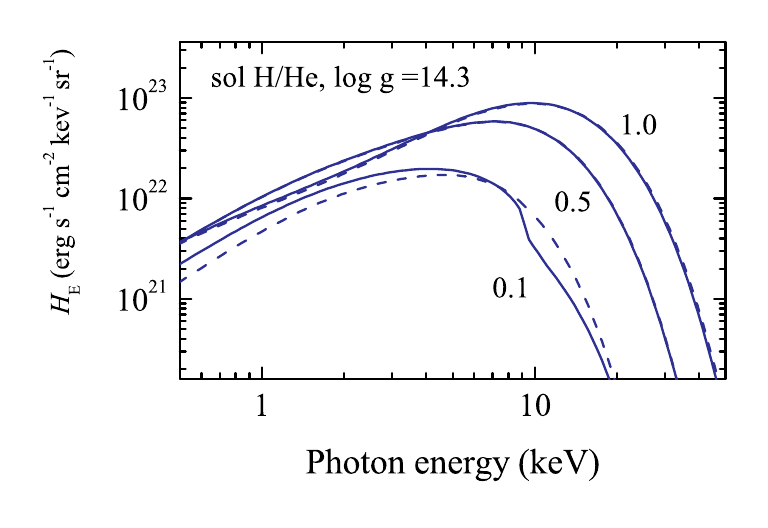}
}
\caption{Model spectra of pure helium (left) and solar H/He mix (right) atmospheres for $\log g = 14.3$ and  relative luminosities $\ell =$\,0.1, 0.5, and 1.0. The right panel shows the model spectra for solar (solid curves) and hundred times less heavy element abundances (dashed curves). 
}
\label{sv_fig3}   
\end{center}
\end{figure}

At present, extended grids of hot neutron star atmospheres have been computed using the Kompaneets equation for the Compton scattering description \citep{SPW11}, and the fully relativistic angle-depended redistribution function \citep{SPW12}. The latest grids have been computed for different chemical compositions: pure hydrogen, pure helium and mixed H/He with a solar H/He ratio with different fractions of solar metallicities (i.e. heavy element abundances). The models are computed for nine surface gravities, with $\log g$ from 13.7 to 14.9 in steps of $\delta \log g=0.15$. The relative luminosity $\ell$ is  used instead of the effective temperature as a model parameter, $\ell = L/L_{\rm Edd}$, or $T_{\rm eff} = \ell^{1/4} T_{\rm Edd}$. Here $L_{\rm Edd}=L_{\rm Edd,\infty}(1+z)^2 = 4\pi R^2\,gc/\kappae$  and $T_{\rm Edd} = T_{\rm Edd,\infty}(1+z) = (gc/\sigma_{\rm SB}\kappae)^{1/4}$ are the intrinsic Eddington luminosity and the intrinsic critical effective temperature, respectively. For every chemical composition and surface gravity, about 20 models with $\ell$ varying from 1.1. to 0.1 were computed. For three $\log g$ values (14.0, 14.3, and 14.6) the grids were extended down to $\ell=0.001$. Some examples of the emergent model spectra are shown in Figure\,\ref{sv_fig3}. 

\begin{figure}
\begin{center}
\resizebox{1.\textwidth}{!}{%
  \includegraphics{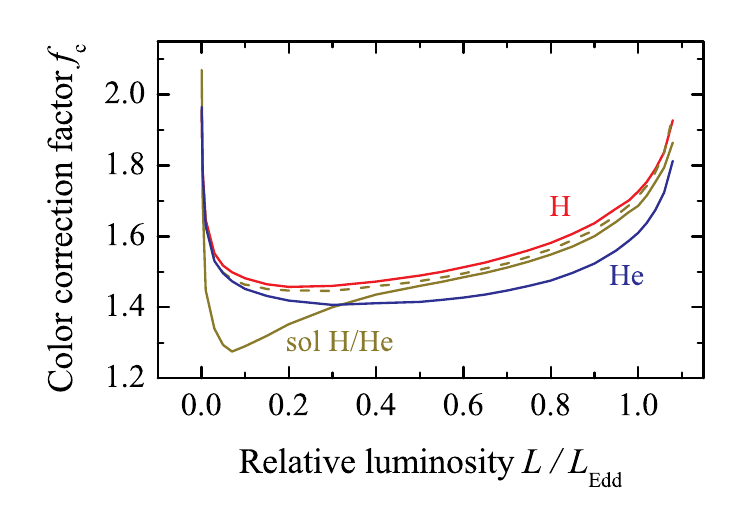}
    \includegraphics{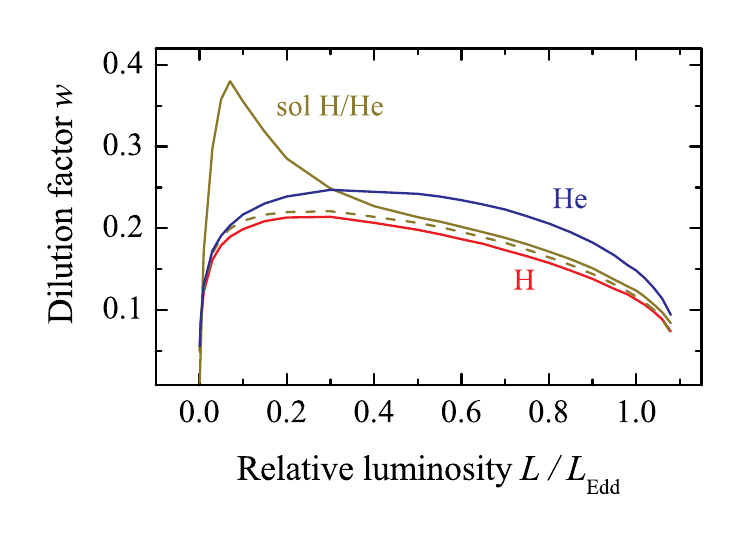}
}
\caption{The model dependences of color correction factors $f_c$ (left) and dilution factors $w$ (right), computed for pure hydrogen (red curves), pure helium (blue curves) and solar H/He mix (dark-yellow curves) atmospheres. The model curves for solar H/He mix are presented for two heavy element abundances: solar (solid curves) and reduced solar abundance (dashed curves). All these model curves were computed for $\log g = 14.3$. 
}
\label{sv_fig4}   
\end{center}
\end{figure}

To provide an anchor point to compare the theoretical behavior of X-ray bursts to actual data, the model atmosphere spectrum can be fitted by a diluted black body. Examples of the dependence on the relative luminosities are shown in Figure\,\ref{sv_fig4}. The values of the fit parameters depend significantly on the relative luminosity and the chemical composition, whereas the dependence on the surface gravity is less significant. The color correction factor $\fc$ and the dilution factor $w$ both evolve rapidly when the relative luminosity $\ell$ decreases from 1 to 0.5. Formally super-Eddington model atmospheres can occur because the effective electron scattering cross-section decreases with increasing temperature \citep[Klein-Nishina reduction; see][]{Pacz:83, SPW12, Pout:17}.

\subsubsection{Methods for obtaining mass and radius measurements from X-ray bursts}\label{subsubsec:measurebursts}
The potential of using X-ray bursts to obtain mass and radius measurements of neutron stars was recognized many years ago \citep{Ebisuzaki87, vP:90, Damen:90,LvPT93}. However, the low quality of the early data and lack of sufficiently extended model atmosphere grids prevented to obtain meaningful results. With the availability of much better data (mainly from \rxte), the touchdown method has been applied to 6 different LMXBs in recent years \citep[Table~\ref{tab:sources}; see e.g. ][]{Ozel06,ozel:09,guver2010}. The results obtained via this approach were recently reviewed by \citet{ozel2016_review}, and yield a combined radius preference of $R=9.8-11.0$~km for a mass of $M=1.4~\Msun$. This method does not take into account the spectral evolution of X-ray bursts. The cooling-tail method has been laid out by \citet{SPRW11, Poutanen.etal:14, Sul.etal:17} and has been applied to 5 sources to date. Since this method is very promising but has not been described in a book review yet, we here focus on that method. 

Before continuing to describe the cooling-tail method and its application in detail, we point out that there is a third approach where observable properties such as the recurrence time, peak flux and decay time of X-ray bursts are compared with theoretical light curve models to infer the surface redshift of the neutron star \citep[e.g.][]{heger2007_2}. However, the only neutron star LMXB (out of $\approx 110$ known X-ray bursters) that shows X-ray bursts with light curves that match the theoretical calculations, is GS 1826--24 (also called ``the clocked burster''). Therefore, this method has only been applied to this particular source, yielding $R<6.8-11.3$~km for $M<1.2-1.7~\Msun$ \citep[][]{zamfir2012}. This is consistent with results obtained from the other X-ray burst analysis types as well as those obtained for quiescent neutron star LMXBs (see Section~\ref{subsubsec:globclust}). For this approach the X-ray bursts do not have to show PRE (i.e. reach Eddington).

In the cooling-tail method, a black-body model is used to fit both the observed spectra and the model atmosphere spectra. This shows that the observed normalization $K_{\rm BB}$ depends only on the dilution factor when the late burst phases (after the touchdown point) are considered:
\be \label{kw}
      K_{\rm BB} = w\,\frac{R^2(1+z)^2}{D^2}=w\,\Omega.
\ee

\begin{figure}
\begin{center}
\resizebox{0.7\textwidth}{!}{%
  \includegraphics{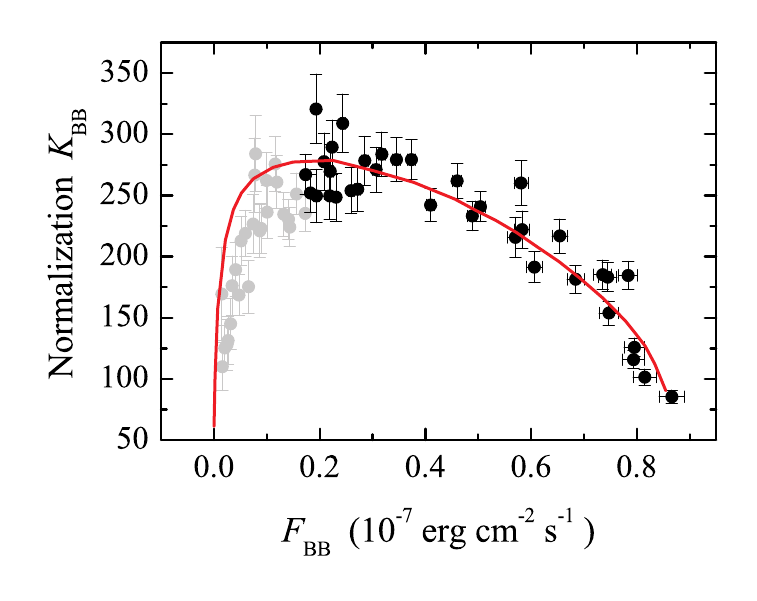}
}
\caption{The observed $K_{\rm BB}-F_{\rm BB}$ dependence (circles with error bars) for
the cooling tail (after touchdown) of a PRE burst from SAX\,J1810.8--2609 \citep[see][]{nattila16, Sul.etal:17}. 
The red solid  curve is the best-fitting theoretical dependence $w-w\fc^4\,\ell$ for solar H/He mix ($Z=0.01 Z_\odot$) 
and $\log g=14.3$, corresponding to the bright data points. The fitting parameters are $\Omega = 1261$\,(km/10\,kpc)$^2$
and $F_{\rm Edd} = 0.776\times 10^{-7}~\flux$.
}
\label{sv_fig5}   
\end{center}
\end{figure}

This means that all changes in the black-body normalization occurring after the touchdown point, which we can see in Figure\,\ref{sv_fig1} (bottom), are connected to changes in the dilution factor. It is more convenient to demonstrate the evolution of the normalization as an observed 
$K_{\rm BB} - F_{\rm BB}$ dependence (see Figure\,\ref{sv_fig5}) and approximate it with the model curves $w - w\fc^4\,\ell$ with two fit parameters\footnote{Initially the suggested method was using the assumption $w=\fc^{-1/4}$ and the observed curves $K_{\rm BB}^{-1/4} - F_{\rm BB}$ were fitted with the model  $\fc - \ell$ ones.}: $F_{\rm Edd}$ and the solid angle occupied by the neutron star on the sky $\Omega$. Both fit parameters depend on the (poorly known) distance to the neutron star but their combination, the apparent critical surface temperature $T_{\rm Edd,\infty}$, is independent of distance:
\be \label{tefit}
      T_{\rm Edd,\infty} = \left(\frac{F_{\rm Edd}}{(\sigma_{\rm SB}\Omega}\right)^{1/4}=
      \left(\frac{GMc}{\sigma_{\rm SB}\,0.2(1+X)\,R^2(1+z)^3}\right)^{1/4}.
\ee

\begin{figure}
\begin{center}
\resizebox{0.7\textwidth}{!}{%
  \includegraphics{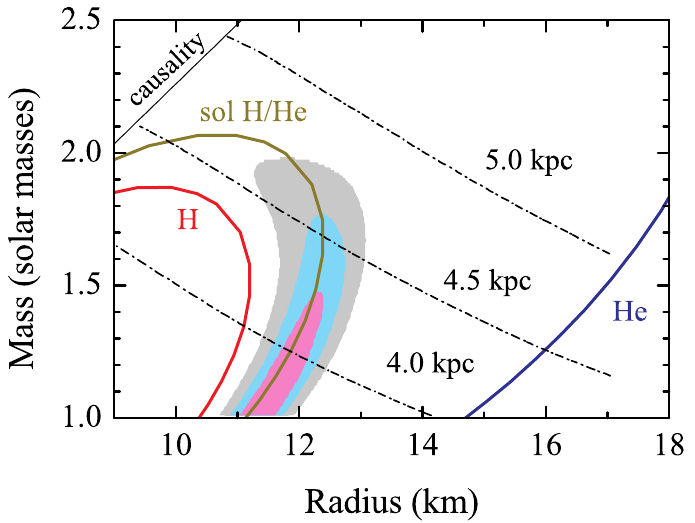}
}
\caption{
The $\chi^2$ confidence regions (68, 90 and 99\% probabilities) in the $M-R$ plane for SAX\,J1810.8--2609, obtained using the cooling-tail method for solar-mix H/He with reduced heavy element abundance \citep[$Z=0.01 Z_\odot$][]{Sul.etal:17}. The black dashed-dotted curves correspond to distances of 4.0, 4.5 and 5.0 kpc. The solid curves correspond to the best-fitting $T_{\rm Edd,\infty}$ obtained for $\log g = 14.3$ and three chemical compositions: pure hydrogen (red curve), solar H/He mix with the reduced heavy element abundance (dark-yellow curve), and pure helium (blue curve). The fit parameters are $\Omega = 1310$\,(km/10\,kpc)$^2$, $F_{\rm Edd} = 0.774\times 10^{-7}~\flux$ for the pure hydrogen models, and $\Omega = 1088$\,(km/10\,kpc)$^2$, $F_{\rm Edd} = 0.757\times 10^{-7}~\flux$ for the pure helium models. The fit parameters for the solar H/He mix are the same as in Figure\,\ref{sv_fig5}.
}
\label{sv_fig6}   
\end{center}
\end{figure}

The curve of equal $T_{\rm Edd,\infty}$ on the neutron star $M-R$ plane gives the possible values of the neutron star mass and radius, as shown in Figure\,\ref{sv_fig6}. However, this depends on the surface gravity of the model curves used and the assumed chemical composition. It is clear that for a given neutron star mass, the pure helium models ($X=0$) will give a larger neutron star radius than pure hydrogen models ($X=1$). Formally, the curve of constant $T_{\rm Edd,\infty}$ gives the correct $M$ and $R$ at the point were $\log g$ is equal to the value used for the model curve computation. Therefore, we have to assume some chemical composition of the neutron star atmosphere and interpolate the model curves $w - w\fc^4\,\ell$ computed for nine surface gravities for every $M-R$ pair. Then we can fit the observed curve  $K_{\rm BB} - F_{\rm BB}$ by this specific model curve and obtain the $\chi^2$ map on the $M-R$ plane, allowing to define confidence regions (68, 90 and 99\%). The distance to the source is the only free parameter for a fixed chemical composition, and its value is unique for every $M-R$ pair. 
In Figure~\ref{sv_fig5} we also illustrate the importance of the chemical composition. The observed curve $K_{\rm BB} - F_{\rm BB}$ was fitted with the model curves $w - w\fc^4\,\ell$ computed for pure hydrogen and pure helium model atmospheres, and the corresponding curves of constant $T_{\rm Edd,\infty}$ are drawn. The pure hydrogen atmospheres are hard to sustain because of the significant mass accretion rate of bursting neutron stars. Realistically only two possibilities exist: solar H/He mix for normal LMXBs and pure helium atmospheres for ultracompact LMXBs in which the secondary star is a He white dwarf (such as 4U\,1820--30). The derived neutron star radii are so different in these two cases that this allows us to distinguish helium-rich and solar-mix atmospheres using the cooling-tail method. 

\subsubsection{Obtained results from the cooling-tail method}\label{subsubsec:burstresults}
Just like other approaches, the described cooling-tail method is only correct for an isolated, passively cooling neutron star. Of course, the X-ray bursting neutron stars accrete matter and the influence of the accretion flow could muddy with the neutron star mass and radius measurements. To reduce any bias from the presence of the accretion flow, the X-ray bursts for the analysis have to be carefully chosen and ideally should occur at low accretion rate. A detailed discussion of this issue was given by \citet{Poutanen.etal:14, Kajava.etal:14,Suleimanov16EPJA}. A careful inspection of the X-ray bursts detected by \rxte\ resulted in three LMXBs that show X-ray bursts at low mass-accretion rates:  SAX\,J1810.8$-$260, 4U\,1702$-$429, and 4U\,1724$-$307. These were therefore selected to apply the cooling-tail method \citep{nattila16}. In addition to the results from the burst analysis itself, limits from the nuclear physics were imposed \citep{Steiner.etal:15}, and two different approximations for the model EOS were applied \citep[see][for details]{nattila16}. Moreover, the results from the three sources were combined by assuming that all these neutron stars have the same EOS and hence lie on the same theoretical $M-R$ curve. This resulted in a predicted radius of $R=10.5-12.8$~km for $M=1.4~\Msun$, with some dependence on the assumed composition. Interestingly, the cooling-tail analysis of \citet{nattila16} showed that SAX\,J1810.8$-$260 and 4U\,1724$-$307 require atmospheres with solar mixed H/He abundances, whereas 4U\,1702$-$429 needs to accrete helium-rich matter. Applying the same method to the ultra-compact LMXB 4U~1820--30 yielded $R=10-12$~km for $M<1.7~\Msun$ and $R=8-12$~km for $M=1.7-2.0~\Msun$ \citep[using a pure-He atmosphere;][]{suleimanov2017}. The constraints obtained for this source are not very strict because the cooling tail phase that can be described by the theoretical atmosphere models is only very short (about a second). Finally, applying the cooling tail method to 4U~1608--52 yielded a lower limit on the radius of $R>12$~km for $M=1.0-2.4~\Msun$ and a constraint of $R=13-16$~km for $M=1.2-1.6~\Msun$ \citep[][]{Poutanen.etal:14}. The rather wide range for this source is mainly caused by the large distance uncertainty.

The most recent development regarding the cooling-tail method is that rather than fitting a diluted black body to the observed X-ray burst spectra and comparing that with similar fits to the model spectra, the model spectra are fitted directly to the observational data \citep[][]{nattila2017}. This new approach was applied for 4U 1702--429 and resulted in a preferred radius and mass of $R=12.4\pm0.4$~km and $M=1.9\pm0.3~\Msun$ for $5.1 < D <6.2$~kpc and a hydrogen mass fraction of $X<0.09$ \citep[confirming previous suggestions of a H-poor atmosphere;][]{nattila16}. An interesting aspect of this application is that it allows to put strong constraints on the atmosphere composition, due to the strong dependence of the spectral evolution on the element abundances. This first application is promising and the method can be further developed \citep[][]{nattila2017}.

Comparing results from the cooling-tail and touchdown method suggests that the latter yields systematically lower radii \citep[cf.][]{ozel2016}. Indeed, both techniques have been applied to 4U 1608--52 \citep[][]{guver2010_4u1608,Poutanen.etal:14}, 4U 1728--34 \citep[][]{SPW11,ozel2016}, and 4U 1820--30 \citep[][]{ozel2016,suleimanov2017} with inconsistent results. A concern for the X-ray bursts sample used in the touchdown analysis is that the accretion luminosity may be a significant source of contamination \citep[see][for a review]{degenaar2018}, which may be the reason that the spectral evolution of these bursts deviates significantly from the theoretically expected cooling tails \citep[e.g.][]{Poutanen.etal:14,Kajava.etal:14}. On the other hand, the cooling-tail method received criticism for the poor quality of the black-body fits for some sources \citep[][]{guver2012_bursts,ozel2015}. 

In the past few years, results from X-ray burst studies have also been combined with those obtained from quiescent LMXBs in a Bayesian formalism. We further discuss that in Section~\ref{subsubsec:globclust}. 

\subsubsection{Biases and uncertainties in X-ray burst studies}\label{subsubsec:burstbias}
Apart from the influence of the accretion flow mentioned in the previous section, there are other systematic uncertainties in the X-ray burst analysis that can bias the results.\footnote{We note that the cooling-tail method is independent of distance: see Equation~(\ref{tefit}).} %This first is the distance to X-ray burst sources, which is often subject to large uncertainties (unless the sources are in globular clusters). Since to first order we constrain the quantity $R/D$, uncertainties in the distance directly translate into uncertainties in $R$. 
For instance, the rapid rotation of the neutron star is typically neglected. We know that many neutron stars in LMBXs rotate at frequencies of several hundred Hz \citep[][]{patruno2017_spin}. The resulting oblateness causes the equatorial radius to be larger than that of a non-rotating neutron star with the same mass \citep{AlGM14}. As a result the apparent area of the neutron star surface increases so that we obtain a larger neutron star radius. We refer to \citet{Baubock.etal:15} for a discussion on this topic. \citet{Poutanen.etal:14} applied the cooling-tail method to the fastest spinning neutron star LMXB 4U\,1608$-$52 (620~Hz) and argued that the obtained radius limit would be $\sim$10\% lower if its rapid rotation was taken into account.

\begin{figure}
\begin{center}
\resizebox{0.7\textwidth}{!}{%
  \includegraphics{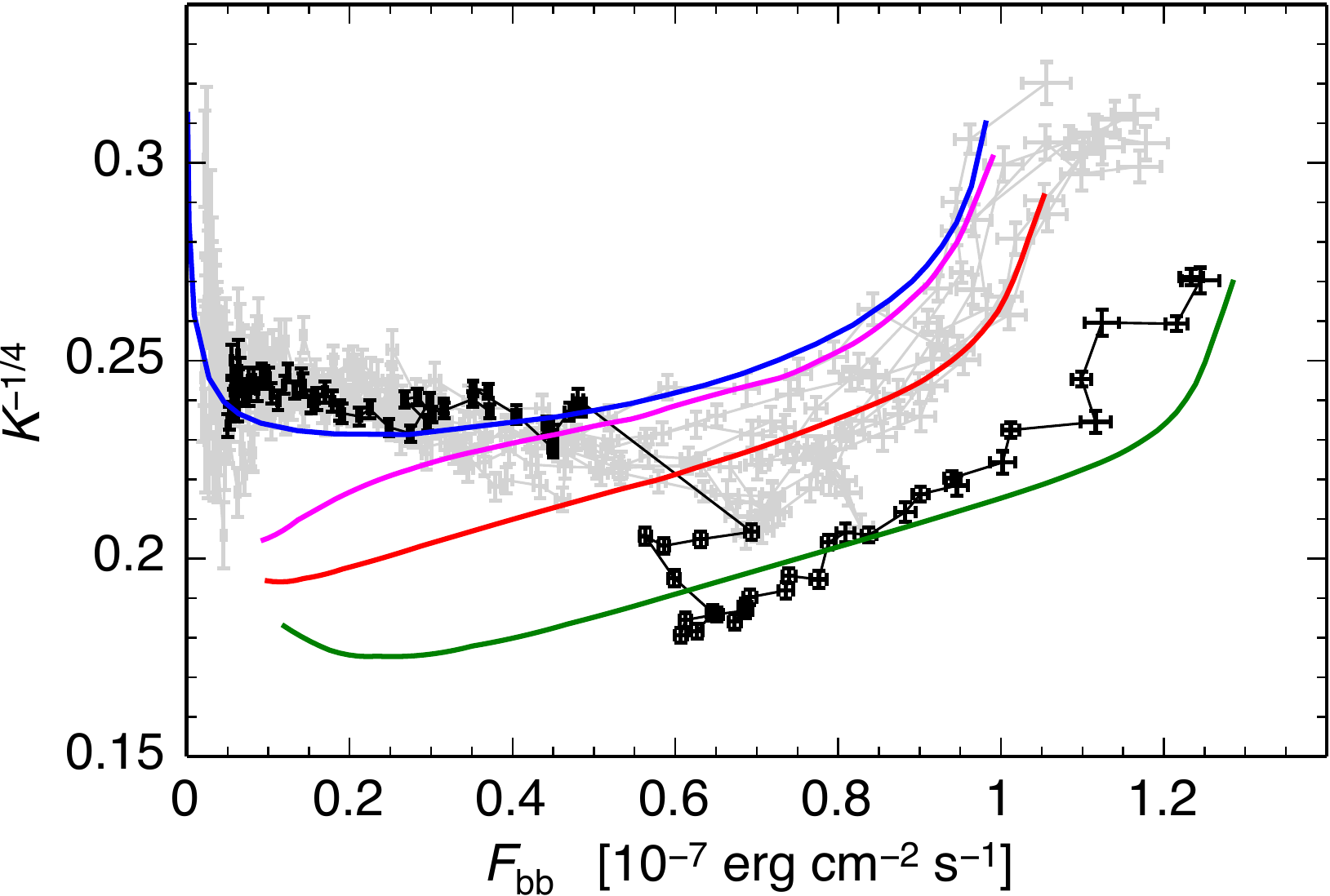}
}
\caption{
Black-body normalization versus flux for X-ray bursts from HETE\,J1900.1$-$2455, together with the model curves $\fc -\ell$. The models were computed for $M=1.4~\Msun$ and $R=12$\,km neutron star atmospheres with metal enhancement factors $\zeta=$ 0.01 (blue curve), 1 (purple curve), 10 (red curve) and 40 (green curve). The track with the largest deviations is highlighted in black \citep[see][]{Kajava.etal:17}. 
 }
\label{sv_fig7}   
\end{center}
\end{figure}

Another important factor is the heavy element abundance in the atmospheres of X-ray bursters. Thermonuclear ashes could appear at the neutron star surface during early stages of the burst due to convection, and due to ejection of the surface layers during a strong PRE phase \citep[e.g.][]{PP86,weinberg06,intZW10}. Model atmospheres of hot neutron stars enriched with heavy elements were computed by \citet{Netal15}. This work showed that for increasing heavy element abundances, the color correction factor decreases and the dilution factor increases (see Figure\,\ref{sv_fig7}). Interestingly, a burst from HETE\,J1900.1$-$2455 showed a highly unusual cooling tail (Figure\,\ref{sv_fig7}). This odd behavior can be explained if an almost pure heavy metal atmosphere just after touchdown was covered by solar H/He mix matter at the later cooling phase, perhaps due to accretion \citep{Kajava.etal:17}. The heavy metal enrichment could be less apparent if the heavy element abundance is lower, e.g. ten solar abundances, but this cannot change the result significantly (cf. the results of \citet{SPRW11} and \citet{nattila16}). 

Important and unsolved questions in X-ray burst analysis are whether the emitting radii measured during a single or during multiple bursts are identical and equal to the neutron star radius \citep[e.g.][]{guver2012_bursts,galloway2012}, whether the bursts consistency reach the Eddington limit \citep[e.g.][]{boutloukos2010,guver2012_bursts}, if the burst emission is by approximation isotropic \citep[e.g.][]{zamfir2012}, and whether the chemical composition varies \citep[e.g.][]{bhattycharyya2010}.

%%%%%%%%%%%%%%%%%%%%%%%%%%
%%%%% Quiescent LMXBs 
%%%%% 
\subsection{Radius constrains from thermally-emitting quiescent LMXBs}\label{subsec:thermal}  
Using sensitive X-ray satellites, thermal surface radiation can also be detected from neutron star LMXBs during their quiescent episodes. If the neutron star is radiating uniformly, one can derive constraints on the mass and radius by fitting the thermal surface radiation (Section~\ref{sec:surface}) with an appropriate neutron star atmosphere model (see Section~\ref{subsubsec:qatmosphere}). Neutron stars in LMXBs generally have low magnetic fields, and these are not expected to cause surface temperature inhomogeneities. However, even if their magnetic field may not cause temperature anisotropies in itself, when residual accretion takes place even a weak magnetic field may channel the accreted matter on to a fraction of the stellar surface and hence heating it non-uniformly (see Section~\ref{subsubsec:qbiases}). We know that at least in some LMXBs gas is still accreting onto the neutron star in quiescence. Evidence for such low-level accretion comes from irregular X-ray variability on a time scale of hours to years \citep[e.g.][]{cackett2010_cenx4,degenaar2012_exo1745,bernardini2013,wijnands2013_saxj1750,cotizelati2014}. Moreover, the presence of a non-thermal emission component in the quiescent spectrum can be evidence for continued accretion \citep[e.g.][]{fridriksson2011,chakrabarty2014,dangelo2015}. 

\begin{figure}
\centering
\includegraphics[width=0.65\textwidth]{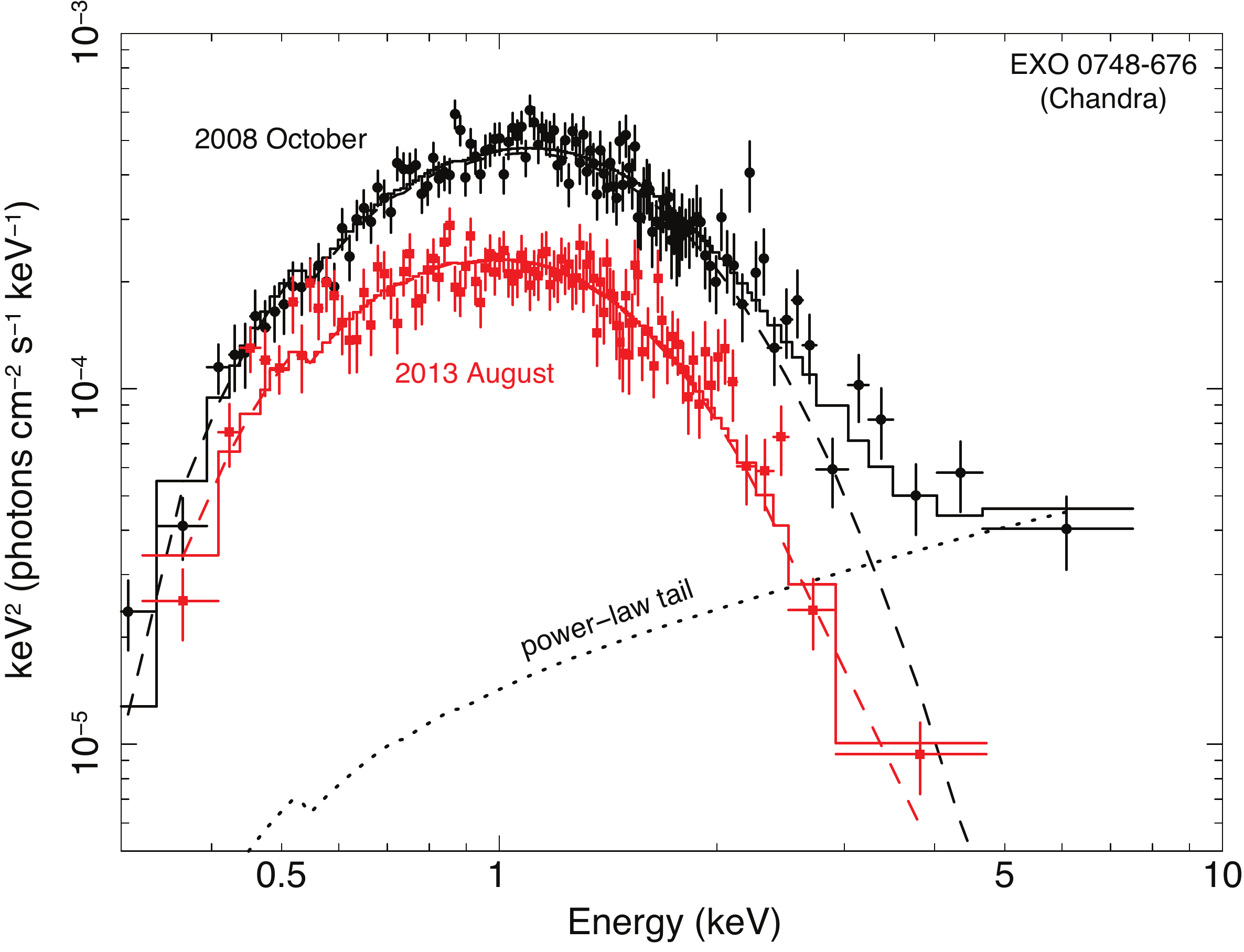}
\caption{
Unfolded X-ray spectra of the quiescent neutron star LMXB EXO 0748--676. Shown are two observations taken $\sim$2 months (black, top) and $\sim$5~yr (red, bottom) after the end of its $\sim$24-yr long accretion outburst \citep[][]{degenaar2014_exo}. Whereas the 2013 spectrum only consisted of thermal emission \citep[dotted line, fitted here with a neutron star atmosphere model \textsc{nsatmos};][]{Heinke.etal:06}, the 2008 spectrum exhibited an additional hard power-law tail (dashed curve) that contributed $\sim$20\% to the total unabsorbed 0.5--10 keV flux. A decrease in the thermal emission component is apparent and is thought to result from the thermal relaxation of the accretion-heated crust (see also Section~\ref{subsubsec:crustcool}).
} 
\label{fig:spec}
\end{figure}

A few dozen neutron star LMXBs have been observed in quiescence \citep[e.g.][for a recent overview]{wijnands2017}. Most of these show a distinctive soft spectral component that is ascribed to thermal surface emission of the neutron star that was heated during previous accretion episodes (see Section~\ref{subsec:thermalevo}). When fitted with a black body, temperatures of $kT_{\rm BB}\sim 0.1-0.2$~keV are typically obtained, whereas neutron star atmosphere models yield $kT\sim 50-150$~eV. Many neutron star LMXBs also show a hard emission tail in their quiescent spectra that is typically modeled as a simple power-law with an index of $\sim 1-2$ (see Figure~\ref{fig:spec}). 

The fractional contribution of the power-law component to the 0.5--10 keV flux varies widely between different sources \citep[with some being fully dominated by the non-thermal emission, whereas it is absent in others; e.g.][]{jonker2004}, but can also vary for a single source \citep[e.g.][see also Figure~\ref{fig:spec}]{cackett2011,fridriksson2011}. Although this non-thermal emission is often taken as evidence of ongoing accretion, it could perhaps also be associated with processes that involve the magnetic field of the neutron star (e.g. a pulsar wind or a shock from where the gas runs into the magnetosphere). This idea springs from the fact that the AMXPs, which show X-ray pulsations and hence a dynamically important magnetic field during outburst episodes, are often fully dominated by the non-thermal emission component \citep[e.g.][and references therein]{degenaar2012_amxp}. A non-thermal component in the quiescent spectrum could thus imply that the surface temperature is not homogeneous, which is a problem for measuring the mass and radius. Moreover, such a high-energy tail increases the uncertainties on the fit parameters of the thermal component.

Since uncertainties in the distance translate directly into uncertainties in the measured masses and radii (see Section~\ref{subsubsec:qbiases}), neutron star LMXBs in globular clusters are often exploited for studying their quiescent thermal emission. There are numerous thermally-emitting quiescent neutron star LMXBs in globular clusters \citep[e.g.][]{heinke2003,guillot2009,maxwell2012} and some of these are sufficiently bright to obtain good-quality spectra that allow for $M$ and $R$ constraints from detailed spectral fitting (see Section~\ref{subsubsec:globclust}). Moreover opposed to neutron star LMXBs in the field, the quiescent X-rays of globular cluster sources are often strikingly non-variable and have strong constraints on the absence of any hard X-ray emission component \citep[e.g.][]{guillot2011,Heinke.etal:14,bahramian2015,walsh2015,bogdanov2016}.

%%%%
% atmosphere models for qneutron stars
\subsubsection{Model atmospheres for quiescent neutron stars}\label{subsubsec:qatmosphere}
Since a strong stellar magnetic field can alter the ionization energies and opacities of the atmosphere, it is easier to study neutron stars that are expected to have a relatively low surface magnetic field (i.e. to reduce the number of free parameters involved in shaping the spectrum), like those in LMXBs (see Section~\ref{subsec:classes}). The atmosphere models usually applied to quiescent neutron star LMXBs assume a uniform and isotropic surface temperature, a negligible magnetic field strength ($B\lesssim 10^{9}$~G), and ignore the effect of rotation (see Section~\ref{sec:surface}). \citet{rutledge1999,rutledge2001_aqlx1,rutledge2001_cenx4} presented the first studies where a non-magnetic neutron star atmosphere model \citep[][]{Zavlin.etal:96} was applied to thermal spectra of quiescent neutron star LMXBs. This delivered the first broad constraints on the stellar radii. 

Due to the rapid gravitational settling of heavy elements \citep[][]{romani1987}, for accretion rates of $\dot{M} \lesssim 10^{14}~\mdotgs$ ($\lesssim 10^{-12}~\mdot$), the atmosphere of quiescent neutron stars is thought to consist of light elements \citep[][]{bildsten1992}. However, if mass is still supplied to the neutron star at a sufficiently high rate of $\dot{M}\gtrsim 10^{14}~\mdotgs$, heavy elements may be dumped into the atmosphere fast enough not to stratify and hence a (potentially measurable) metal abundance may exist \citep[e.g.][see also Section~\ref{subsec:other}]{bildsten1992}. 
%if spallation does not occur, which depends on the accretion geometry; 
Moreover, some neutron star LMXBs are known to accrete from H-poor companions \citep[e.g.][]{bildsten2004,ivanova2008}. In particular in globular clusters a significant fraction of the neutron star LMXBs have small orbits that can only fit small, H-deficit donor stars \citep[e.g.][for a recent list]{bahramian2014}. 

The impact of He-rich atmospheres on radius determinations has been investigated in a number of works \citep[e.g.][]{servillat2012,catuneanu2013,Heinke.etal:14}. This revealed that He-atmosphere models for low magnetic field neutron stars resulted in significantly larger masses and radii than fits with H atmospheres \citep[see also][]{lattimer2014}. This can be ascribed to the larger difference between the effective and color temperatures for He atmospheres, and underlines again the importance of the atmosphere composition in these studies. Nevertheless, as shown by \citet{bogdanov2016}, even minute traces of H in a He-rich donor can still lead to a H-dominated atmosphere. Furthermore, if residual accretion occurs and spallation is effective, H could also be produced in the atmosphere \citep[][]{bildsten1992,bildsten1993}.

%%%%
% constraints from globular cluster sources
\subsubsection{Observational constraints from globular cluster sources}\label{subsubsec:globclust}
Individual mass-radius measurements of quiescent LMXBs typically have uncertainties that are too large to provide meaningful constraints on the EOS \citep[e.g.][]{Heinke.etal:06,webb2007}. However, fairly tight constraints can be obtained when statistical techniques are applied to an ensemble of sources under the assumptions that these neutron stars exhibit the same radius. This approach is motivated by the fact that for many of the most plausible EOSs, the radius remains constant for different masses \citep[][see Figure~\ref{fig:eosmr} right]{lattimer2001}. A statistical Bayesian analysis technique was applied by \citet{guillot2013} to a sample of 5 quiescent neutron star LMXBs in globular clusters. This resulted in a joint radius of $R=9.1^{+1.3}_{-1.5}$~km for a $M=1.4~\Msun$ neutron star. This work was extended by \citet{guillot2014} after adding one more globular cluster source and including new data for a previously analysed cluster, arriving at a similar measurement of $R=9.4\pm 1.2$~km, which would reject several EOSs.  However, not all statistical uncertainties were explored in full \citep[see Section~\ref{subsubsec:qbiases};][]{Heinke.etal:14,lattimer2014}. Perhaps the most important source of bias in this work is the assumption that all neutron stars have pure-H atmospheres. As discussed in Section~\ref{subsubsec:qatmosphere}, it is likely that some globular cluster sources accrete from H-poor companions, which may have a big impact on the resulting radius measurements. Most recently, \citet{steiner2017} analysed a sample of 8 globular cluster sources applying a Bayesian formalism and scrutinizing several uncertainties such as distances, atmosphere composition and surface temperature inhomogeneities (see Section~\ref{subsubsec:qbiases}). With this conservative treatment, it was found that a $M=1.4~\Msun$ neutron star most likely has a radius of $R=10-14$~km. Furthermore, this work showed that tighter constraints are only possible when stronger assumptions are made about the atmosphere composition of the neutron stars, the systematics of the observations, or the nature of dense matter \citep[][]{steiner2017}.

Statistical approaches have also been applied to obtain mass-radius constraints from combining samples of quiescent globular cluster sources with X-ray bursters. \citet{steiner2010} used a Bayesian framework to combine the mass-radius results of 3 globular clusters with that of 3 X-ray bursters obtained from the touchdown method (see Section~\ref{subsec:bursts}). Additional constraints implied by for instance causality and a theoretical minimum neutron star mass were included in this analysis, which led to preferred radii of $R=11-12$~km for a $M=1.4~\Msun$ neutron star. This analysis was followed-up by using an extended sample of 6 globular cluster sources \citep[from][]{guillot2014}, accounting for the discovery of a $M\approx2~\Msun$ neutron star \citep[][]{demorest2010}, and allowing for the possibility of He atmospheres. This led to a preferred radius of $R=10.5-12.7$~km, ruling out a number of hard EOSs \citep[][]{steiner2013}.

An alternative Bayesian formalism was developed in which measured masses and radii are mapped to pressures at 3 fiducial radii \citep[e.g.][]{read2009,ozel:09}. This approach makes use of the fact that the radii of neutron stars are only sensitive to the EOS in a fairly narrow range of densities ($\rho \sim 2-7.5~\rho_0$), so that the relevant equations can be mathematically described by only sampling a small number of points in this density range. This method was applied to a combined sample of globular cluster sources \citep[the same as in][]{guillot2014} and 5 X-ray bursters (touchdown method) by \citet{ozel2016}. Additional constraints from other types of analysis were also imposed, including the requirement to allow for a $M=2~\Msun$ neutron star, and results of laboratory experiments in the vicinity of the nuclear saturation density \citep[][and references therein]{tsang2012,lattimer2013}. Taken together, this led to a narrow preferred radius range of $R=10-11$~km (assuming pure H atmospheres for all globular cluster sources). This approach was further extended by \citet{bogdanov2016}, who increased the globular cluster sample with new data from two more sources. The empirical EOS that is inferred from this most recent analysis is consistent with relatively small radii of $R=9.9-11.2$~km around $M=1.5~\Msun$. This would suggest a fairly soft EOS, with a lower pressure above $\rho = 2 \rho_0$ than predicted by a number of hard, nucleonic EOSs \citep[][]{bogdanov2016}.

\subsubsection{Observational constraints from field LMXBs}\label{subsubsec:field}
There have also been a few attempts to constrain masses and radii of neutron stars in field LMXBs. These efforts have concentrated on sources that are relatively bright in quiescence. One of such sources is EXO 0748--676 (shown in Figure~\ref{fig:spec}), which was recently studied by \citet{cheng2017} in an attempt to constrain its mass and radius. However, this analysis exposed a worrisome dependence on the energy range over which the fits were performed with a best-fitting  mass and radius of $M\approx 2~\Msun$ and $R\approx11$~km for 0.3--10 keV and $M\approx 1.5~\Msun$ and $R\approx 12$~km for 0.3--10 keV. This was ascribed to the strong energy-dependence of the applied neutron star atmosphere model \citep[][]{cheng2017}. A similar type of analysis was recently attempted for the well-studied transient LMXB Aql X-1, leading to $M\approx 1.2~\Msun$ and $R\approx 10.5$~km \citep[][]{li2017_aqlx1}. However, these measurements were only marginally consistent with those obtained from analysing its X-ray bursts \citep[via both the touchdown and the cooling-tail method;][]{li2017_aqlx1}. This is possibly related to the large number of systematic biases that come into play for the quiescent method, as described in Section~\ref{subsubsec:qbiases}. For instance, Aql X-1 exhibits a power-law emission component in its quiescent spectra \citep[e.g.][]{cackett2011} and the neutron star likely continues to accrete at a low level \citep[e.g.][]{cotizelati2014}.

%%%%
% systematic uncertainties for the thermal qneutron stars approach
\subsubsection{Biases and uncertainties in studies of quiescent LMXBs}\label{subsubsec:qbiases}
Several works have discussed the different sources of systematic uncertainties for inferring masses and radii of quiescent LMXBs in detail \citep[e.g.][]{Heinke.etal:14,ozel2015,bogdanov2016}. We summarize these here:

\vspace{+0.2cm}
\noindent \textbf{Distance:} Since to first order we constrain the quantity $R/D$, uncertainties in the distance directly translate into uncertainties in $R$. For transient LMXBs in the field, distances are usually inferred from X-ray burst analysis (by assuming that the peak of an X-ray burst reaches the Eddington limit). However, the Eddington limit depends strongly on the atmosphere composition (see Section~\ref{subsubsec:atmosburst}) and uncertainties can easily be as large as $\sim 20-50$\% \citep[e.g.][]{kuulkers2003}. For globular clusters distances can be more reliably measured through a number of techniques, yielding smaller uncertainties of $\sim 5-10$\%. 

\vspace{+0.2cm}
\noindent \textbf{Atmosphere composition:} The thermal emission observed from neutron stars is shaped by the atmosphere. In case of quiescent neutron star LMXBs, it was demonstrated that a different composition (H versus He) has a strong effect on the inferred masses and radii \citep[see Section~\ref{subsubsec:qbiases}; e.g.][]{servillat2012,catuneanu2013,Heinke.etal:14}. For neutron star LMXBs that display accretion outbursts, information on the chemical composition of the accreted matter can be obtained e.g. from their X-ray burst properties or from their optical spectra. However, the quiescent LMXBs in globular cluster typically have never shown an accretion outburst \citep[e.g.][for a discussion]{wijnands2013} and hence the exact composition of their atmospheres is unknown. The exceptions are $\omega$Cen, which has strong H features in its spectrum, and 47 Tuc X5, which has a long orbital period measured from X-ray eclipses and must thus contain a H-rich donor.

\vspace{+0.2cm}
\noindent \textbf{Modeling of the interstellar absorption:} The soft, thermal X-ray emission received from LMXBs is altered by interstellar extinction and a reliable estimate of the absorption column is therefore required for accurate $M$--$R$ determinations. Particularly if the absorption column is high, absorption edges due to metals become prominent and modeling the X-ray spectrum is then sensitive to the assumed abundances for the ISM \citep[e.g.][]{juett2004,pinto2013,schulz2016}. The latter issue \citep[explored by][]{Heinke.etal:14} can be circumvented by targeting objects with low interstellar absorption. % Heinke+2014 specifically addressed the strong influence of ISM abundances on the radius measurements. 
\citet{bogdanov2016} studied the effect of the changing absorption column in the high-inclination source 47 Tuc X5. It was found that the inclusion of episodes of strong absorption lowered the inferred neutron star radius.

\vspace{+0.2cm}
\noindent \textbf{CCD pile-up:} Even for dim quiescent LMXBs, the received count rate might be higher than the readout time of the CCD, causing multiple events to be recorded as single events with an artificially high energy. In general, pile-up causes a hardening of the X-ray spectrum. Spectra that suffer from pile-up can be corrected at the expense of loss of counts by ignoring the piled-up pixels in the data extraction, or by modeling the pile-up with a dedicated spectral model. The effect of CCD pile-up on mass and radius measurements of quiescent LMXBs was addressed specifically by \citet{bogdanov2016}. This analysis showed that even a pile-up fraction as low as $\sim$1\% can have a huge impact; not only are the confidence contours enlarged due to the statistical uncertainty in the pile-up model parameter, but there is also a displacement to lower $M$ and $R$ values if pile-up is not accounted for (caused by the artificial hardening of the spectrum). Some older works typically left a pile-up fraction up to a few percent uncorrected for, and hence may need to be revisited.

\vspace{+0.2cm}
\noindent \textbf{Instrument calibration uncertainties:} Since inferring the mass and radius requires an absolute determination of the thermal flux from the neutron star, it directly relies on the calibration of the instrument that is used for the measurement. To account for this, a systematic error of a few percent is often included in the analysis \citep[e.g.][]{guillot2013}.

\vspace{+0.2cm}
\noindent \textbf{Energy range considered for spectral fitting:} A study of the quiescent neutron star LMXB EXO 0748--676 recently highlighted that the energy range over which the thermal fits are performed (0.3--10 or 0.5--10 keV) can have a profound impact on the resulting mass and radius measurement \citep[][]{cheng2017}. 

\vspace{+0.1cm}
\noindent \textbf{The occurrence of residual accretion:} If accretion continues in quiescence, the assumptions of a steady state and passively cooling atmosphere, a purely thermal flux, or a uniform surface temperature may no longer be valid. \citet{elshamouty2016} performed a dedicated study of the impact of surface temperature inhomogeneities on radius measurements for quiescent LMXBs. For this study the sources X5 and X7 in the globular cluster 47 Tuc were used, as well as the field LMXB Cen X-4. Assuming that residual accretion would cause pulsed X-ray emission, current limits on the pulsed fraction of these quiescent neutron stars (on the order of $\sim$10\%) allow the radii to be underestimated by $\sim 10\%-30\%$ \citep[][]{elshamouty2016_2}. Based on this result, it has been argued that improving constraints on the presence of pulsations from quiescent LMXBs may be essential for progress in constraining their radii.

\vspace{+0.2cm}
\noindent \textbf{The presence of a hard spectral component:} The presence of a non-thermal emission component, even if it cannot be detected, can potentially bias the mass and radius measurements. In particular, the presence of a power-law emission tail would harden the spectrum and thus lead to a higher temperature measurement and underestimated radius. This issue  was specifically addressed by \citet{bogdanov2016} for X5 and X7 sources in 47 Tuc. It was found that for these two particular objects not accounting for the presence of an undetectable power-law emission component (with contributions to the total unabsorbed 0.5--10 keV flux constrained to be $\lesssim 0.2$\% and $\lesssim 1.6$\% for X7 and X5, respectively) would result in a $\sim 0.5$\% change in the neutron star radius confidence limits.

%%%%%%%%%%%%%%%%%%%%%%%%%%%%%%
%%%%% X-ray timing 
%%%%%
\subsection{X-ray pulse profile modeling}\label{subsec:timing}   
In Sections~\ref{subsec:bursts} and~\ref{subsec:thermal}, we presented different ways to obtain radius (and mass) measurements by modeling the X-ray spectra of neutron star LMXBs during X-ray bursts and quiescent episodes. A critical assumption in those approaches is that the neutron star is spherically symmetric and homogeneously emitting. However, there is a sizable number of neutron stars that display pulsed X-ray emission, modulated at their spin period, from surface hotspots (see Section~\ref{sec:surface}). The exact shape of these \textit{pulse profiles} (or waveforms) is affected by relativistic Doppler shifts, aberration, and light bending and hence depends on the compactness of the neutron star. Accurate modeling of the pulse profiles can thus provide $M$ and $R$ constraints \citep[e.g.][]{lo2013,ML15}. 

There are different circumstances in which surface hotspots are produced. For instance, the radio pulsar mechanism is thought to produce energetic electrons and positrons that collide the polar caps \citep[e.g.][]{ruderman1975,arons1981,harding2001} and thereby create hot spots that give rise to pulsed thermal emission. For accreting neutron stars, hotspots may occur when the stellar magnetic field concentrates the accretion flow onto the polar caps \citep[e.g.][]{pringle1972,rappaport1977,finger1996,wijnands1998}. In addition, unstable thermonuclear burning is sometimes confined to specific parts of the neutron star and produces so-called \textit{burst oscillations} at/near the spin of the neutron star \citep[][]{strohmayer1996}.\footnote{The oscillations seen during the rise are thought to come from spreading of the burning front that is modulated by the neutron star spin period \citep[e.g.][]{strohmayer1997}, whereas the rapid variability seen during the cooling tails are thought to be associated with oscillatory behavior of the surface \citep[``surface modes''; e.g.][]{muno2002,piro2005}.} The temperatures of these various types of hotspots are such that these can be observed at X-ray wavelengths. 
%{muno2002,muno2003,heyl2004,piro2005,narayan2007}

Pulsed X-ray emission is also observed for some isolated (slowly rotating) neutron stars, where the hotspots are possibly due to preferential leakage of heat from the crust/core along paths with a certain magnetic field orientation \citep[e.g.][]{potekhin2001}. However, the most accurate constraints from X-ray pulse timing can be obtained for neutron stars that spin rapidly, because Doppler boosting becomes more pronounced with increasing spin and this helps to reduce degeneracies in the data. Therefore, the technique of X-ray pulse profile modeling has mainly focussed on LMXBs and millisecond radio pulsars. The basic model and methodology of this technique was recently reviewed by \citet{bogdanov2016}. Here, we briefly discuss the concept, main results and its challenges.

There is a large number of parameters that shape the pulse profile. Some of these are geometrical factors, such as the angle between the rotation axis and our line of sight, the angle between the rotation axis and the center of the hotspots, as well as the geometry of the hotspots (see Figure\,\ref{sv_fig11}, left). Another important parameter is the angular distribution of the emergent radiation. While this is not important for a homogeneously emitting spherical star, 
% because the emergent flux is averaged over the stellar surface
hotspots are observed at different angles for different rotational phases. These parameters together set the observed
pulse profile of slowly rotating stars in Newtonian gravity, where their influence scales with the stellar radius and does not
depend on the mass. However, due the compactness and rapid spin of neutron stars, relativistic effects can become important and this introduces a mass dependence. The masses and radii of rapidly rotating neutron stars can thus be inferred from their X-ray pulse profiles. We consider the main relativistic effects separately.

\subsubsection{Relativistic effects: light bending}\label{subsubsec:releffects}
\begin{figure}
\begin{center}
\resizebox{0.99\textwidth}{!}{
  \includegraphics[width=0.8\textwidth]{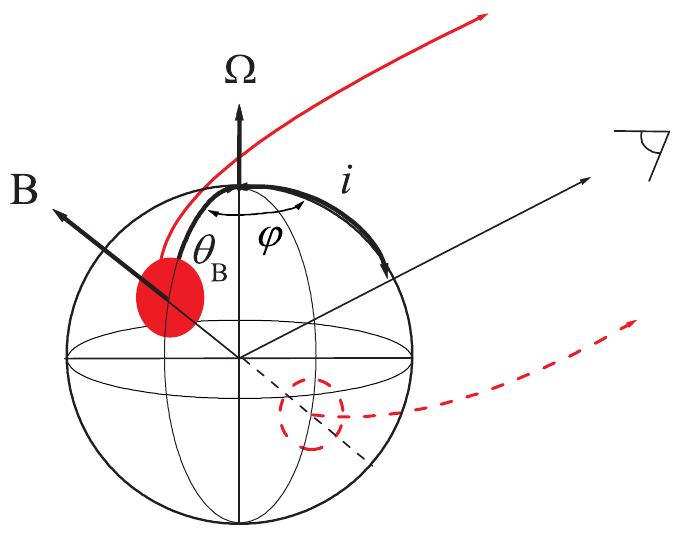}
  \includegraphics{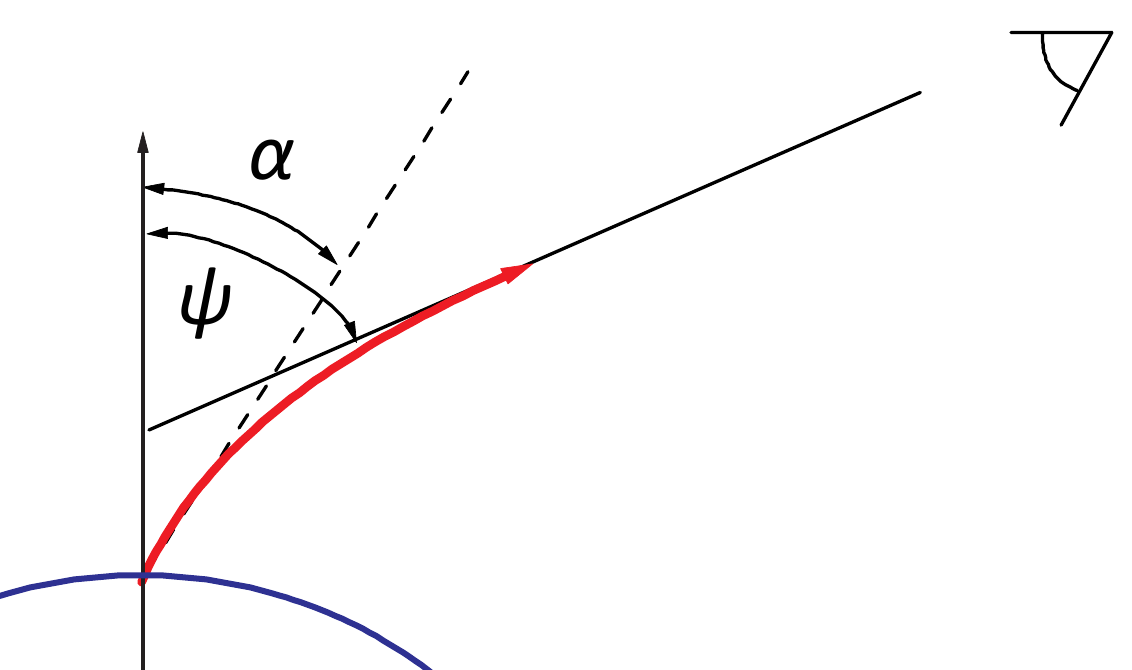}
}
\caption{Left: Geometry of a hotspot on the neutron star surface and relevant angles. $B$ and $\Omega$ indicate the magnetic and rotational axis, respectively, $i$ is the inclination angle of the rotation axis to the line of sight, $\theta_{\rm B}$ is the hotspot co-latitude, and $\varphi$ is the rotational phase angle. Right:  The emitted $\alpha$ and the observed $\psi$ angles of a light ray from the neutron star normal. 
}
\label{sv_fig11}   
\end{center}
\end{figure}

In general relativity, light rays do not travel in straight lines but rather along geodesic curves. The shape of these light trajectories depends on the geometry of space-time. We start with considering light bending in the Schwarzschild geometry (i.e. ignoring spin). In this
case the light trajectory lies in one plane and only two angles need to be connected to the surface normal: the emitted
angle $\alpha$ and the observed angle $\psi$ (see Figure\,\ref{sv_fig11}, right).  The correct connection is given by the
integral \citep{PFC83}:
\be
  \psi=\int_{R}^{\infty}\,\frac{dr}{r^2}\left[\frac{1}{b^2}-\frac{1}{r^2}\left(1-\frac{R_{\rm S}}{r}\right)\right]^{-1/2},
\ee
where
\be \label{imp_f}
   b=R(1+z)\sin\alpha
\ee
is the impact parameter, and $R_{\rm S}=2GM/c^2$ is the Schwarzschild radius. 
This integral allows us to compute the angle between the light ray and the surface normal at every
distance $r$ from the neutron star. A simple and useful analytical approximation of this integral was made by \citet{B02}:
\be \label{lb_b}
    1-\cos\psi \approx (1-\cos\alpha)(1+z)^2. 
\ee

This relation is sufficiently accurate for  $R> 2R_{\rm S}$ and is widely used for light bending computations. 

\begin{figure}
\begin{center}
\resizebox{0.8\textwidth}{!}{
  \includegraphics[width=0.8\textwidth]{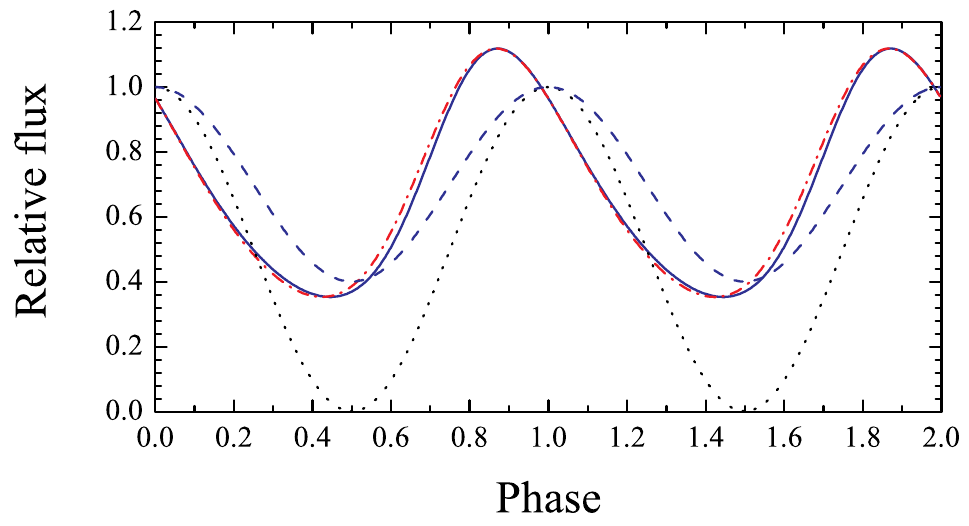}
}
\caption{The observed bolometric flux versus phase computed for a small hotspot on the surface of a rotating ($\nu_{\rm rot}$=400 Hz) neutron star with $M=1.4~\Msun$, $R=2.5~R_{\rm S}$, and angles $i=\theta_{\rm B}=45\deg$ (solid curve).
Also shown are the light curves computed without any relativistic effects (dotted curve), with the light bending only (dashed curve), and without the time delay taken into account (dash-dotted curve). We created these curves using the same methods and parameters that produced figure~2 of \citet{PB06}. 
}
\label{sv_fig12}   
\end{center}
\end{figure}

\subsubsection{Rapidly rotating spherical neutron stars}
There are three principal effects caused by the rapid rotation of neutron stars: the Doppler effect, the time delay and the oblateness of a neutron star. The first two effects can be taken into account even assuming a spherical form of a rapidly rotating neutron star. In what follows, we use the description of \citet{PB06}. The equation for the spectrum of the unit surface $dS'$ is:
 \begin{equation} \label{spot_sp}
  dF_{\rm E} = \frac{\delta^4}{1+z}\,I'_{\rm E'}(\alpha')\cos\alpha\,\frac{{\rm d}\, \cos\alpha}{{\rm d}\, \cos\psi}
  \frac{{\rm d}S'}{D^2},
\end{equation}
where $D$ is the distance to the neutron star and $\delta$ is the Doppler factor, given by:
\be
  \delta=\frac{1}{\gamma\,(1-\beta\cos\xi)}.
\ee

Here $\gamma=1/\sqrt{1-\beta^2}$, $\beta={\rm v}/c$, and $\xi$ is the angle between the direction of the velocity vector ${\rm v}$
and the line of sight.  The velocity is determined as
\be \label{vel}
   {\rm v} = 2\pi\nu_{\rm rot}\,R\,\sin\theta\,(1+z),
\ee
where $\nu_{\rm rot}$ is the neutron star spin frequency, and $\theta$ is co-latitude of the given point.
The observed photon energy $E$ is shifted relative to the emitted energy $E'$ both due to the gravitational redshift and the Doppler effect:
\be
    E = \frac{\delta}{1+z}\,E'.
\ee
The aberration, which changes the observed inclination of the surface unit 
to the line of sight, is taken into account using the Doppler factor:
\be
       \cos\alpha'=\delta\,\cos\alpha.
\ee
The transformation factor is (${\rm d}\, \cos\alpha$)/(${\rm d}\, \cos\psi) \approx (1+z)^{-2}$, if Beloborodov's  approximation (\ref{lb_b}) is used. The bolometric flux of the surface unit is:
\be \label{spot_bl}
  dF = \frac{\delta^5}{(1+z)^2}\,I'(\alpha')\cos\alpha\,\frac{{\rm d}\, \cos\alpha}{{\rm d}\, \cos\psi},
  \frac{{\rm d}S'}{D^2}.
\ee

where $I'(\alpha')$ is the emergent radiation intensity. Equations\,(\ref{spot_sp}) and (\ref{spot_bl}) allow us to compute the light trajectory of a small hotspot on a rapidly rotating spherical neutron star, if the connection between its spherical coordinates, $\theta$ and $\phi$, and the angle $\psi$ can be established:
\be
  \cos\psi = \cos i \cos\theta + \sin i \sin\theta\cos\phi,
\ee
where $i$ is the inclination angle of the rotation axis to the line of sight. The angle $\xi$ can be also computed:
\be
     \cos\xi = -\frac{\sin\alpha}{\sin\psi} \sin i \sin\phi,
\ee 

where the coordinate $\phi$ is a rotational phase as well. The travel time of the spot emission  to the observer depends on the spot position on the neutron star surface. The difference is small, but it could be significant if the neutron star rapidly rotates and turns over a large angle during a tiny time step. Therefore, the observed phase will differ from the intrinsic neutron star rotational phase
\be
\phi \approx \phi_{\rm obs}- \Delta \phi_{\rm obs},
\ee
where the phase delay depends on the time delay $\Delta t$
\be
     \Delta \phi_{\rm obs} =  2\pi\nu_{\rm rot}\,\Delta t.
\ee
This delay time depends on the impact parameter (\ref{imp_f})
\be
  c\Delta t(b) = \int_R^\infty \frac{dr}{1-R_{\rm S}/r}\left\{\left[1-\frac{b^2}{r^2}\left(1-\frac{R_{\rm S}}{r}\right)\right]^{-1/2}-1\right\},
\ee
and there is a sufficiently accurate approximation for this expression:
\be
  c\Delta t = (1-\cos\psi)\,R.
\ee 
The influence of all described effects on the bolometric light curve of the rapidly rotating neutron star is shown in Figure\,\ref{sv_fig12}.

\begin{figure}
\begin{center}
\resizebox{0.9\textwidth}{!}{
  \includegraphics[width=0.99\textwidth]{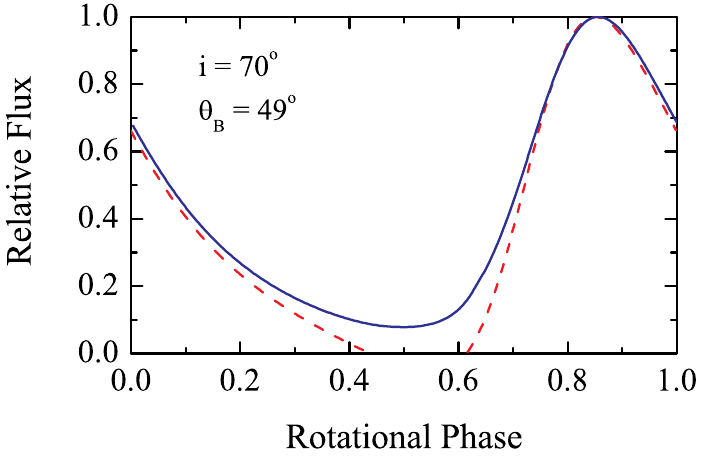}
  \includegraphics[width=0.99\textwidth]{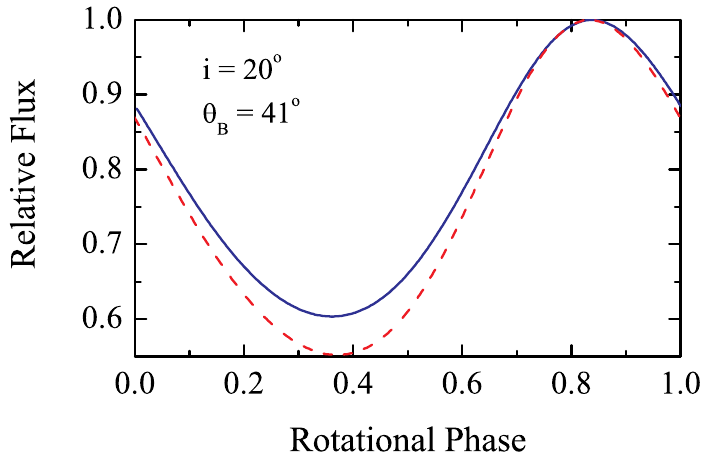}
}
\caption{The observed bolometric flux versus phase computed for one small hotspot on the surface of a rotating ($\nu_{\rm rot}$=600 Hz) oblate neutron star with $M=1.4~\Msun$, an equatorial radius $R_{\rm eqv}=16.4$\,km, and angles $i=70\deg$, $\theta_{\rm B}=49\deg$ (left), and $i=20\deg$, $\theta_{\rm B}=41\deg$ (right). The profiles for spherical neutron stars with  radii equal to the neutron star radius at the hotspot colatitudes (15.1 and 14.8 km) are also shown by dashed curves.
We created these plots using the same methods and parameters that produced figures 3 and 4 of \citet{MLCB07}.
}
\label{sv_fig13}   
\end{center}
\end{figure}

\subsubsection{Rapidly rotating oblate neutron stars}
Rapidly rotating neutron stars are oblate, i.e. their shape is not perfectly spherical. Theoretical models of rapidly rotating neutron stars were computed by many authors \citep[see, e.g.][and references therein]{CST94}. Their shape weakly depends on the details of their inner structure, and can be fitted using a few basic parameters only \citep[see, e.g.][]{MLCB07}. The most simple fit was suggested by \citet{AlGM14}:
 
\be
    R(\theta) = R_{\rm eqv}\left(1-\bar\Omega^2(0.788-1.03x)\cos^2\theta)\right),
\ee  

where $x=GM/(c^2R_{\rm eqv})$ and $\bar\Omega = 2\pi\nu_{\rm rot}\,(R_{\rm eqv}^3/GM)^{1/2}$. Here $R_{\rm eqv}$ is 
the equatorial radius. These authors suggested approximate formulae for the neutron star moment of inertia, the 
quadruple momentum, and the surface gravity distribution with the centrifugal force taken into account.

The space-time in the vicinity of a rapidly rotating neutron star differs from the Schwarzschild metric, and is usually considered in the form suggested by \citet{BI76}:
\begin{eqnarray}
  ds^2 &=&-e^{2\nu}c^2dt^2 +\bar r^2\sin^2\theta\, B^2\,e^{-2\nu}(d\phi -\varpi cdt)^2 \\ \nonumber
    && + e^{2(\zeta-\nu)}(d\bar r^2+ \bar r^2\, d\theta^2),
\end{eqnarray}

where $\varpi$ is an angular velocity of a local inertial frame at the neutron star surface. The radial coordinate $\bar r$ is connected with the circumferential coordinate $r$ as $r=B\,e^{-\nu}\,\bar r$  \citep{Friedmanetal.86}. \citet{AlGM14} have also suggested approximate formulae for the coordinate function $B, \nu$ and $\zeta$, which depend on $M$, $R_{\rm eqv}$, 
and $\nu_{\rm rot}$ only. Another approach was taken by \citet{BPOJ12}, using the Kerr metric with a quadrupole 
correction. A correct consideration of the light travel paths in such complicate metrics is not easy. It can be done directly by applying a ray tracing method \citep[see e.g.][]{BPOJ12}, which can be used to consider spectral changes \citep{Baubock.etal:15}. 
For light curve computations a different, simplified approach was used in which the shape  of the rapidly rotated neutron star is treated correctly, but the Schwarzschild metric is used to account for light bending \citep[see, e.g.][]{MLCB07, ML15}. Whereas this is much simpler than the ray tracing method, it gives acceptable results \citep{MLCB07}. In this approach the main effect gives the angle between the radius-vector and the normal to the surface. Figure\,\ref{sv_fig13} shows a comparison of the light curves obtained by this method and those computed for spherical neutron stars. The effects of finite hotspot size were considered by \citet{BPO15}.

\begin{figure}
\begin{center}
\resizebox{0.9\textwidth}{!}{
  \includegraphics[width=0.99\textwidth]{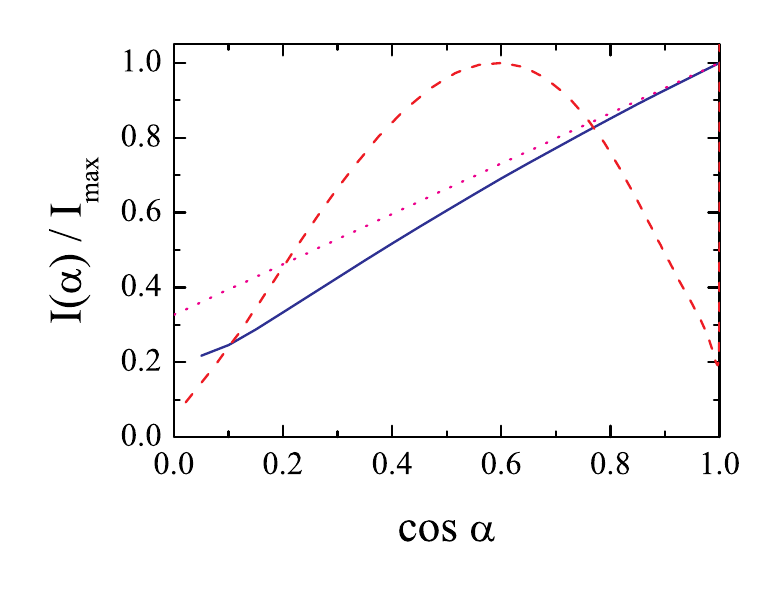}
  \includegraphics[width=0.99\textwidth]{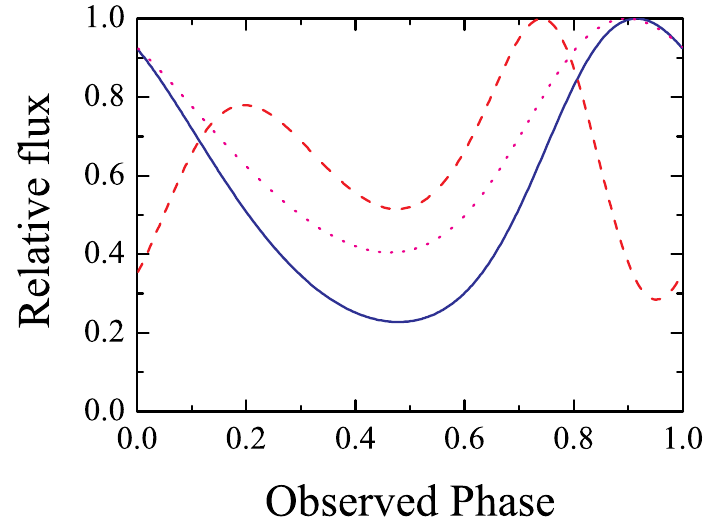}
}
\caption{Left: The model angular distributions computed at a photon energy of $E=1$\,keV for a pure hydrogen
non-magnetized atmosphere \citep[$T_{\rm eff}= 2$\,MK, solid curve;][]{Suleimanov.etal:17} and a highly magnetized
 atmosphere \citep[$T_{\rm eff}= 1.2$\,MK, $B=1.2\times10^{13}$\,G, dashed curve curve;][]{Suleimanov.etal:10}.
The angular distribution for a pure electron scattering atmosphere is also shown by the dotted curve.
Right: The observed flux at an energy of $E=1$~keV versus the observed phase computed for 
a small bright spot on the surface of a rotating ($\nu_{\rm rot}$=600 Hz) oblate neutron star with a mass $M=1.4~\Msun$, equatorial radius $R=16.4$\,km,
and angles $i=45\deg$ and $\theta_{\rm B}=45\deg$. The light curves computed using the model angular distributions for a pure hydrogen non-magnetized atmosphere
(solid curve), pure hydrogen magnetized atmosphere (dashed curve), and for an isotropic angular distribution (dotted curve) are shown.   
}
\label{sv_fig14}   
\end{center}
\end{figure}

\subsubsection{Angular distribution of the emergent radiation}
A hotspot on a neutron star surface is seen by a distant observer at different angles for different rotational phases.
This implies that the angular distribution of the emergent radiation determines the observed flux variation of a rotating neutron star
with a surface hotspot. It is not possible to observationally resolve the angular distribution of the surface radiation, hence theoretical models need to be employed to describe it. Existing neutron star atmosphere models give the required angular distributions of the emergent radiation.

The model angular distribution depends mainly on the temperature structure of the atmosphere 
and the dominant opacity sources at a given photon energy. 
The angular distribution of the emergent radiation computed for hot neutron star atmospheres ($kT_{\rm eff}  > 1$\,keV), 
where the dominant source of electron scattering is close \citep{SPW12} to the Sobolev-Chandrasekhar angular distribution
 derived for a pure electron scattering atmosphere \citep{Chandra60, Sobol63}. 
 We note that this is correct only for sufficiently high photon energies, because the angular distribution at low photon energies, where free-free opacity dominates, becomes close to an isotropic distribution.   

The angular distribution of the radiation emerging from relatively cold neutron stars is more complicated and cannot be described by any 
analytical relations. In this case only numerical distributions given by the model atmospheres have to be used 
\citep[see, e.g.][and references therein]{sb16}.  Recently, the angular distributions for pure hydrogen model atmospheres 
in an effective temperature range of $T_{\rm eff}=0.5 - 10$~MK and for nine surface gravity values were computed and implemented into the X-ray spectral fitting package \textsc{XSpec}  \citep[model \textsc{hatm};][]{Suleimanov.etal:17}. The same was done for pure carbon atmospheres using an effective temperature range of $T_{\rm eff}=1 -4$~MK (\textsc{carbatm} in \textsc{XSpec}).  

The radiation angular distributions  emerging from non-magnetized neutron star atmospheres are peaked relative to the atmosphere normal
({\it pencil beam}). This is not true for highly magnetized neutron star atmospheres. In that case the emergent radiation has a relatively narrow 
(a few degrees, depending on the magnetic field strength) peak near the normal, and a second broad smoothed peak
at inclinations of $\alpha \approx 40 - 60\deg$   \citep{PSVZ94}.  The total amount of energy radiated in the normal peak
is relatively low, and the magnetized model atmospheres produce instead {\it fan-beamed} radiation.
Examples of the angular distributions of non-magnetized and strongly-magnetized neutron star hydrogen atmospheres, as well as corresponding light curves, are shown in Figure\,\ref{sv_fig14}.
  
\subsubsection{Application to millisecond pulsars}
Millisecond radio pulsars with thermally emitting X-ray hotspots at their polar caps are attractive objects
for constraining neutron star radii and masses via the pulse profile modeling technique. So far this has been attempted for three radio pulsars: PSR J0437$-$4715, PSR J0030+0451 and PSR J2124--3358. The procedure was applied for the nearest \citep[156.3$\pm$1.3\,pc;][]{Deller.etal:08} millisecond radio pulsar PSR J0437$-$4715 by \citet{PZ97}. This pulsar has a relatively low spin period of 5.75\,ms  ($\nu_{\rm rot} = 174$\,Hz), so that the effects of oblateness and Doppler boosting can be ignored.  \citet{PZ97} evaluated the neutron star compactness
 $M/\Msun \approx 1.4 - 1.6~R/10$\,km using \textit{ROSAT} observations and fixed angles $i=40\deg$ and $\theta_{\rm B} = 35\deg$ \citep{MJ95}. The obtained neutron star radius is $R\approx10-15$\,km taking into account the mass measured for this neutron star from radio pulse timing \citep[$M=1.76\pm0.2~\Msun$;][]{Verbiest.etal:08}. Studies of PSR\,J0437$-$4715 were continued by using new observations performed with {\it XMM-Newton} and applying hydrogen model atmospheres \citep[][]{sb07,sb13}. In these studies no angles were fixed and the resulting constraint on the neutron star radius is not very strict ($R > 11$\,km). Similar studies were also performed for the next nearest radius pulsar, PSR J0030+0451, and another radio pulsar PSR J2124--3358 \citep[][]{bogdanov2008,bogdanov2009}. Unfortunately no mass measurement is available for these objects and only lower limits on their radii have been obtained so far ($R > 10.7$ and $>7.8$\,km for PSR J0030+0451 and PSR J2124--3358, respectively).

The X-ray pulse profile modeling technique was also applied to three AMXPs: SAX\,J1808.4$-$3658, XTE\,J1814$-$338 and XTE\,J1807$-$294. Since these have higher spin frequencies \citep[up to 600\,Hz; e.g.][]{patruno2017_spin}, the relativistic Doppler effect has to be taken into account. \citet{PG03} modeled the soft and hard X-ray pulse profiles of the first discovered and frequently active AMXP SAX\,J1808.4$-$3658 
($\nu_{\rm rot} \approx 400$\,Hz) assuming a spherical shape and Schwarzschild spacetime, which led to a relatively small neutron star radius ($R\approx8 -11$\,km for $M= 1.4 - 1.6~\Msun$). The main source of uncertainties in these studies is the applied approximations for the angular distributions of the emergent radiation. \citet{LMC08} and \citet{ML11} took into account the time delay effect as well as the neutron star oblateness in analysing the
 X-ray pulse profiles of this source. The obtained mass and radius limitations remain rather wide, $R\approx5-13$\,km for $M=0.8 - 1.7~\Msun$. Using the same technique, the masses and the radii of two other AMXPs were also constrained: XTE\,J1814$-$338 \citep[$\nu_{\rm rot} \approx 314$\,Hz;][]{leahy2009} and XTE\,J1807$-$294 \citep[$\nu_{\rm rot} \approx 191$\,Hz;][]{leahy2011}. The derived confidence regions in the $M-R$ plane are also large for these objects, with radii in the range of $R\approx8-24$\,km and masses of $M\approx1-2.8~\Msun$ for XTE\,J1807$-$294, and $R\approx11-20$\,km and $M\approx1-2.6~\Msun$ for XTE\,J1814$-$338. We note that neutron star radii of $R\approx12$\,km are compatible for all three found confidence regions.  Finally, pulse profile modeling has been performed for two sources with burst oscillations, 4U 1636--536 \citep[][]{nath2002} and XTE J1814--338 \citep[][]{bhattacharyya2005}, but these have not yielded strong constraints either \citep[e.g.][]{weinberg2001,muno2002,muno2003}. Although current mass-radius constraints from X-ray pulse profile studies of different types of neutron stars yielded only loose constraints, there are very good prospects for improving this (Sections~\ref{subsubsec:timingchallenge}~and~\ref{sec:future}).

\subsubsection{Challenges of X-ray pulse profile modeling}\label{subsubsec:timingchallenge}
The reason why this modeling has so far not yielded stringent constraints is clear. Apart from the mass and radius, the pulse profiles depend on geometrical factors such as the size and location of the hotspots as well as the inclination angle between the observer's line of sight and the rotation axis of the neutron star. These geometrical parameters are difficult to determine and also introduce degeneracies with $M$ and $R$. %For instance, whereas increased compactness reduces the pulse fraction, a decrease in the product ($\sin \alpha \sin i$) has the same effect. 
Modeling of AMXPs in particular is complicated due to the contaminating emission from the accretion disk, variations in their pulse profiles and Comptonization in the accretion column \citep[e.g.][]{ozel2013,miller2016_MRreview}. It is expected, however, that these dependencies can be resolved so that $M$ and $R$ can be recovered from detailed modeling of the pulse profile \citep[e.g.][]{psaltis2014,lo2013,ML15}.

Another important factor is the beaming pattern of the hotspot radiation, as high beaming can to some extent mimic the effects of decreased gravitational light bending and Doppler boosting. For X-ray bursts this is well understood from theoretical modeling \citep[e.g.][]{madej:91,SPW12} and not a significant source of uncertainty \citep[][]{miller2013_eos}. The beaming pattern of the hotspots of accreting pulsars, however, is much less understood and diminishes the constraints on $M$ and $R$ that can be obtained through this method \citep[e.g.][]{PG03,leahy2009,leahy2011,ML11}. This is less an issue for the magnetic hotspots of rotation-powered pulsars, although there are some uncertainties about the beaming patterns from hydrogen atmospheres heated by the bombardment of relativistic particles in the magnetosphere \citep[e.g.][]{bogdanov2008}. Another complication of pulse profile modeling for accreting pulsars is the presence of an accretion disk that may block part of the hotspot radiation and introduce harmonic structure in the pulse profile \citep[][]{poutanen2008,poutanen2009,ibragimov2009,kajava2011}.

Despite the existing challenges, the technique of X-ray pulse profile modeling is very promising and there are good prospects for obtaining much better constraints in the (near) future. Firstly, the recently launched \textit{NICER} mission is expected to allow for detailed X-ray pulse profile modeling (see Section~\ref{sec:future}) and accurate methods of constraining neutron star masses and radii using \textit{NICER} data have been developed \citep[see, e.g.][]{ML15, Miller16, SFLM16, ozel2016_nicer, watts2016_review}. Secondly, the relevant geometrical angles may be constrained using X-ray polarization, which could be achieved with several concept missions currently under investigation (Section~\ref{sec:future}). For resolving the problem with the angular distributions, extended theoretical computations have to be performed that allow to include the radiation-dominated shock and the radiation transport through it self-consistently, taking the energy balance into account as well. 

As discussed in Sections~\ref{subsubsec:burstbias} and~\ref{subsubsec:qbiases}, X-ray spectroscopic measurements of accreting neutron stars are subject to a number of systematic uncertainties and biases. This does not appear to be the case, however, for X-ray pulse profile modeling. How to apply the pulse profile method in practice and what is required to obtain meaningful constraints on $M$ and $R$ was investigated by \citet{psaltis2014,lo2013,ML15}. By calculating synthetic pulse profiles under various assumptions, e.g. for the hotspot size, stellar rotation frequency, inclination, the effect of the key pulse profile parameters on $M$, $R$, and other relevant parameters was studied. This showed that the uncertainties in $M$ and $R$ are most sensitive to the stellar rotation (with more rapid rotation resulting in smaller errors), spot inclination and observer inclination, and to a much lesser extent to various background components (instrumental, astrophysical, other emission components). A key result of the parameter estimation studies of \cite{lo2013} and \citet{ML15} is that the mass and radius can be reliably obtained without any strong systematic errors. Currently this is the only method for determining radii for which this is the case.

%%%%%%%%%%%%%%%%%%%%%%%%%%
%%%%% CCOs
%%%%% 
\subsection{Radius constraints from thermally-emitting isolated neutron stars}\label{subsec:cco}
Part of the neutron stars located in the centers of SNRs are magnetars \citep[e.g.][for a review]{MPM15}, whereas others form the separate, small class of CCOs \citep[e.g.][]{Pavlov:etal:02, Pavlov:etal:04}. These objects have thermal X-ray spectra that can be roughly described by black bodies with temperatures less than a few MK, or $kT_{\rm BB} \approx 0.2 -0.6$\,keV. CCOs have typical X-ray luminosities of $L_{\rm X} \sim 10^{33} -10^{34}$\,erg\,s$^{-1}$ and are not detected at other wavelengths. These slowly spinning neutron stars show no sign of radio pulsar activity \citep[e.g. no radio pulsations, no pulsar wind nebula or no non-thermal X-ray emission; see short reviews by][]{GH07, deLuca08, HG10}. 

Individual CCOs are not completely identical \citep[see e.g.
table~1 of][for a list of currently known properties]{Gotthelf:etal:13}. For instance, three CCOs display X-ray
pulsations with periods $P \approx 0.1 -0.5$\,s and their measured period derivatives $\dot P$ suggest dipole magnetic field strengths $B\lesssim10^{11}$\,G \citep[e.g.][and references therein]{Gotthelf:etal:13}. The X-ray spectrum of 1E\,1207.4-5209 shows absorption features associated with the cyclotron line and harmonics \citep{Sanwal.etal:02,Bignami.etal:03, SPW10, SPW12a}. The lowest absorption feature ($\approx 0.7$\,keV) is consistent with the magnetic field estimation $B \sim 10^{11}$\,G if gravitational redshift is taken into account \citep[][]{Gotthelf:etal:13}.

Many of the CCOs have relatively good distance estimates from their associated SNRs, which make them attractive candidates for measuring the neutron star radii from their thermal X-ray emission \citep[see ][for reviews]{Pavlov:etal:02, Pavlov:etal:04, Gotthelf:etal:13}. The apparent radii obtained by fitting black bodies to the observed spectra are only a few km \citep{Pavlov:etal:04}. For the pulsating CCOs the small sizes of the emitting area can be easily explained by the existence of relatively small hotspots on the neutron star surface. However, fitting
the spectra with pure-hydrogen atmosphere models yields reasonable neutron star sizes in
some cases. For example, the radius of 1E\,1207.4$-$5209 obtained with
a black-body fit is $R\approx1-3$\,km, while the radius derived from hydrogen atmosphere
models is $R\approx10$\,km \citep[][]{Zavlin.etal:96}. This underlines the importance of considering appropriate neutron star model atmospheres for inferring stellar radii. 

In this section we focus on the results of neutron star radii determinations of two CCOs, located in the SNRs Cas~A and HESS\,J1731$-$347. These are interesting cases because there are indications that their atmospheres may be carbon rich, and their observed surface temperatures and thermal history give additional information on the properties of the dense matter in their cores.

\subsubsection{Model atmospheres of CCOs}\label{subsubsec:atmoscco}
The technique of determining neutron star radii from the thermal X-ray spectra of CCOs again uses relation (\ref{norm}) of Section~\ref{sec:surface}. The key uncertainty of distance is less of a problem for nearby neutron stars (for which parallaxes can be measured) and those associated with SNRs as their distance can usually be determined reliably and accurately \citep[see, e.g. ][]{Pavlov.etal:00, Rutledge.etal:02}. However, the applied model spectrum also plays a significant role. Fortunately, the atmospheres of non-accreting neutron stars should be chemically homogeneous 
due to gravitational separation \citep{AI80}. As a result, the lightest chemical element of the neutron star envelop dominates the atmosphere chemical composition.  

\begin{figure}
\begin{center}
\resizebox{0.99\textwidth}{!}{%
  \includegraphics{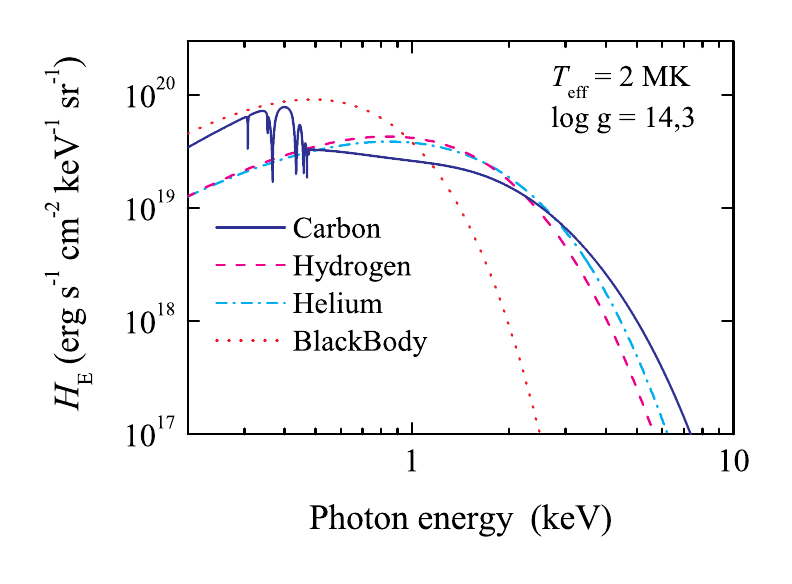}
  \includegraphics{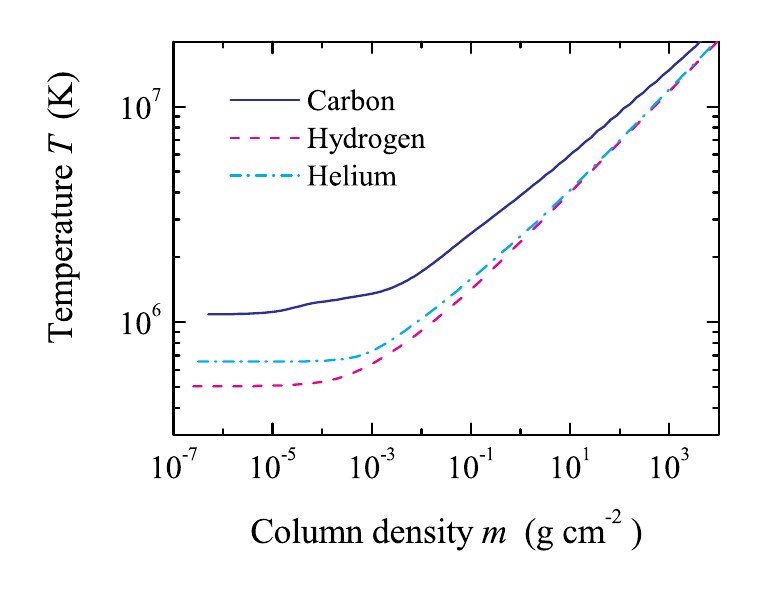}
}
\caption{Model spectra (left) of neutron star atmospheres with various chemical compositions  and their temperature structures (right), all computed for $T_{\rm eff} = 2$\,MK and
$\log g = 14.3$. The corresponding black-body spectrum is also shown in the left panel (red dotted curve).
 }
\label{sv_fig8}   
\end{center}
\end{figure}

As mentioned before, model spectra of fully ionized H and He atmospheres with relatively low effective temperatures (a few MK and less) 
are harder than black-body spectra of the same effective temperatures \citep[see reviews by][and Figure\,\ref{sv_fig8}]{ZP02, Z09}. Moreover, neutron star atmosphere spectra are wider than black-body spectra; when high quality data are available, applying a simple black-body model often requires two temperature components (Figure~\ref{sv_fig8}). When fitting H-atmosphere models instead of black bodies, this leads to larger neutron star radii \citep[see details in ][]{Z09}. Helium model spectra are even harder and slightly more diluted than hydrogen model spectra (Figure\,\ref{sv_fig8}), and therefore result in larger radii still \citep[e.g.][]{Heinke.etal:14}. This is because the bremsstrahlung helium opacity is larger due to its $Z^2$ dependence, and in addition the temperature of the atmosphere is higher because of the increased opacity. Both factors lead to harder spectra than that of H-atmospheres.

The above mentioned effects are even stronger if an atmosphere dominated by C is considered. Carbon is not completely ionized at the typical temperatures of neutron stars, and the photo-ionization opacity is very strong near the photo-absorption edge of the hydrogen-like carbon ion ($\approx0.49$~keV). The presence of this edge leads to a lower flux at the blue side of the edge, and energy prefers to escape even at higher photon energies than in H or He atmospheres \citep{Ho:Heinke:09}. As shown in Figure\,\ref{sv_fig8}, the C-atmosphere spectra are harder and more diluted than the H ones. This leads to larger neutron star radii obtained from the spectral fit for a given distance \citep{Ho:Heinke:09}. 

\subsubsection{Cooling of isolated neutron stars}
As mentioned in Section~\ref{subsec:thermalevo}, neutron stars gradually cool down after being formed in a supernova explosion. The cooling trajectory after birth can be divided into two main stages. During the first $\sim 10^5 - 10^6$\,yr a neutron star cools mainly due to neutrino emission from its dense core. Once the core temperature drops below $T_{\rm B}\sim 10^8$~K, the neutrino emission processes become inefficient and in this second stage a neutron star cools mainly due to radiative losses from its surface. Young neutron stars, which are still in the neutrino cooling stage, are of particular interest because their surface temperatures are high enough (on the order of 1\,MK) to be detected with sensitive X-ray satellites. There are a number of excellent works that detail neutron star cooling theory \citep[e.g.][]{YP04,page2006,Weiskopf.etal:11}. Here we briefly summarize the key elements of the thermal evolution of neutron stars.  

Neutrinos can be produced in a variety of particle interactions in the dense neutron star core. The efficiency at which neutrinos are produced depends sensitively on its interior density and composition. In general, the more massive a neutron star is, the more rapidly it cools \citep[e.g.][]{lattimer2001}. Neutrons and protons in the core are likely in a superfluid state, which affects the efficiency of neutrino cooling \citep[e.g.][]{Gusakov.etal:04,YP04,page2009,wijnands2013}. On the one hand, the pairing of nucleons in Cooper pairs decreases the primary neutrino emission process, but on the other hand the forming and breaking of Cooper pairs itself can generate a strong splash of neutrino emission. This can lead to short-lived, fast cooling episodes in the thermal history of neutron stars, which possibly have been observed in the CCO Cas~A \citep[e.g.][and references therein]{Shternin.etal:11}. 

The temperature of neutron star cores cannot be directly measured, but need to be inferred from the observable surface temperature. This is not an easy task. For instance, during the first $\sim 100$~yr, neutron stars are not isothermal because the core cools faster than the crust. As a result the surface temperature stays relatively constant, tracing the temperature of the hot crust, and then drops fast when the crust starts to cool and reaches thermal equilibrium with the core. This phase is called the initial thermal relaxation \citep[e.g.][]{lattimer1994}. During the following neutrino cooling stage, the connection between the core temperature and the surface temperature is governed by the electron thermal conductivity in the neutron star ocean/envelop \citep[see e.g.][]{PCY97}. The efficiency of the thermal conductivity depends particularly on the envelop chemical composition: for the same interior temperature, a light element envelop will yield a higher surface temperature than a metal-rich envelop due to their different thermal conductivities.  Figure\,\ref{sv_fig10} (left) displays examples of neutron star thermal evolution curves for different interior properties, illustrating the different stages of cooling. 

Although the core temperature of neutron stars does not provide direct constraints on the EOS, the efficiency of neutrino cooling gives some insight into its density. Furthermore,  studying neutron star cooling can give additional information about the behavior of ultra-dense matter, particularly the superfluidity of neutrons and protons (Section~\ref{subsubsec:corecool}). Very strong magnetic fields such as encountered in magnetars can change the thermal evolution significantly, but this is not relevant for the moderate magnetic field strengths of CCOs \citep[e.g.][]{Pons.etal:09}.

\subsubsection{The neutron star in Cas A}
\begin{figure}
\begin{center}
\resizebox{0.99\textwidth}{!}{%
  \includegraphics[width=0.75\textwidth]{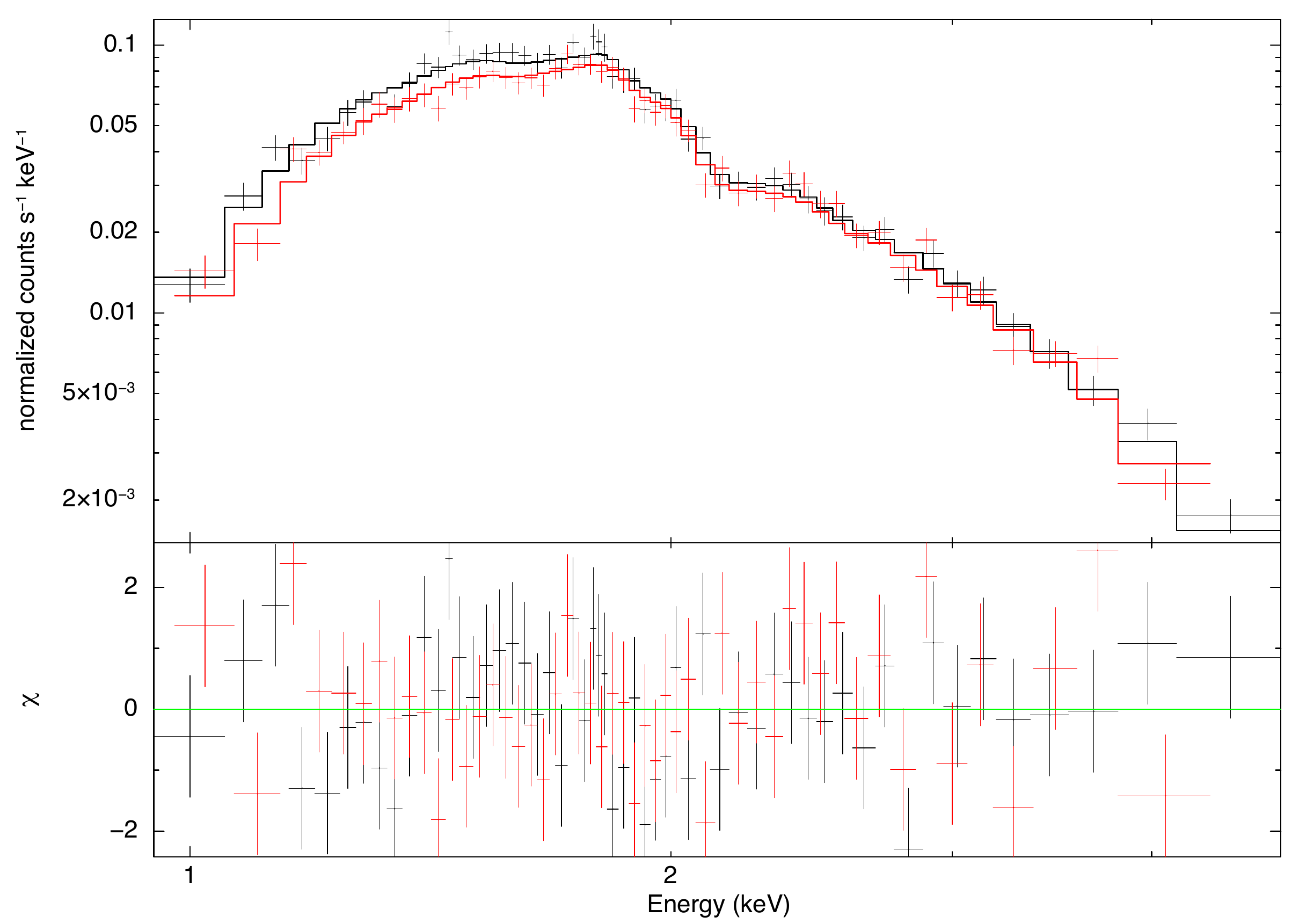}
  \includegraphics{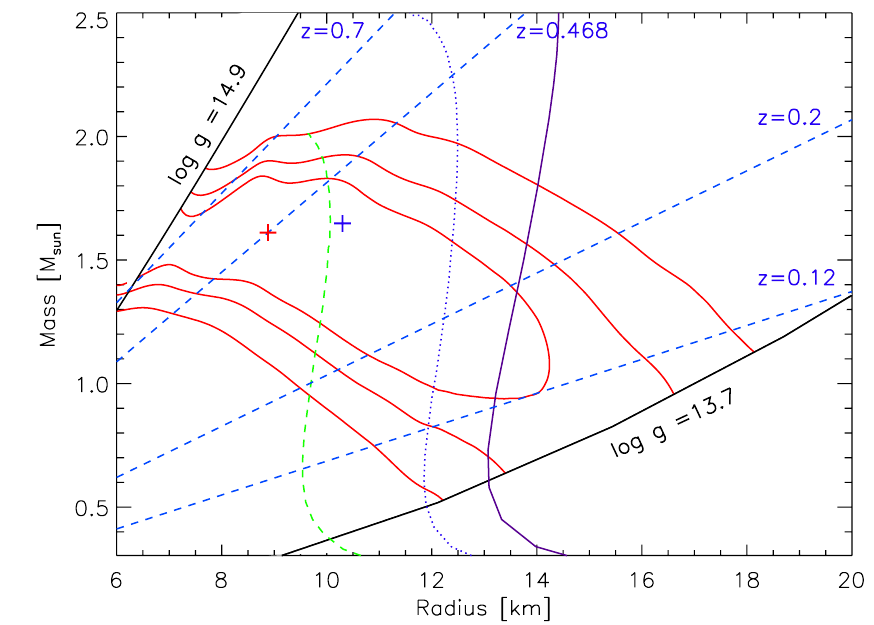}
}
\caption{Left: Spectra of the CCO in Cas A along with fits to a C-atmosphere model ($\log g =14.45$ and $z=0.375$) for the \chan\ observations of 2006 (black) and 2012 (red).  It is assumed that the entire neutron star surface is emitting and that $D$$=$3.4\,kpc \citep[from ][]{Posselt:etal:13}. The bottom panel shows the chi-square contributions. Right: $M-R$ confidence contours (68\%, 90\%, 99\%) for the CCO in Cas A obtained using the C-atmosphere model fit. The two  black curves indicate the range of the considered $\log g$ values, while the dashed blue lines indicate different gravitational redshift parameters. The over-plotted curves are the $M-R$ relations for three possible EOSs according to \citet{Hebeler.etal:13}. The best-fit model is marked with a red cross and the best-fit model obtained by \citet{Heinke:Ho:10} is marked with a blue cross. This plot was kindly provided by B. Posselt \citep[adapted from;][]{Posselt:etal:13}.
}
\label{sv_fig9}   
\end{center}
\end{figure}

The CCO in Cas A was discovered by \chan\ during its first-light observations \citep{T99}. Black-body fits to its featureless X-ray spectrum yielded a high temperature ($T_{\rm BB}=6-8$\,MK) and a small emitting radius ($R_{\rm BB} = 0.2-0.5$\,km) for a distance of 3.4\,kpc \citep{Pavlov.etal:00}. A one-component black-body fit did not provided a statistically acceptable description of a later obtained \chan\ spectrum. The higher-quality data required at least two thermal components, both fitted with H atmospheres, with significantly different temperatures and sizes \citep[$T_1\approx 4.5$\,MK, $R_1\approx 0.4$\,km and $T_2 \approx 1.6$\,MK, $R_2\approx 12$\,km;][]{Pavlov:Luna:09}. Interpreting the smallest and hottest of the two components as a hotspot may not be consistent with the lack of pulsed X-ray emission. The $3\sigma$ upper limit on the pulsed fraction is $\approx16$\%, assuming a sinusoidal pulse shape, and a homogeneous H atmosphere model gives an emitting size of $R\approx 4 - 5.5$\,km.\footnote{For reference, the pulsed fractions of other CCOs are $\approx11$\% for RX~J0822.0--4300 in SNR Puppis A, $\approx9$\% for 1E~1207.4--5209 in SNR PKS~1209--51/52, and $\approx64$\% for  CXOU~J185238.6+004020 in SNR Kes 79 \citep[][]{Gotthelf:etal:13}.} 
% that is only compatible a strange quark star only. 
A possible solution for this discrepancy was provided by \citet{Ho:Heinke:09}, who found that a pure-C atmosphere model provides a good fit to the data and yields a reasonable neutron star size of $R\approx10 - 14$\,km. The observed X-ray spectra of the CCO in Cas~A are shown in Figure\,\ref{sv_fig9} (left) together with the C-atmosphere model fits. The resulting confidence contours on the $M-R$ plane are shown in Figure~\ref{sv_fig9} (right). 

What also makes Cas~A particularly interesting is that it was found to display a significant temperature decrease over $\sim$10~yr time, which would point to unusually fast cooling of the neutron star \citep[][]{Heinke:Ho:10}. The presence of a significant temperature evolution was confirmed in subsequent studies \citep{Shternin.etal:11,Elshamouty.etal:13}, but has been questioned by \citet{Posselt:etal:13}. If the observed rapid cooling in Cas~A is real, it provides very important insight into  the physics of neutron star cores.  In particular, such a fast cooling stage can be accounted for by a neutrino emission splash that results from the transition to a neutron superfluidity phase \citep{Page.etal:11,Shternin.etal:11}, and would be direct evidence that the neutrons in the core are superfluid.

\subsubsection{The neutron star in HESS\,J1731--347}
The CCO at the center of the TeV-emitting supernova remnant HESS~J1731$-$347, also known as G\,353.6$-$0.7, 
was discovered with \emph{XMM-Newton} in 2007 \citep{Acero:etal:09,Tian:etal:10,HESS:2011}. This neutron star emits a black-body like X-ray spectrum with $kT\approx 0.5$~keV and is strongly absorbed at low energies  
\citep[hydrogen column density of $N_{\rm H}\approx 1.5\times 10^{22}$\,cm$^{-2}$;][]{Acero:etal:09,Halpern:Gotthelf:10:b,Bamba:etal:12}. X-ray timing studies yield an upper limit on the pulsed fraction of $\approx$ 7--8\% \citep[for sinusoidal pulsations;][]{Klochkov.etal:15}. The host SNR is most likely located either in the
Scutum-Crux arm at $D$$\approx$3\,kpc or in the Norma-Cygnus arm at $D $$\approx$$4.5$\,kpc, whereas the measured X-ray absorption and $^{12}$CO emission suggest $D>3.2$~kpc \citep[][]{HESS:2011}. 

\begin{figure}
\begin{center}
\resizebox{0.99\textwidth}{!}{%
  \includegraphics[width=0.8\textwidth]{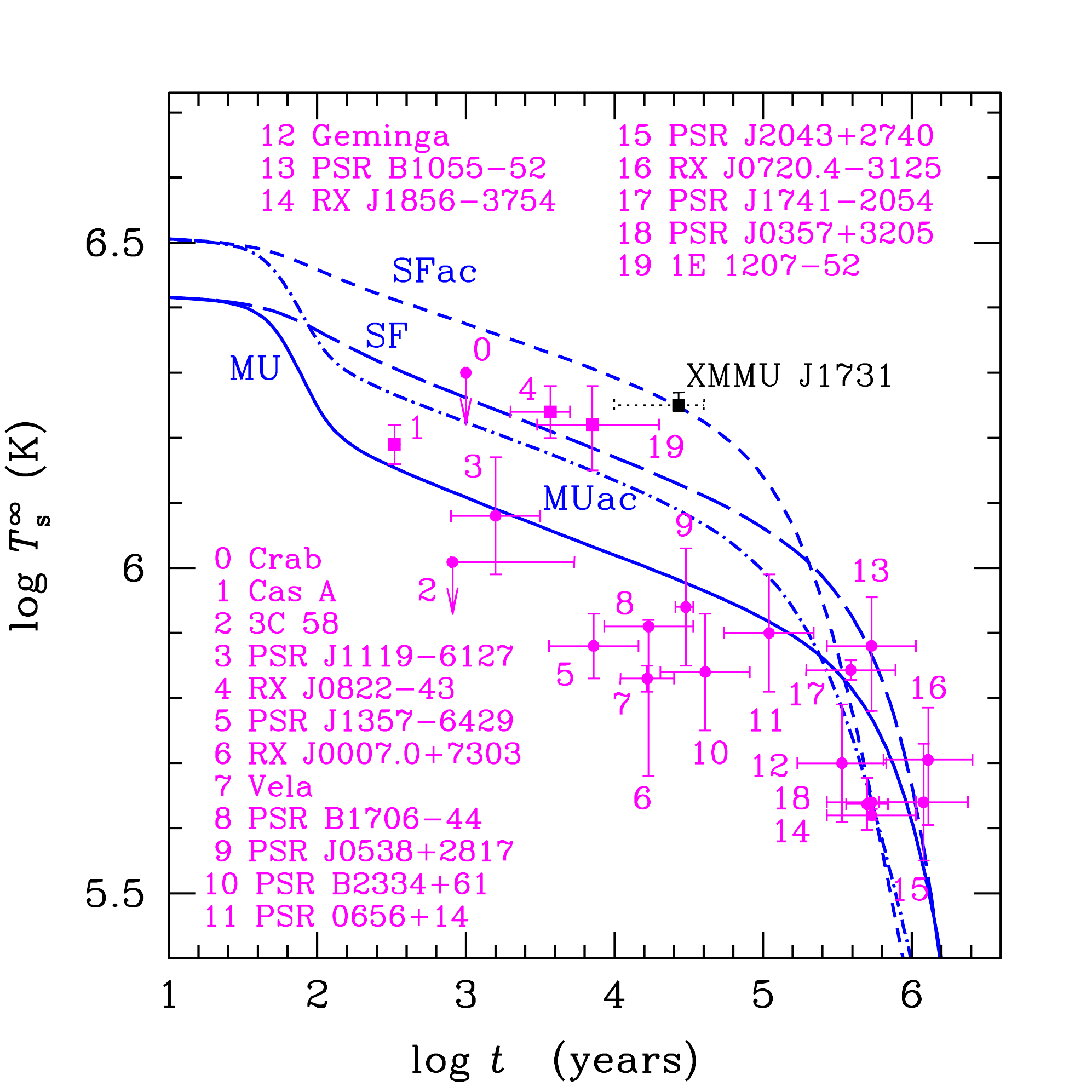}
  \includegraphics{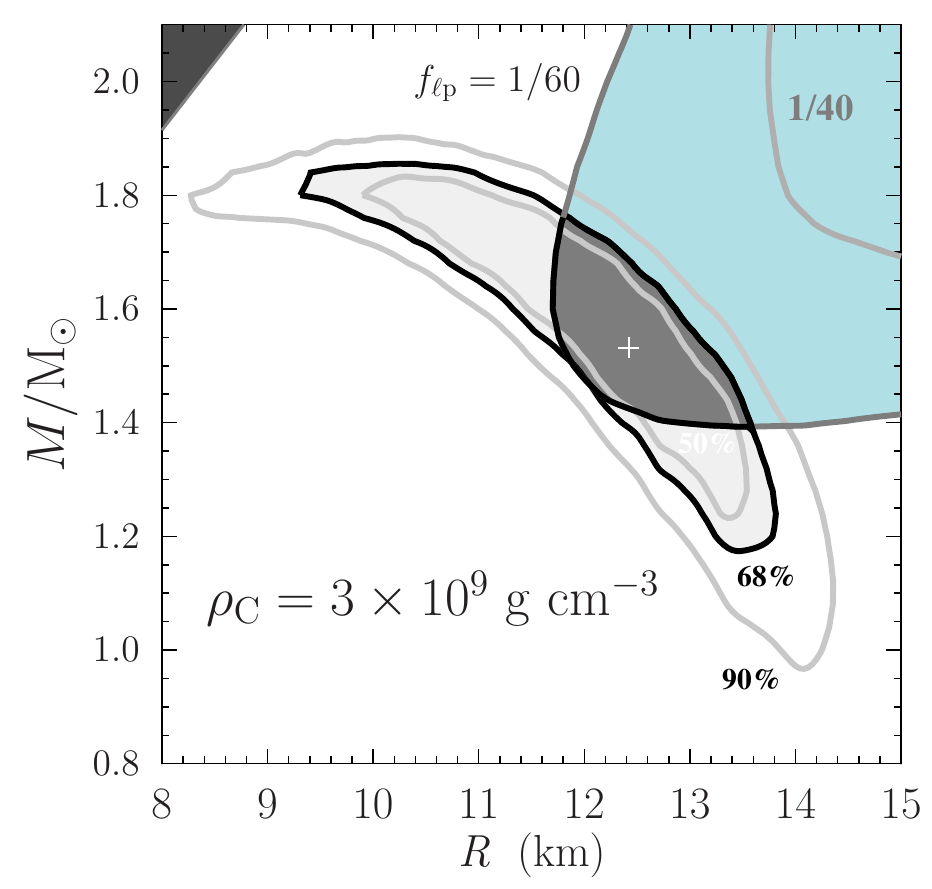}
}
\caption{Left: Surface temperatures  for a number of
cooling isolated neutron stars, including the CCOs in Cas~A (\#1) and HESS\,J1731$-$347 (XMMU\,J1731), versus their ages compared with theoretical  cooling  curves for  a  $M=1.5~\Msun$ neutron star.  Here, MU refers to a non-superfluid star with slow core cooling via modified Urca processes and an envelop of iron, SF is for strong proton superfluidity in the core and
a similar envelop, while MUac and SFac refer to the same models as MU and SF but with a pure-C envelop.  Right: $M$ and $R$ constraints for the CCO in HESS\,J1731$-$247 from fitting the observed spectra and applying cooling theory. It was assumed that the C envelop extents to $\rho=3\cdot 10^{9}$\,g\,cm$^{-3}$ and that the suppression factor for the superfluidity of protons is $f_{\rm \ell p}=$\,1/60. The dark shaded area is the resulting confidence $M-R$ region. For further details we refer to \citet{Ofengeim.etal:15}.
}
\label{sv_fig10}   
\end{center}
\end{figure}

Fitting the X-ray spectrum of the CCO in HESS~J1731$-$347 with an absorbed black-body model for $D=3.2$\,kpc leads to an unrealistic neutron star radius of $R\approx$0.5\,km \citep{Klochkov:etal:13,Klochkov.etal:15}. Fitting the spectra with H-model atmospheres yield too small radii for this distance as well, but C atmospheres give acceptable radii \citep{Suleimanov.etal:14}. This results in $M = 1.55^{+0.28}_{-0.24}~\Msun$ and $R= 12.4^{+0.9}_{-2.2}$~km for $D=3.2$~kpc 
\citep[see Figure~\ref{sv_fig10};][]{Klochkov.etal:15}. %At larger distances the neutron star mass and radius increase, but these values remain preferable.
 
The neutron star in HESS\,J1731$-$347 also stands out in its thermal properties. The inferred temperature is much higher than that of other CCOs  ($T_{\rm eff,\infty} =1.78^{+0.04}_{-0.02}$\,MK), and unusual for its estimated age of $\approx 27$\,kyr \citep[see Figure\,\ref{sv_fig10};][]{Tian:etal:08}. Modern neutron star cooling theory 
\citep[e.g.][]{Weiskopf.etal:11} limits the radius to $R>12$~km, because smaller radii cannot produce the observed neutron star temperature for this age \citep{Klochkov.etal:15}. We note that agreement between the observed temperature and measured age of the CCO can only be achieved if the neutron star is covered by a thick C envelop ($\Delta M \sim 10^{-8}~\Msun$), and if the protons in the core are superfluid. Detailed consideration of  neutron star cooling theory leads to even stronger constraints of the neutron star parameters  \citep[see Figure\,\ref{sv_fig10} right;][]{Ofengeim.etal:15}.

 %%%%%%%%%%%%%
%%%%% Other types of constraints from EM observations  
\subsection{EOS constraints from measuring fast spin rates}\label{subsec:spin} 
Apart from mass and radius measurements, the spin of neutron stars can potentially also provide interesting constraints on the EOS. This is because very fast spin rates constrain the maximum neutron star radius \citep[e.g.][]{CST94}. %As with mass measurements, the tightest constraints are obtained from extremes.
The spin frequency of the neutron star must be lower than the Keplerian frequency, otherwise it would shed mass at its equator due to centrifugal forces. The mass-shredding limit depends on the EOS, $M$ and $R$ \citep[][]{haensel2009}:

\begin{equation}
f_{\rm max} = C_{\rm dev} \left ( \frac{M}{\Msun} \right ) ^{1/2} \left ( \frac{R}{\mathrm{10~km}} \right ) ^{-3/2}~\mathrm{kHz}.
\end{equation}
 
Since the deviation factor $C_{\rm dev}$ depends on the EOS, the above equation can provide a limit on $R$. The more compact the neutron star (i.e. the smaller $R$ for a given $M$), the higher the supported rotation can be. Therefore, softer EOSs allow for a higher spin rate \citep[][]{haensel2009}. 
 
The fastest spinning radio pulsar currently known is MSP J1748--2446ad, which has a rotation frequency of 716~Hz \citep[1.396 ms;][]{hessels2006}. The fastest known AMXP in an LMXB is IGR J00291+5938, which spins at 599~Hz \citep[][]{galloway2005}, whereas the highest rotation rate for a neutron star LMXB inferred from X-ray burst oscillations is 620~Hz for 4U 1608--52 \citep[][]{gallow08}. The centrifugal break-up frequency predicted by most EOSs is $f_{\rm max}\approx 1.5-2$~kHz \citep[e.g.][]{lattimer2011}. Rotation speeds $< 1$~ms would rule out certain families of hard EOSs, and would be particularly constraining in combination with a large mass. 

The lack of more rapidly spinning neutron stars has been taken as evidence that there is some mechanism that limits their spin-up \citep[e.g. spin-equilibrium, spin-down by magnetic dipole radiation, gravitational wave emission;][]{papaloizou1978,wagoner1984,bildsten1998_gravwav,melatos2005,disalvo2008,burderi2009,haskell2011,patruno2012_gravwav,patruno2017_spin}. However, there are also physical reasons that might make it difficult to find very rapidly spinning neutron stars, even if they exist. For instance, since accretion is thought to spin up neutron stars, high spin may be naively expected in sources that accrete at high rates. However, high accretion rates are also thought to suppress the magnetic field and in absence of channeled accretion no pulsations are expected to be produced \citep[e.g.][]{cumming2001,romanova2008}. Moreover, it has been suggested that strong accretion promotes spin alignment, which would weaken the pulsations \citep[e.g.][]{ruderman1991,lamb2009}. Furthermore, in case of X-ray bursters, very rapid spin may suppress flame spreading and make bursts shorter and weaker \citep[e.g.][]{spitkovsky2002,cavecchi2013}.

 %%%%%%%%%%%%%
%%%%% Other types of constraints from EM observations  
\subsection{Constraints from other types of electromagnetic observations}\label{subsec:other} 
The radio, optical, X-ray observations and analysis techniques discussed in Sections~3.1--3.7 provide the most direct constraints on neutron star masses and radii to date. However, there are several other observational phenomena that can also put some interesting constraints on the dense matter EOS, particularly with upcoming facilities (see Section~\ref{sec:future}). These are briefly discussed below.

\subsubsection{Mass measurements from glitches in young radio pulsars}\label{subsubsec:glitches} 
The most accurate mass measurements have been obtained for rapidly spinning radio pulsars in binary systems (Section~\ref{subsec:pulsars}). However, some constraints on the masses of  young, slowly spinning radio pulsars can be obtained from observing \textit{glitches}. These are a sudden increase in the spin period of young, slowly spinning radio pulsars \citep[e.g.][for a review]{espinoza2011}. This is thought to be caused by the fact that unlike the normal matter in the neutron star crust, the superfluid component does not slow down due to the electromagnetic energy losses \citep[][]{anderson1975}. Superfluids rotate by forming vortices that are usually ``pinned'' to the normal matter and the area density of these vortices determines the spin rate of the superfluid. The superfluid therefore acts as a reservoir of angular momentum. As the neutron star spins down, an increasing lag develops between the normal matter in the crust (rotating at the stellar spin rate) and the superfluid (rotating faster); once it reaches a critical value, the superfluid vortices will suddenly ``unpin'' and transfer angular momentum to the normal matter, explaining the observed jump in spin frequency that is referred to as a glitch.

Recently, two different approaches have been developed to measure neutron star masses from the angular momentum reservoir inferred from glitches \citep[e.g.][]{ho2015,pizzochero2017}. Firstly, \citet{ho2015} uses observable quantities inferred from X-ray and radio data such as the pulsar spin and its time derivative, the glitching activity and the temperature of the neutron star, and couples these to theoretical models for (temperature sensitive) superfluidity and the EOS to infer neutron star masses for about a dozen glitching radio pulsars. Secondly, \citet{pizzochero2017} show how the maximum observed amplitude and recurrence time of glitches can constrain the mass of nearly two dozen glitching radio pulsars when combined with microphysical models of the interactions between the normal and superfluid matter. This analysis showed that lower-mass neutron stars produce larger-amplitude glitches.

The mass measurements obtained via these means depend on the assumed EOS and are subject to a number of systematic uncertainties. However, future advances in theoretical modeling and radio/X-ray observing (see Section~\ref{sec:future}) allow to further develop these methods. In particular, if an independent mass measurement for a glitching pulsar can be obtained, that can be used to tightly constrain the EOS even if the neutron star mass is not extreme. This seems particular promising with the discovery of young radio pulsars in binaries \citep[e.g.][]{lyne2015}. For neutron stars that have a smaller mass, the moment of inertia in the crust will be larger and hence stronger amplitude glitches can be produced.

\subsubsection{Gravitationally redshifted lines and edges during X-ray bursts}\label{subsubsec:lines}
As discussed in Section~\ref{subsec:bursts}, the neutron star surface is visible during X-ray bursts. As the radiation from the X-ray burst, ignited in the accreted ocean/envelop, passes through a metal-rich atmosphere, this can potentially create absorption lines or edges \citep[e.g.][]{rajagopalromani1996,brown2002}. The rotational broadening of such a line depends on $R$, whereas its centroid energy depends on the ratio $M/R$. If the inclination is known, this leads to an independent measure of $M$ and $R$ \citep[e.g.][]{ozelpsaltis2003,Ozel06}. 

Narrow atomic features can be detected with high-resolution X-ray spectrographs such as e.g. the gratings aboard the currently active missions \chan\ and \xmm\ (see Section~\ref{subsec:xarm} for prospects with future missions). Since rapid rotation will further broaden narrow spectral features through Doppler smearing, this would be most promising for slowly spinning neutron stars. Several attempts in this direction have been made \citep[e.g.][]{kong2007,galloway2010_4u1728,intzand2013,intzand2017}, but only in one case narrow spectral features were claimed to be seen \citep[][]{cottam2002}. However, the later discovery of a high spin rate for this particular neutron star \citep[552~Hz;][]{galloway2010} rules out that the putative lines originated from the stellar atmosphere \citep[][]{lin2010}. Moreover, attempts to solidify the result by performing new observations failed to detect the features claimed in the initial study \citep[][]{cottam2008}.

\subsubsection{Radius lower limits from mHz QPOs}\label{subsubsec:mhzqpos} 
Quasi-Periodic Oscillations (QPOs) at mHz frequencies are detected for a handful of LMXBs and are thought to be associated with quasi-stable burning on the neutron star surface \citep[e.g.][]{revnivtsev2001,yu2002,heger2007,altamirano2008,linares2012,keek2014,lyu2014,lyu2016}. As recently argued by \citet{stiele2016}, the maximum black-body emitting radius measured during a QPO cycle provides a lower limit on the radius of the neutron star (it is uncertain whether the entire surface should be emitting). Applying this approach to 4U 1636--536 using \rxte\ data resulted in a lower limit of $R> 11$~km, after accounting for various uncertainties. Better constraints could be obtained using data from new missions such as \nicer\ and \hxmt\ (Section~\ref{sec:future}). A possible advantage is that the mHz QPOs are not expected to cause significant changes in the accretion flow that can complicate the radius measurements as in case of X-ray bursts \citep[e.g.][]{vanparadijs1986,ballantyne2005,intzand2013,ji2014_4u1608,worpel2015,degenaar2018}. 

\subsubsection{Radius upper limits from accretion disk reflection}\label{subsubsec:reflection}
The X-ray spectra of many neutron star LMXBs show broad emission lines near 6.5~keV that are interpreted as radiation that is reflected off the inner edge of the accretion disk \citep[e.g.][]{george1991,matt1991,fabian2010}. Accurate modeling of this Fe-K line, as well as the corresponding Compton hump near 20--30 keV, is widely used to infer the location of the inner edge of the accretion disk. In principle this provides an upper limit on the radius of the neutron star, since the disk must truncate at the stellar surface if not before \citep[e.g.][]{bhattacharyya2007,cackett2008_iron}. In several LMXBs the inner disk radii measured from reflection analysis appear to be significantly truncated away from the neutron star, which is attributed for instance to the presence of a geometrically thick boundary layer where the accretion flow impacts the stellar surface \citep[e.g.][]{dai2014_reflection,ludlam2017_rxj1709}, the magnetic field of the neutron star \citep[e.g.][]{degenaar2014,vandeneijnden2017}, or evaporation of the inner accretion disk at low mass-accretion rates \citep[e.g.][]{papitto2013_hete}. However, in about a dozen sources the inferred inner disk radii are small and may suggest that the disk is running into the neutron star surface, hence providing an upper limit on the stellar radius \citep[e.g.][]{miller2013,degenaar2015,chiang2016,ludlam2016,ludlam2017}. This information can be useful when combined results obtained from other techniques (see Section~\ref{subsec:combined}).

\subsubsection{Radius and mass upper limits from kHz QPOs}\label{subsubsec:khzqpos} 
Like disk reflection, another way to obtain an upper limit on the neutron star radius from studying the accretion flow properties are kHz QPOs, which are detected for about two dozen neutron star LMXBs. Although the origin of this rapid variability (which can be as fast as $\sim 10^3$~Hz) is not well understood, it is typically linked to the very inner accretion disk. Requiring an observed kHz QPO frequency to be lower than the Keplerian frequency at the neutron star surface places an upper limit on the neutron star radius \citep[][]{miller1998}. For most EOSs, however, the neutron star radii lie within the innermost stable circular orbit (ISCO); the Keplerian frequency at the ISCO is set by the neutron star mass, and therefore requiring that the kHz QPO frequency is lower than that at the ISCO (i.e. assuming that the QPO is produced at or outside the ISCO) places an upper bound on the neutron star mass as well \citep[][]{kluzniak1990,miller1998}. However, the limits obtained in this way are not particularly constraining \citep[e.g.][]{vanstraaten2000}. More precise mass measurements would in principle be possible  \citep[e.g.][]{stella1999,psaltis1999,barret2006}, but heavily relies on the specific interpretation of the kHz QPO, of which there is no consensus.

\subsubsection{Core cooling of transiently accreting neutron stars}\label{subsubsec:corecool}
As discussed in Section~\ref{subsec:cco} in the context of thermally-emitting isolated neutron stars, the temperature of the neutron star core is set by the rate at which it is cooling through neutrino emissions. This is related to its mass since more massive neutron stars should have higher interior densities that lead to more efficient neutrino cooling  \citep[rendering more massive neutron stars colder; e.g.][]{YP04,page2006}. Comparing the inferred temperatures and ages of a number of isolated neutron stars with theoretical calculations of their thermal evolution shows that their cores are likely not dense enough to allow for very efficient cooling mechanisms and hence does not point to particularly massive neutron stars \citep[e.g.][see also Figure~\ref{sv_fig10}]{YP04,page2009}. This is in sharp contrast with neutron stars in LMXBs for which similar types of tests (see below) suggest that efficient core cooling is taking place and would thus point to more massive neutron stars. However, it is possible that the lack of very cool objects among the isolated neutron stars is a selection effect, since relatively high temperatures are required to detect and identify them as neutron stars. 

When located in LMXBs, neutron stars can be reheated through the nuclear reactions that are induced in the crust during accretion episodes (Section~\ref{subsec:accretion}) and can heat the core on a thermal timescale of $\sim 10^4$~yr \citep[][]{brown1998,colpi2001,wijnands2013}. In such systems, knowledge or estimates of the accretion history can then be compared to their core temperature \citep[e.g.][]{PCY97,brown2002,YP04}, to determine the rate of neutrino cooling. This suggests that enhanced cooling mechanisms should be operating in several neutron stars and hence that these objects should be relatively massive \citep[e.g.][]{YP04,heinke2009,wijnands2013,han2017}. Although this approach does not allow for accurate mass measurements and there are many systematic uncertainties both in the observations and the models \citep[see][for discussions]{wijnands2013,han2017}, it does provide means to pick out potentially massive neutron stars that could be interesting objects for other types of studies (e.g. optical dynamical mass measurements; see Section~\ref{subsec:opticalnew}).

\subsubsection{The potential of crust cooling of transiently accreting neutron stars}\label{subsubsec:crustcool}
As discussed in Section~\ref{subsec:accretion}, the crust of a neutron star is heated during accretion phases due to nuclear reactions. Dedicated X-ray monitoring of $\approx 10$ transient LMXBs in quiescence following accretion outbursts have revealed a steady decrease in the thermal X-ray flux and inferred neutron star temperature over the course of about a decade. This has been ascribed to the thermal relaxation of the accretion-heated crust \citep[e.g.][for reviews]{wijnands2013,wijnands2017}. Such a cooling trajectory depends on the properties of the outburst (which determines how long and intense the crust was heated), the microphysics of the crust such as its structure and composition (which set the thermal conductivity and the nuclear reactions), and the thickness of the crust. The latter is of particular interest, since it is determined by the surface gravity. Therefore, if all the microphysics of neutron star crusts were understood and the outburst is closely monitored (as is often the case nowadays), the only free parameter determining the cooling trajectory is the compactness of the neutron star \citep[e.g.][]{wijnands2013,deibel2015}. %, as is the case for the cooling crust of a newly born neutron star \citep[][see also Section~\ref{subsec:cco}]{lattimer1994}. 

Although there are still many uncertainties about the thermal and transport properties of neutron star crusts \citep[e.g.][for a review]{page2012}, studies to improve this are well under way and are alongside providing interesting constraints on other physical properties of neutron stars. For instance, these studies can also reveal the presence of non-spherical shapes for the nucleii at the bottom of the crust \citep[referred to as nuclear pasta; e.g.][]{horowitz2015,deibel2017}. Moreover, some interesting constraints on the core heat capacity have recently been obtained: this approach can potentially lead to more stringent constraints in the future and limit the number of baryons in the core that can be bound in a superfluid \citep[][]{cumming2017,degenaar2017_hete}. Crust cooling studies are therefore an interesting avenue to learn more about neutron star crusts as well as their cores, and in principle also have the potential to lead to constraints on the compactness of these neutron stars as well as the dense matter EOS.

\subsubsection{Constraints from magnetars}\label{subsubsec:magnetars}
Although the heating mechanism is likely different, it appears that crust cooling is also observed in transient magnetars \citep[of which about two dozen are known; e.g.][for a recent overview]{cotizelati2017}. 
%lyubarsky2002,kouveliotou2003,guver2007,pons2012,rea2009,rea2012,rea2013,scholz2012,scholz2014,cotizelati2015,
Such studies may provide similar prospects as the crust cooling studies of transiently accreting neutron stars (Section~\ref{subsubsec:crustcool}). Furthermore, magnetars show glitches \citep[e.g.][]{dib2008,dib2014}, which can potentially be used to measure their masses in a similar fashion as done for young radio pulsars (Section~\ref{subsubsec:glitches}). Finally, on rare occasions QPOs have been detected during active flaring episodes of magnetars and ascribed to seismic vibrations \citep[e.g.][]{israel2005,watts2006,huppenkothen2013}. It has been proposed that these magnetar QPOs offer a view into the neutron star $M$ and $R$ \citep[e.g.][]{strohmayer2005,watts2007,steiner2009,gabler2012}, but the mode frequencies also depend on unknown factors such as the magnetic field strength, superfluid properties, and crust composition. Moreover, there is some controversy over the interpretation of magnetar QPOs \citep[see e.g.][for a discussion]{watts2016_review}.

%%%%%%%%%%%%%%%%%%%%%%
%% COMBINING METHODS
%%%%%%%%%%%%%%%%%%%%
\subsection{Combining different methods for improved constraints}\label{subsec:combined}  
Radio pulsar timing has been combined with optical observations to obtain very accurate mass constraints for a number of radio pulsars (see Section~\ref{subsec:dynamical}). Moreover, constraints from samples of X-ray bursters and quiescent LMXBs have been combined in statistical frameworks to obtain accurate radius constraints (see Section~\ref{subsubsec:globclust}). This underlines the power of combining different, complementary techniques for mass and radius measurements. In this section we briefly explore which other methods can be combined to obtain improved EOS constraints. A cross-comparison between techniques is also very important to identify and better understand the systematic uncertainties subject to each approach.

It would be valuable to obtain $M$ and $R$ for a single neutron star both from its quiescent thermal emission and its X-ray bursts. However, for the quiescent method mostly globular cluster sources have been used that have never been seen to exhibit an accretion outburst (and hence X-ray bursts). Conversely, the quiescent emission of X-ray bursters is often either contaminated by a hard spectral component, or so dim that no accurate constraints can be obtained. Nevertheless, a few attempts have been made, e.g. for the prolific transiently accreting neutron star Aql X-1, which has been observed in quiescence numerous times \citep[e.g.][]{cackett2011,campana2014}. Analysis of its quiescent spectra were compared to that of a number of PRE bursts \citep[][]{li2017_aqlx1}. Both methods led to reasonably constrained $M-R$ confidence intervals, but these overlap only marginally: the constraints from the quiescent analysis are shifted to lower radius and mass compared to the constraints inferred from the burst analysis. This can likely be attributed to systematic effects (see Section~\ref{subsubsec:qbiases}). Another source for which quiescent and X-ray burst measurements can potentially be combined is 4U 1608--52. It displays an accretion outburst every few years during which it shows (PRE) X-ray bursts \citep[e.g.][]{Poutanen.etal:14}, and is relatively bright in quiescence (albeit exhibiting a power-law spectral component). Moreover, 4U~1608--52 also exhibits a number of other phenomena that can potentially lead to $M$ and $R$ constraints such as disk reflection and mHz QPOs (see below).

It can also be interesting to combine lower limits on the neutron star radius obtained from the mHz QPOs (Section~\ref{subsubsec:mhzqpos}) with upper limits inferred from disk reflection modeling (Section~\ref{subsubsec:reflection}) or kHz QPOs (Section~\ref{subsubsec:khzqpos}). For instance, reflection studies constrain the radius of the neutron star in 4U~1636--536 to $R\lesssim$11~km \citep[e.g.][]{ludlam2017}, whereas the lower limit inferred from its mHz QPOs is $R\gtrsim$11~km \citep[][]{stiele2016}. Without scrutinizing the systematic errors and assumptions of both methods, it is striking that these two independent approaches come together at the same value for the radius. It is conceivable that during the mHz QPO the entire surface was radiating and that the disk in 4U~1636--536 is truly truncating near the neutron star surface, hence that the stellar radius is $R\approx$11~km. Another example of a neutron star that displays mHz QPOs \citep[][]{yu2002} and possibly has the disk running in the neutron star surface \citep[][]{degenaar2015} is 4U~1608--52. Combining these different X-ray techniques can thus possibly bracket the radius measurement for this neutron star. It also displays PRE bursts \citep[][]{Poutanen.etal:14} and burst oscillations \citep[at 620~Hz;][]{gallow08}, and is X-ray bright in quiescence. It could therefore also be a good target for pulse profile modeling with the new mission \nicer\ (Section~\ref{subsec:nicer}) and X-ray polarization studies (Section~\ref{subsec:polarization}). Finally, the companion stars of some transient neutron star LMXBs may also be bright enough to be studied in quiescence at optical wavelengths, especially with the future generation of instruments (Section~\ref{subsec:opticalnew}). This can provide independent and complimentary constraints on the neutron star mass.

For accreting neutron stars, X-ray pulse profile modeling can be performed in two different ways. For the AMXPs, we observe emission from hotspots at the magnetic poles (in addition to emission from a shock that forms just above the surface as the rapidly in-falling material is abruptly decelerated). For burst oscillations, on the other hand, hotspots arise due to unstable thermonuclear burning zones on the stellar surface (i.e. not confined to the magnetic poles). Several neutron stars show both coherent X-ray pulsations and burst oscillations; ideally, one would want to use both types of hotspots to model the resulting pulse profile and obtain $M-R$ constraints, to check for consistency and to calibrate both methods \citep[][]{watts2016_review}. An important breakthrough could also be provided by the detection of surface atomic lines (Section~\ref{subsubsec:lines}) in the hotspot emission of a neutron star for which the pulse profile (Section~\ref{subsec:timing}) can also be accurately modeled \citep[e.g.][]{rauch2008}. Combining these two pieces of information yields complementary and independent measurements of $M$ and $R$. Lastly, some radio pulsars with accurate $M$ measurements show thermal emission that may allow for a $R$ measurement through X-ray pulse profile modeling with \nicer\ (see Section~\ref{subsec:nicer}).

%%%%%%%%%%%%%%%%%%%%%%%%%%%%
% CHALLENGES + FUTURE 
%%%%%%%%%%%%%%%%%%%%%%%%%%%%
\section{Future prospects with new and upcoming instrumentation}\label{sec:future}
Over the past decade we have witnessed significant developments in inferring the dense matter EOS from electromagnetic observations of neutron stars. We have started to gain significant constraints on the pressure-density relation and our knowledge of the superfluid properties of their interiors is steadily growing. However, it is at present not yet possible to infer the composition of the dense core of neutron stars (i.e. nuclear versus exotic matter). The most important challenges that presently limit tighter constraints are systematic uncertainties, limited data quality, and small number statistics. Below we give an overview of the exciting prospects of new and future instrumentation  to continue and improve determinations of neutron star masses and radii in the next decade and beyond.

\vspace{-0.4cm}
\subsection{The future generation of radio telescopes}\label{subsec:futureradio} 
\subsubsection{Significantly increasing the number of mass measurements of radio pulsars}\label{subsubsec:ska}
The Square Kilometre Array (\ska) will be the world's largest radio telescope and is expected to begin science operations in the early 2020s. The \ska\ can provide major breakthroughs in neutron star research, including EOS constrains \citep[][]{watts2015_review}. In particular, it is expected to discover significant numbers of radio pulsars. This allows for an increased number of mass measurements, including those with extreme properties that put the most stringent constraints on the EOS \citep[e.g.][see also Section~\ref{subsec:highspin}]{keane2015,hessels2015}. Furthermore, the precise timing techniques that are being developed for pulsar timing arrays (PTAs) have started to yield accurate masses for more pulsars and may eventually lead to finding more extreme ones \citep[e.g.][]{reardon2016,fonseca2016}.

\subsubsection{Measuring the moment of intertia of radio pulsars}\label{subsubsec:inertia}
Radio and gamma-ray observations of radio pulsars have the potential to measure their moment of inertia. Since this is a function of both $M$ and $R$, it allows the radius to be measured if the mass can be determined independently (e.g. via radio pulsar timing and/or optical studies of the companion star). Attempts have been made for PSR J1614--2230, one of the two radio pulsars with an accurate high mass measurement of $M\approx 2~\Msun$ \citep[][]{demorest2010,fonseca2016}. Unfortunately, the spin-down luminosity inferred from its gamma-ray emission does not provide very tight constraints and this is not likely to improve with new instrumentation \citep[][]{watts2015_review}. However, long-term observations of the double pulsar system PSR J0737--3039 will eventually lead to a measurement of the moment of inertia of one of the pulsars, resulting in a radius constraint with $\approx$5\% accuracy \citep[][]{lyne2004,lattimer2005,kramer2009}. As mentioned by \citet{watts2015_review}, the \ska\ (Section~\ref{subsubsec:ska}) can possibly discover more pulsar systems for which the moment of inertia will be measurable, but this is likely challenging due to the highly restrictive requirements for the system geometry.

\subsection{The new and future generation of optical telescopes}
\subsubsection{Accurate distance determinations with \gaia}\label{subsec:gaia} 
As discussed in Section~\ref{subsubsec:qbiases}, one of the most important systematic uncertainties in inferring radii from neutron stars in LMXBs, is that their distance is usually not accurately known. \gaia, launched in 2013, is an astrometry mission dedicated to measure the parallaxes of stars with unprecedented precision \citep[][]{gaia2016}. In a few years, this may provide accurate distances for a number of neutron star LMXBs, either located in the field or in globular clusters. This will strongly reduce the systematic uncertainties in the radii determined for these objects from X-ray observations.

\subsubsection{Dynamical mass measurements with the next generation optical telescopes}\label{subsec:opticalnew}
The highly improved sensitivity of upcoming optical facilities such as the European Extremely Large Telescope (\eelt), the Thirty Meter Telescope (\tmt) and the Large Synoptic Survey Telescope (\lsst) will be transformable in obtaining dynamical mass measurements for neutron stars in binary systems. Synergies with the \ska\ are expected to be particularly promising \citep[e.g.][]{antoniadis2015}. 

% LSST \citep[e.g.][]{ivezic2008}

\subsection{Advances from X-ray astronomy}
\subsubsection{Radius measurements from X-ray bursts with new X-ray telescopes}\label{subsec:burstsnew}
Radius constraints from X-ray bursting neutron star LMXBs have all been obtained with the wealth of data provided by NASA's \rxte, which was decommissioned in 2012. However, in 2015 the Indian satellite \astrosat\ was launched \citep[][]{singh2014} and in 2017 the Chinese Hard X-ray Modulation Telescope (\hxmt) was brought into orbit \citep[][]{zhang2014}. Their X-ray detectors have similar capabilities to those of \rxte, and both missions thus provide continued opportunities to measure neutron star radii from X-ray bursts. In addition, the \nicer\ mission installed in 2017 (see Section~\ref{subsec:nicer}) is a very promising tool for X-ray burst studies \citep[e.g.][]{keek2016_future}, and can potentially also lead to more accurate radius constraints for bursting neutron stars. 

\subsubsection{Radius constraints of dim thermally-emitting neutron stars with \athena}\label{subsec:athena}
Further in the future, currently planned for launch in the late 2020s, we will have access to the ESA mission \athena\ \citep[][]{barcons2017}. Although dense matter is not a core science goal of \athena, its very high collective area, soft X-ray coverage and good spectral resolution allows for more accurate modeling of the thermal spectra of quiescent neutron stars, studying the cooling tails of X-ray bursts\footnote{Although pile-up can be an issue \citep[e.g.][]{keek2016_future}.}, 
modeling the pulse profiles of different kinds of pulsars, and searching for gravitationally redshifted lines in the surface emission of neutron stars \citep[][]{motch2013}.  Furthermore, \athena\ might provide more stringent limits on the pulsed fraction of quiescent neutron star LMXBs, which is of high importance to assess the possible presence of temperature inhomogeneities  \citep[][]{elshamouty2016_2}.

\subsubsection{High-resolution X-ray spectroscopy to search for atomic spectral features}\label{subsec:xarm}
Detecting gravitationally redshifted atomic features during X-ray bursts or low-level accretion activity, which would constrain the compactness of a neutron star (Section~\ref{subsubsec:lines}), requires high-resolution X-ray spectrographs. This is incorporated in the design of \athena\ (the X-ray Integral Field Unit, X-IFU) and was provided by the short-lived mission \hitomi, which may find follow-up with the X-ray Astronomy Recovery Mission (\xarm). There are thus some future prospects for searching for narrow, gravitationally redshifted features from neutron star atmospheres.

\subsubsection{Accurate X-ray pulse profile modeling with \nicer}\label{subsec:nicer} 
The Neutron Star Interior Composition ExploreR (\nicer) is a NASA mission that was successfully installed on the International Space Station in 2017 June \citep[][]{Gendreau:2012,Arzoumanian:2014}. It covers the energy range from 0.2--12\,keV, has a very high effective area and unprecedented timing precision (absolute time-tagging of $<$300\,ns). \nicer\ is dedicated to achieve precise ($\approx$5\%) mass and radius measurements for a few selected neutron stars through highly accurate pulse-profile modeling of pulsars \citep[e.g.][see also Section~\ref{subsec:timing}]{sb16,Miller16,ozel2016_nicer}. Moreover, neutron stars that display rapid oscillations during the rise of an X-ray burst are very promising targets to obtain EOS constraints with \nicer\ \citep[e.g.][]{watts2016_review}. Obtaining independent constraints on geometrical factors significantly improves the constraints obtained from such pulse profile modeling. This can uniquely be achieved from X-ray polarization studies, of which several concepts are currently being investigated (Section~\ref{subsec:polarization}).

\subsubsection{Geometrical constraints from X-ray polarization}\label{subsec:polarization}
As discussed in Section~\ref{subsec:timing}, pulse profile modeling is a very promising technique to obtain mass and radius constraints without systematic errors, if a number of geometrical factors can be determined. Fortunately, the hotspot emission that is used by this technique is expected to be polarized \citep[e.g.][]{rees1975,meszaros1988,viironen2004}. Phase-resolved measurements of the angle and degree of polarization can then constrain both the inclination angle of the observer and the hotspot, thereby breaking degeneracies that current limit pulse profile modeling. AMXPs and X-ray burst oscillation sources are prime targets for such polarimetry studies. The concept missions \extp\ \citep[China/Europe;][]{zhang2016_extp}, \ixpe\ \citep[NASA;][]{weisskopf2016_ixpe}, and \xipe\ \citep[ESA;][]{soffitta2016_xipe} have X-ray polarimeters in their design concepts that would facilitate such studies.

\subsection{Searching for the most rapidly spinning radio and X-ray pulsars}\label{subsec:highspin}
As discussed in Section~\ref{subsec:spin}, very high spin rates also have the potential to put interesting constraints on the neutron star EOS. The future holds great prospects for continuing and enhancing searches for rapidly spinning neutron stars. At radio wavelengths, such advances will be brought through timing studies with \ska\ and pulsar timing arrays, whereas at X-ray wavelengths such studies are currently facilitated by searching for coherent X-ray pulsations and X-ray burst oscillations with \astrosat, \hxmt\ and \nicer. Several mission concepts currently under investigation would also facilitate such studies. For instance, in the current design both \extp\ and \strobex\ would have a higher effective area than \rxte, and hence allow for detecting weaker pulsations \citep[e.g.][]{zhang2016_extp,wilsonhodge2016}.

%%%%%%%%%%%%%%%%%%%%%%%%%%%%
% CONCLUSIONS 
%%%%%%%%%%%%%%%%%%%%%%%%%%%%
\section{Conclusions}\label{sec:conclusions}
Neutron stars are unique, natural laboratories to constrain the EOS of ultra-dense matter. In particular, measuring the mass and radius of several neutron stars with $<$10\% errors can place strong constraints on the EOS \citep[e.g.][]{ozel2010,steiner2010}. Such measurements are facilitated by observations at radio, optical and X-ray wavelengths. Precise and reliable mass measurements have so far only been obtained for radio pulsars ($\sim$40 objects with masses ranging between $M \approx 1.2-2.0~\Msun$), whereas radius measurements have only been obtained for about two dozen neutron stars in LMXBs ($\approx$15 objects leading to an overall estimate of $R \approx$10--12~km for an assumed mass of $M\approx 1.4-1.5~\Msun$). 

Whereas the masses of radio pulsars are determined with high accuracy, the radius measurements of neutron star LMXBs are more strongly model dependent and subject to a number of systematic uncertainties. Nevertheless, over the past decade much progress was made in obtaining masses and radii from the thermal X-ray emission of neutron star LMXBs. Alongside, there have been important theoretical developments, including a detailed assessment of spin effects \citep[e.g.][]{BPOJ12,Baubock.etal:15}, advanced calculations of atmosphere models \citep[e.g.][]{Ho:Heinke:09,SPW11,SPW12,Netal15} and the development of statistical analysis methods \citep[e.g.][]{ozel2010,ozel2015,steiner2010,steiner2013}. Moreover, there are many different techniques that can ultimately lead to constraints on the neutron star EOS. Several approaches are still under development and can be improved with new and upcoming observatories. There lies great power in applying different techniques to individual neutron stars: orthogonal constraints can be obtained or cross-checks can be performed that allows us to address the systematic uncertainties of different methods.

Finally, it is important to note that although probing the behavior of ultra-dense matter is typically the main scientific driver to try and constrain the neutron star EOS, there are much wider implications. The behavior of matter near and beyond the nuclear density is governed through the strong interactions and understanding this fundamental force is important for several other areas of astrophysics. For instance, it plays an important role in core-collapse supernova explosions, the dynamics of compact object mergers that involve a neutron star and the formation timescale of black holes, as well as the precise gravitational wave and neutrino signals produced in these processes, the resulting mass loss and nucleosynthesis, and associated $\gamma$-ray bursts and hypernovae. The recent discovery of gravitational wave signals and electromagnetic counterparts of the merger of two neutron stars has ever more increased interest in constraining the neutron star EOS.

%%%%%%%%%%%%%%%%%%%%%%%%%%%%
% ACKNOWLEDGEMENTS
%%%%%%%%%%%%%%%%%%%%%%%%%%%%
\begin{acknowledgements}
The authors acknowledge support from NewCompStar COST Action MP1304. ND is supported by a Vidi grant from the Netherlands Organization for Scientific Research (NWO). VS is supported by Deutsche Forschungsgemeinschaft (DFG) grant WE 1312/51-1. For the creation of Figure~\ref{fig:eosmr}, the authors thank Morgane Fortin for providing unified EOSs and Feryal \"{Ozel} for making her compilation of EOSs publicly available online.\footnote{http://xtreme.as.arizona.edu/NeutronStars}
\end{acknowledgements}

\vspace{-0.8cm}
\bibliographystyle{apj}

\end{document}